# Research Report

**on**

Assessment of Consumers' Awareness of Nutrition, Food Safety and Hygiene in the Life Cycle for Sustainable Development of Bangladesh: A Study on Sylhet City Corporation.

Professor Dr. Md. Nazrul Islam

Professor, Department of Business Administration

**June-2021**

# Assessment of Consumers' Awareness of Nutrition, Food Safety and Hygiene in the Life Cycle for Sustainable Development of Bangladesh: A Study on Sylhet City Corporation [1]

[1] Funded by University Research Centre, Shahjalal University of Science and Technology through Promotional Research Grant–One year

**Submitted to**

**University Research Centre**

**Shahjalal University of Science and Technology, Sylhet-3114**

Project code: **SS/2020/1/11**

Project category: Promotional Research Grant – One year

**Submitted by**

**Principal Investigator: Professor Dr. Md. Nazrul Islam**
Department of Business Administration,
Shahjalal University of Science and Technology, Sylhet-3114

**Research Assistant**
**Fazlur Rahman**
MS in Statistics, Department of Statistics,
Shahjalal University of Science and Technology

**June, 2021**

# Acknowledgement

The project was funded by University Research Centre , Shahjalal University of Science and Technology Sylhet-3114, so my first I show my highest gratitude to the Director of the center and also its members for making research opportunity by granting the fund. I am also indebted to the data collectors and research assistant their sincere cooperation in conducting the research.

(Professor Dr. Md. Nazrul Islam)

Professor, Department of Business Administration

**and**

**Principal Investigator of the project.**

# Assessment of Consumers' Awareness of Nutrition, Food Safety and Hygiene in the Life Cycle for Sustainable Development of Bangladesh: A Study on Sylhet City Corporation

## Abstract

The objective of the study is to explore the knowledge of consumers on nutrition, food safety and hygiene of the sampled respondents in the Sylhet City Corporation (SCC) in Bangladesh in terms of different demographic characteristics. This study includes the different types of consumers in the life cycle viz. baby, child, adolescent, young and old so that an overall awareness level can be measured in the urban area of Bangladesh. The study is confined to SCC area in which all types of respondents has been included and findings from this study will be used generally for Bangladesh in making policy In conducting the study the population has been divided into six group as: (i) Baby; (ii) child; (iii) adolescent; (iv) parental; (v) unmarried adult young and (vi) married adult matured. We find that the average score of awareness of food nutrition and hygiene of unmarried adult is higher than that of married adults. The study suggested it is needed to increase awareness in of the parents for feeding babies. The average awareness of parents to their child's eating behavior between 5-9 years is 3.36 out of 5. The awareness is around 67% so we should be more careful in this regard. The average awareness adolescent (10-19 years) food habit is 1.89 on three points scales which about 63% only. Therefore, the consciousness of adolescent (10-19 years) has been to increase in taking food. The average feeding styles of parents is 4.24 out of 5 to their children up to 9 years and in percentage it is 84%.

# Executive Summary

The food systems approach initiated in the 2011 First Country Investment Plan (CIP-1) under its third component of food utilization through program later in the Second Country Investment Plan which focused on Nutrition Sensitive Food Systems reflects the multi-sector collaboration for ensuring Nutrition, Food safety and Hygiene (NFSH). Achieving adequate nutrition and food security for all entails that the right foods are available, and that people are able to physically and financially access those foods. But an underlying assumption is that people are willing to eat the right foods and prepare and to eat them in a manner that conserves their nutritious value and that is hygienic. Thus, Bangladesh still consumes far more rice per capita than recommended by dietary guidelines. Feeding and hygiene practices are often inadequate with dire consequences for small children's nutritional status and development. This is true across socio-economic categories. The World Food Summit-1996 emphasized on three important components for ensuring food security viz. availability of adequate food, stability in food supplies, and access to food and nutrition security. The present government of Bangladesh has given emphasis on poverty alleviation and elimination including all forms of disparity at the forefront of country's development strategy. The ending of poverty and hunger is one of the prime objectives of the Sustainable Development Goals (SDGs) to be implemented by 2030. In consonance with the SDGs, the government of Bangladesh has adopted Seventh Five Year Plan (SFYP), in which one of the targets has been fixed to bring down extreme poverty to around 8.9% by the year 2020. In the SFYP the government has identified that there is persistent micronutrient deficiencies, lack of public awareness, maternal under nutrition, acute malnutrition and lack of dietary diversity are among other challenges in the nutrition program sub-sector. In Bangladesh there are very few studies and all the studies are made on specific group on the awareness of food safety, nutrition and hygiene of the consumers. But the present study is an in-depth study which has included the different types of consumers in the life cycle viz. baby, child, adolescent, young and old so that an overall awareness level can be measured in the urban area of Bangladesh. The study is confined to SCC area in which all types of respondents has been included and findings from this study will be used generally for Bangladesh in making policy. The core objective of the study is to explore the knowledge of consumers on nutrition, food safety and hygiene of the sampled respondents in the Sylhet City Corporation in Bangladesh in terms of different demographic characteristics . In conducting the study the population has been divided into six group as: (i) Baby; (ii) child; (iii) adolescent; (iv) parental; (v) unmarried adult young and (vi) married adult matured. Separate questionnaire for each group has been used for collecting data. Total sample size is 234 under the six group as (i) Baby-40; (ii) child-40; (iii) adolescent-34; (iv) parental-34; (v) unmarried adult young -40 and (vi) married adult matured-40. The collected data has been analysed by different techniques of (Frequency distribution mean and standard deviation of the variables have been computed to get the picture of the data. Tables and Graphs) and inferential statistics (Principal Component analysis (PCA) and two sample t-test have been used for conclusion about the data) have also been used to present the data. The report is divided into three chapters where first chapter has comprehended the introductory aspects of the study such as problem statement, methodology and review of related literature . The second chapter comprises the analysis and presentation of the collected data with statistical methods and techniques. Finally the third chapter has depicted the findings and suggestions of the study.

In order to find the Baby's Eating Behaviour up to 4(Four Year) Responded by Parents 20 mother and 20 father were interviewed with a questionnaire with five point scale as Table- never =1 rarely=2 sometimes =3 "Baby often=4 and Always =5. The average frequency in percentage in the different scales are 23%, 30%, 16%,11%and 11% respectively .There were 18 scaling questionnaire and the average score of the items is 100.5 and the average score of the respondents is 2.51. It means the eating behaviour of baby lies between rarely and sometimes. Reliability of the questionnaire was checked by Cronbach's Alpha test which is 0.710 that means there is consistency of the items or variables. The PCA reduces 21 items into six factors significant for eating behaviours of parents. In assessing the Child's eating behaviour between 5-9 years similarly 20 fathers and 20 mothers were given questionnaire with 35 items or variables in the five point scale like previously. The average percentage of frequency distribution in the five point scales are 20%,27%, 23%, 16% and 19% respectively. The average of items is 117.66 and the average of respondents is 3.36. That means the eating behaviour of children varied between sometimes and often. Through the PCA 35 items are reduced to 10 factors. The food habits of adolescent (10-19) is measured in terms of given items or variables with three point scale as Most like=2; like=1 and Neutral=0; with 23 items. The average percentages of frequency in the scales are 37%, 55% and 8% respectively. The data showed that that the average of score of the items is 43.56 and the average score of respondents is 1.89. It means the choice varies between most like and like. The PCA reduces 21 items to eight factors of food habits of adolescent. Parental Feeding Style to his Children up to 9 years has been measured with five point scale 27 items. The average percentage of frequency distribution in the five point scales are 15%, 10%, 33%, 13% and 30% respectively .The Average of items is 114.52 and the average of respondents is 4.24 that mean the parental Feeding Style to his Children up to 9 years varies between often and always. The total variance explained by the PCA is 78.96% that means the PCA is accepted and based on Eigen values seven items are considered important. The food preference of adults unmarried (20-29) has also been measured by the same five pint scale with 62 items. The average percentages of frequency distribution in the different scales are 21%, 8%, 17% 31% and 24% respectively. The Average of items is 131.35 and the average of respondents is 2.12 that mean the food preference of adults (20-29 years) varies between neither like or dislike and like a little. The Food Preference for Adults (29-64) Married Male and Female has been measured by the same questionnaire as adult unmarried that means 62 items with five point scale. The average percentages of frequency distribution in the different scales are 37%, 16%, 16% 18% and 14% respectively The Average of items is 102.87 and the average of respondents is 1.66 that means the food preference of adults (29-64 years) varies between dislike a little and neither like nor dislike. A Comparison of food preference between adults of 20-29 unmarried and 29-64 married is made with two sample t-test to confirm whether there is any similarity between these two groups. We found that mean score of unmarried and married are 191.10 and 102.87 correspondingly with Std. deviation 89.873 and 39.889 for unmarried and married respectively. Therefore the average score of unmarried adult is higher than that of married adults. Levene's Test for Equality of Variances and t-test for Equality of means were used also. In the first test i.e. Levene's Test for Equality of Variances here our null hypothesis is that the both the variances are equal and alternative is that the variances are different for the two groups. The Levene's Test for Equality of Variances showed the p- value is 0.00 that means there are significant differences of variances of the two groups. The second test is t-test for Equality of Means which is presented in the second row and showed that the observed test statistics is 7.065 with p- value 0 .000. In this case we can also reject the null hypothesis and can conclude that there is significant difference between the means of the two groups. Finally we can conclude that there is significant difference between the food preference of unmarried adults and married adults and their preferences s are not same. The overall score of awareness of food nutrition and hygiene of six groups are : (i) Parents of baby up to 4 years = 2.51(in 5 point scale); (ii) Parents of Child eating behaviour between 5-9 years = 3.36(in 5 point scale); (iii) Adolescent (10-19 years ) Food Habits= 1.89(in 3 point scale); (iv) Parental Feeding Style to his Children up to 9 years= 4.24(in 5 point scale);; (v) Food preference of adults unmarried (20-29)=2.12 (in 5 point scale); and (vi)Food Preference for Adults (29-64) Married Male and Female=1.61(in 5 point scale). The study suggested it is needed to increase awareness in of the parents for feeding babies. The average awareness of parents to their child's eating behaviour between 5-9 years is 3.36 out of 5. The awareness is around 67% so we should be more careful in this regard. The average awareness adolescent (10-19 years) food habit is 1.89 on three points scales which about 63% only. Therefore the consciousness of adolescent (10-19 years) is be to increase in taking food. The average feeding styles of parents is 4.24 out of 5 to their children up to 9 years and in percentage it is 84% . Therefore it is about to satisfactory.

# Table of Contents

# List of Tables

# List of Figures

# List of Acronyms

| | |
|---|---|
| BBS | Bangladesh Bureau of Statistics |
| BHWDB | Bangladesh *Haor* and Wetland Development Board |
| CBN | Cost of Basic Needs |
| CEGIS | Center for Environment and Geographic Information Services |
| DCI | Direct Calorie Intake |
| DID | Difference-in-difference |
| FEI | Food Energy Intake |
| FGD | Focus Group Discussions |
| GoB | Government of Bangladesh |
| HH | Household |
| HIES | Household Income and Expenditure Survey |
| LPL | Lower Poverty Line |
| LPL | Lower Poverty Line |
| MAS | Multi-agent systems |
| MC | Micro-credit |
| MDGs | Millennium Development Goals |
| MFI | Micro Finance Institution |
| MPI | Multidimensional Poverty Index |
| NGO | Non-Government Organization |
| PCA | Principal Component Analysis |
| SACCOS | Savings and Credit Co-operative Societies |
| SFYP | Seventh Five Year Plan |
| SMCP | Savings and Micro Credit Program |
| SPSS | Statistical Package for Social Sciences |
| UP | Union Parishad |

**1.1.Inroduction :** The food systems approach initiated in the 2011 First Country Investment Plan (CIP-1) under its third component of food utilization through program 10, 11 and 12 where total fund required were 736 billion US $ (GoB, 2011). Later in the Second Country Investment Plan (CIP-2) which focused on Nutrition Sensitive Food Systems reflects the multi-sector collaboration for ensuring Nutrition, Food safety and Hygiene (NFSH). Achieving adequate nutrition and food security for all entails that the right foods are available, and that people are able to physically and financially access those foods. But an underlying assumption is that people are willing to eat the right foods and prepare and to eat them in a manner that conserves their nutritious value and that is hygienic. Thus, Bangladesh still consumes far more rice per capita than recommended by dietary guidelines. Feeding and hygiene practices are often inadequate (Manikam *et al.*, 2017; Mahmud, 2016) with dire consequences for small children's nutritional status and development. This is true across socio-economic categories. All these issues materialize in poor nutritional outcomes where high rates of stunting coexist with fast rising levels of obesity. While efforts are made all round to ensure awareness of what adequate nutrition and safe practices are, there is only scattered evidence of the progress so far, which makes a nationwide coherent strategy difficult to develop. In the CIP-2 Investment Plan for Nutrition-Sensitive Food Systems, under its Program III.I proposes measures for 'Enhanced nutrition knowledge, promotion of good practices, and consumption of safe and nutritious diets'. The CIP-2 for this purpose requires investment of 89 million US $. This strategy necessitates a thorough understanding of the situation of awareness of consumers on Nutrition, Food Safety and Hygiene.

**1.2. Statement of the problem:** The World Food Summit-1996 emphasized on three important components for ensuring food security viz. availability of adequate food, stability in food supplies, and access to food and nutrition security. The present government of Bangladesh has given emphasis on poverty alleviation and elimination including all forms of disparity at the forefront of country's development strategy. The ending of poverty and hunger is one of the prime objectives of the Sustainable Development Goals (SDGs) to be implemented by 2030. In consonance with the SDGs, the government of Bangladesh has adopted Seventh Five Year Plan (SFYP), in which one of the targets has been fixed to bring down extreme poverty to around 8.9% by the year 2020. In the SFYP the government has identified that there is persistent micronutrient deficiencies, lack of public awareness, maternal under nutrition, acute malnutrition, climate risks and adaptations, and lack of dietary diversity are among other challenges in the nutrition program sub-sector (GoB. SFYP, p. 519; Finance Division, 2014). In this plan nutrition and hygiene education has been incorporated in the education and ensure the availability of diverse foods quality (GoB. SFYP,p. 520). Food- borne illness is still an important issue in the world. Mishandling of food occur during food preparation, handling and storage and studies shows that consumers have inadequate knowledge about measures to prevent food borne illness in the home. The economic burden resulting from the diseases caused by foods on the health systems and the productivity losses they cause are at significant levels (Gil *et al.*, 2002; Banati, 2003).

In Bangladesh there is no general study on the awareness of NFSH. There are small studies on specific group viz. Karmakar (2017) made a study on a only 200 university students of *Noakhali* region of Bangladesh, Alam *et al.,* (2010) made a study on rural unmarried adolescent girls aged 13–18 years in 708 rural clusters to measure the nutritional status and dietary intake and relevant knowledge of adolescent girls in rural Bangladesh. Iqbal Kabir *et al.,*(2013) developed a Complementary Feeding Manual for Bangladesh and found that two- third of mothers did not have enough knowledge about nutrition. But one-third of them thought that vegetables, fish, egg, meat, water melon, banana, khichuri (one types mixed food of rice, dal, vegetables etc.), seasonal fruits contain a lot of vitamins. Therefore, in Bangladesh there are very few studies and all the studies are made on specific group on the awareness of food safety, nutrition and hygiene of the consumers. But the present study is an in-depth study which has included the different types of consumers in the life cycle viz. baby, child, adolescent, young and old so that an overall awareness level can be measured in the urban area of Bangladesh. Due to budget constraint the study is confined to SCC area in which all types of respondents has been included and findings from this study will be used generally for Bangladesh in making policy.

**1.3. Review of literature**: Safety food is the basic human right when billions of people are in risk of unsafe food and many millions become sick and hundreds of thousand die yearly (Fung, *et al*., 2018). On an average 9.4 million people suffer from food –borne diseases illness and 1351 death in United States of America (Scallan, *et al*., 2011). The estimated total economic loss of foodborne illness ranged between US$5.6 billion and US$9.4 billion in USA and in Australia between $487 million and $1900 million (Cameron, *et al.*, 1995). Food borne and water borne diseases kill about 2.2 million people in each year (WHO, 2004). In developing country like Bangladesh every year 2 million people including children are killed by food-borne disease (Daily star, 2015). The low income and minority people experience grater rate of food-borne illness (Quinlan, 2013). The foods available to consumers in Bangladesh are mostly contaminated as observed by the joint survey Institute of Public Health of Dhaka University and WHO (1994 and 2003). The water is also adulterated in Bangladesh (Saha, *et al.*, 2011). Besides food adulteration irregular intake of breakfast and dinner, eating junk food and fried food regularly, lack of knowledge about balance nutrition and obesity, habits of eating outside of the residence causes illness (Karmakar, *et al.*, 2016). The consumers' knowledge about food safety, nutrition and hygiene is very poor in Bangladesh and also around the world. The knowledge about food safety concern at all levels (family, institution, community and government) may reduce illness of people (Alam & Hossain, 2018; Karmakar, *et al.*, 2016). The consumer must aware of product information, environmental impacts of their buying decisions, food handling behavior and knowing the temperature of the refrigerator and cooking temperature when removing the food from fridge you must reheat the food then only the bacteria disappear from the food (Alam & Zakaria, 2021; Subba rao, *et al.*,2007). Food safety is the most important predictor of attitude while health consciousness appears to be the least important motive in contrast to findings from some previous research and consumers ethically intent to purchase organic foods (Nina and Hassan, n.d) Consumer's awareness of food safety is very essential to avoid the risk of unsafe food and maintain good (Bhuvneswari and Muthukumar, 2015). Most of the peoples have unsatisfactory dietary habit and providing pupils with nutrition knowledge would promote healthy dietary behaviors (Doaa, *et al.*, 2017). A study showed that overall dietary knowledge among the adolescent girls in rural Bangladesh is very low as compare to the semi-urban and small town areas (Islam, Alam, & Afzal, 2021) and more than half could not name the main food sources of energy and protein, and 36% were not aware of the importance of taking extra nutrients during adolescence for growth spurt (Nurul, *et al.*, 2010). Even most of the doctors are lac of nutrition knowledge (Uddin, *et al.*, 2008). The behavioral change communication should improve infant and young child nutrition knowledge (Hoddinott, 2017). It is important to measure the level of consumers' awareness on the safety of food to reduce food

-borne illness. The safety education campaign should raise knowledge about high risk foods and educate persons who will fall high-risk groups about their increased risk for food borne illness (Nesbitt, *et al.*, 2014). In a study on complementary feeding practices living in high income countries showed that promoters influence complementary feeding practices including income, lack of knowledge and incorrect advice (Logan, *et al.*, 2018). Due to ineffective communication of food safety message to consumers, causes of less awareness of food-borne illness and consumers fail to recognize foodborne disease rightly (Bruhn, 1997). The more consumers informed about food safety the more possibility of reducing risk from potential hazards of unsafe foods (Bruhn and Schutz, 1999). The Chinese people have high awareness of safe food but low knowledge on safe food, identification of safe food yet they have positive attitude on safe food (Liu, *et al.*, 2013). Due to increase in awareness of unsafe food the demand of organic fruits and vegetables is increasing and dominating the basket of organic consumers and consumers are ready to pay more prices for organic products (Emmanuel, *et al.*, 2005; Sedef, *et al.*, 2007). There is a need to assess the food safety knowledge, attitude and practice of parents and food preparers as they are more likely to engage in risky eating behaviors (Nevin and Ece, 2012). The hotel and restaurants also include food safety and hygiene as they prepare food for consumers and it is given emphases by Jenny and Serli (2010). The consumers are associated food mostly often with safe food, and food safety is the topic most often discussed while tasty food is the most important food conversation topic (Van, *et al.*, 2003).The assessment of consumers in this respect can be measured in terms of nutrition and food related Knowledge, Attitude and Practice (KAP) as per FAO (FAO, 2104). Therefore, the above literatures tell that the consumers are suffered from various diseases due to consumption of unsafe and un-hygienic food which causes the death of millions of people in Bangladesh and all over the world. This is because the consumers are not aware of unsafe food, required nutrition and hygiene environment. The awareness should be increased about unsafe food, nutrition food habit and hygiene for good health to all the stages of life cycle.

**1.4. Rationale of the Study:** The rationality of the study is discussed from the following perspectives:

**(i)Filling the gap of existing knowledge**: As we observed in the problem statement that there is very few studies on NFSH in Bangladesh and information on this issue is vital for good health of the people in Bangladesh. Further there is no study on SCC and even Sylhet region. The findings of the study will add value to the existing store of knowledge in the field of research in Bangladesh and especially for the Sylhet region.

**(ii) Relationship with the Sustainable Development Goals (SDGs)**: The United Nations has set the 169 targets under 17 SDGs for transforming our world by 2030 for people, planet and prosperity. Among the 17 SDGs, SDG#1 is to end poverty in all its forms everywhere; and SDG# 2 is to end hunger, achieve food security and improved nutrition and promote sustainable agriculture and SDG# 3 is to ensure healthy lives and promote well-being for all at all ages. Through this project we will measure the level awareness of nutrition, food safety and hygiene of consumers so that government can prepare policy and program to ensure balanced nutrition, safety food and hygiene in food preparation and preservation. A large number of people are suffered from food borne ill as stated earlier and suffer from unbalanced nutrition which makes them ill and unhealthy. Further people do not properly make food and preserve them rightly for which they loss food value and suffer from unhygienic illness. The data from such type of an in-depth study will surely help to policy makers in this regard which will help the government to achieve the SDGs 1, 2, and 3 directly. Further SDG# 12 is to ensure sustainable consumption and production patterns. The data will indirectly also help to achieve SDG# 12.

**(iii) Relevant to the Second Country Investment PlanCIP-2):** In the CIP-2 under its Nutrition-Sensitive Food Systems in the Program III.1, has proposed to measures for 'Enhanced nutrition knowledge, promotion of good practices, and consumption of safe and nutritious diets'. Such strategy necessitates a thorough understanding of the baseline situation of the country. But due to limited fund the data of this in-depth study will help to make decision by the concerned authorities in this regard.

**(iv) Relevant to Seventh Five Year Plan (SFYP), vision 2021 and perspective plan (2010-2021)**: The Government seeks to create conditions whereby the people of Bangladesh have the opportunity to reach and maintain the highest attainable level of health. It is the vision of 2021 that recognizes health as a fundamental human right and, therefore, the need to promote health and to alleviate ill health and suffering in the spirit of social justice. The target of vision 2021 and perspective plan (2010-21) are: (a) 85% of the population have standard nutritional food; (b). poor people ensured a minimum of 2122 kilo calories of food; (c) all kinds of contagious diseases eliminated (d)longevity increases to 70 years; and (e) infant mortality comes down to 15 from 54 per thousand at present. All these targets need to assess the awareness of NFSH of consumers in their life cycle.

**1.5 .Objectives of the study**: The core objectives of the study are to:

(i) Explore the knowledge of consumers on nutrition, food safety and hygiene of the sampled respondents in the Sylhet City Corporation in Bangladesh in terms of different demographic characteristics.

(ii) Compare knowledge level among the different groups of respondents ;and

(iii) Identify the factors associated with the levels of awareness and suggestion for betterment of consumers on nutrition, food safety and hygiene.

**1.6. Study Design**: The goal of the study was to make a base line survey on the consumers' awareness of nutrition, food safety and hygiene on different age groups of Sylhet City Corporation (SCC). To this aim the study has collected data from different types of consumers so that whole society is included in the survey. For this purpose the study has divided the population into six group as: (i) Baby; (ii) child; (iii) adolescent; (iv) parental; (v) unmarried adult young and (vi) married adult matured. Separate questionnaire for each group has been used for collecting data .

**1.7. Sample Design**: Bangladesh Bureau of Statistics (BBS) developed primary sampling unit based on Population and Housing Census 2011 in 2012 to conduct various demographic and socio-economic surveys 61 in SCC. The PSU identified by BBS will be the base for the study where each PSU is comprised of 100 to 150 households. The list of PSU is available in the district's statistics office and also in BBS. For conducting the study 240 respondents have been interviewed from 61 PSU conveniently but in practice we collected 234 as given in table 1.1

| Serial No. | Respondents | Number of respondents | Scale of questionnaire |
|---|---|---|---|
| 1. | Parents of baby up to 4 years | 40 | 5.00 |
| 2. | Parents of Child eating behavior between 5-9 years | 40 | 5.00 |
| 3. | Adolescent (10-19 years ) Food Habits | 34 | 3.00 |
| 4. | Parental Feeding Style to his Children up to 9 years | 34 | 5.00 |
| 5. | Food preference of adults unmarried (20-29) | 40 | 5.00 |
| 6. | Food Preference for Adults (29-64) Married Male and | 40 | 5.00 |
| | Female | | |
| | Total | 228 | |

The population up to 64 age is 95.06% of total population in Bangladesh (BBB, HIES, 2016)[1]. The age up to 04 years is baby, age between 5 and 9 is children and age between 10 and 19 is adolescent (UNICEF, 2011)[2]. More than 10% of target population is acceptable in a nation-wide survey.[3]

**1.8. Questionnaire Design:**

Construction of questionnaire: Our target respondents are six groups of people viz. (i) Baby; (ii) child; (iii) adolescent; (iv) parental; (v) adult young (vi) adult matured. The eating behaviour of baby and child will be received from mother and father, feeding behaviour of baby and child has been received by mother and father (parental behaviour). Habit of nutrition, safety food and hygiene of adolescent, young and adult old has been received from themselves respectively by both male and female.

Variables of questionnaire: Each questionnaire will be prepared to tap the factors associated with eating behaviours, feeding behaviours, food preference, food safety, nutrition knowledge, family eating and food buying behaviours, dietary practice, cooking practice, family hygiene knowledge and consciousness. The questionnaire has included demographic and socioeconomic factors of respondents. There are many sub-variables behind each variable so that total consumers awareness of nutrition, food safety and hygiene is explored in the Sylhet City Corporation.

Validity of questionnaire: Validity is characteristics of measurement concerned with the extent that a test measures what the researcher actually wishes to measure and that differences found with a measurement tool reflect true differences among participants drawn from the population (Cooper and Schindler, 2008)[4]. There are three types of validity viz. (i) content validity- it is done through a panel of experts; (ii) criterion related validity- it is done in terms of relevancy, freedom from bias, reliability and availability; (iii) construct validity- is done by factor analysis and each factor is analysed individually using Principal Component Analysis(PCA). It is explained briefly under data analysis. Factor analysis groups' variables (i.e. single question) has been divided into factors based on their common correlation. Those variables which are correlated with other has been grouped together and such a group of variables is called a factor (Saraph *et al.*,1989)[5]. Each factor is assumed as a separate construct and must one or multidimensional.

[1] BBS (2016), Household Income and Expenditure Survey, 2016.
[2] UNICEF (2011), The State of The World's Children
[3] UNICEF (2005), Multiple indicator cluster survey manual, Monitoring the situation of children and woman
[4] Cooper, Donald, R, Schindler & Pamela, S (2008), Business research methods, 9th Ed., New Delhi, The McGraw-Hill, p.24.

Test of internal consistency**:** Testing internal consistency is to test the homogeneity among the items. This is one of the approaches to test reliability of the questionnaire. It is tested by Cronbach's alpha (∝) values. A measure of scale reliability namely Cronbach's alpha (∝) measures the internal consistency of responses, that is, how closely related a set of variables as a group. It can be written as a function of the number of test variables and the average inter correlation among the variables. Mathematically, it can be written as:

$$\propto = \frac{\mathrm{N}\,\bar{\mathrm{c}}}{\bar{\mathrm{v}} + (\mathrm{N}-1)\bar{\mathrm{c}}}$$

Where, N is the number of items in the data set, $\bar{c}$ is the average inter-correlation among the items and $\bar{v}$ is the average variance .The required value of Cronbach's alpha (∝) for reliability is 0.70 (Nunnaly, 1967).[6]

Measurement scale of questionnaire: Scaling is the procedure for the assignment of numbers or other symbols to a property of objects in order to impart some of the characteristics of numbers to the properties in question (Cooper and Schindler, 2008). To measure the intensity of the responses in respect of the statements pertaining to each of the criterions 5 point and 3 point rating scales have been used The Likert scales are probably more reliable and provide greater volume of data than many other scales (Edwards & Kenney, 1946).[7] To answer the demographic question on characteristics, food choice, nutrition intake simple category scale, multiple choices, single response scale, and multiple response scale have also been used.

---

[5] Saraph, J. V., Benson, P. G. and Schroeder, R. G. (1989) 'An instrument for measuring the critical factors of quality management'. Decision Sciences, Vol.20, No.4, pp.810-829.
[6] Nunnaly, JC(1967), Psychometric theory, McGraw-Hill, New York .
[7] Edwards, A. L., & Kenney, K. C. (1946). A comparison of the Thurstone and Likert techniques of attitude scale construction. Journal of Applied Psychology, 30, 72-83.

**1.9. Analysis o:** Data has been analyzed in the following techniques:

**1.91. Descriptive statistics:** Frequency distribution mean and standard deviation of the variables have been computed to get the picture of the data. Tables and Graphs have also been used to present the data.

**1.9.2: Inferential statistics:** Principal Component analysis (PCA) and two sample t-test have been used for conclusion about the data.

**1.10. Conclusion:** This chapter has presented the introductory aspects of the study viz. problem of the study, literature review and methodology of the study. The next chapter will present the analysis of data.

# CHAPTER –TWO

# ANALYSIS OF DATA

**2. Introduction**: In the previous chapter we have discussed about the introductory aspects of the study such as statement of problem, related literature and methodology of the study. In this chapter we have analyzed the collected data following the methodology using different statistical techniques and methods.  It is divided into eight sections and there are different sub sections under each section.

## 2.1. Analysis of Baby's Eating Behavior up to 4(Four Year) Responded by Parents:

The eating behavior of baby is analyzed by the dummy of their parents. The analysis is discussed through the following aspects.

### 2.1.1 Demographic profile   and descriptive statistics of the respondents and the babies:

Demographic profile   and descriptive statistics of the respondents and the babies are presented in table 2.1.1

Table-2.1.1: Demographic profile   and descriptive statistics of the respondents and the babies

| Variable | Number | Variable | Minimum in years | Maximum years | Mean | Std. Deviation |
|---|---|---|---|---|---|---|
| Sex of respondent | | Age of baby | 0.3 | 4 | 3.035 | 1.1160 |
| Father | 20 | Descriptive Statistics of number of Baby | | | | |
| Mother | 20 | Total baby | 1 | 4 | 2.10 | .955 |
| Total | 40 | Up to 4 years | 1 | 2 | 1.10 | .304 |
| Descriptive statistics of  parents occupation | | | | | | |
| Occupation | | | Frequency | % | Valid % | Cumulative % |
| Business | | | 17 | 42.5 | 42.5 | 42.5 |
| Service | | | 14 | 35.0 | 35.0 | 77.5 |
| both service and business | | | 1 | 2.5 | 2.5 | 80.0 |
| Others including housekeeping | | | 8 | 20.0 | 20.0 | 100.0 |
| Total | | | 40 | 100.0 | 100.0 | |
| Descriptive Statistics of Family's annual income and expenditure | | | | | | |
| | | | Minimum | Maximum | Mean | Std. Dev. |
| annual income( in thousand taka) | | | 120 | 1200 | 502.00 | 249.134 |
| annual income( in thousand taka) | | | 95 | 1000 | 390.40 | 217.485 |
| Descriptive Statistics of  Education  of Parents | | | | | | |
| Education  in terms of Schooling year | | | 8.0 | 17.0 | 13.675 | 2.9733 |

Table 2.1.1 showed that out of 40 respondents 20 mother and 20 fathers. The age of the babies varied between 0.3 year and 4.00 and on average 3.035 with standard deviation 1.1160 .The number of total baby fluctuated is between 1 and 4 while number of baby under 4 years is ranged from 1 to 2. The maximum occupation of the parents are business (17) following service (14), others including housekeeping (8) and both service and business (1). The minimum, maximum and average of annual income of the responding household are 120 thousands, 1200 thousands and 502 thousands respectively and the minimum, maximum and average of education in terms of schooling year of the responding parents are 8, 17 and 14 respectively.

**2.1.2.Frequency of items to find out Baby's' eating behavior:** in order to assess the eating behavior of babies we have given 18 items of behavior as in the following table with 5 point Likert Scale 1=Never , 2=Rarely, 3=Sometimes , 4= often to 5=Always. The frequency of the variables is given in table 2.1.2

Table-2.1.2: Frequency of distribution baby's eating behavior

| Sl. No. | Variables/ items of behavior | 1=Never | 2=Rarely | 3=Sometimes | 4= often | 5=Always | Total |
|---|---|---|---|---|---|---|---|
| 1 | Baby seemed contented while feeding | | 2 | 20 | 8 | 10 | 40 |
| 2 | Baby frequently wanted more milk than given | 9 | 15 | 10 | 2 | 4 | 40 |
| 3 | Baby loves milk | 3 | 4 | 15 | 7 | 11 | 40 |
| 4 | Baby had a big appetite | 8 | 7 | 9 | 4 | | 40 |
| 5 | Baby finished feeding quickly | 8 | 11 | 16 | 3 | 2 | 40 |
| 6 | Baby became distressed while feeding | 0 | 9 | 15 | 12 | 4 | 40 |
| 7 | Baby got full up easily | | 6 | 15 | 11 | 8 | 40 |
| 8 | If allowed to, my baby would take too much milk | 18 | 11 | 5 | 6 | 0 | 40 |
| 9 | Baby took more than 30 minutes to finish feeding | 4 | 14 | 12 | 5 | 5 | 40 |
| 10 | Baby got full before taking all the milk | 1 | 4 | 17 | 12 | 4 | 40 |
| 11 | Baby fed slowly | 2 | 5 | 10 | 11 | 12 | 40 |
| 12 | Even baby had just eaten he/she was happy to feed again if offered | 20 | 10 | 7 | 3 | | 40 |
| 13 | Baby found it difficult to manage a complete feed | 2 | 15 | 12 | 4 | 7 | 40 |
| 14 | Baby was always demanding a feed | 12 | 4 | 13 | 9 | 2 | 40 |
| 15 | Baby sucked more and more slowly during the course of a feed | 6 | 13 | 13 | 5 | 3 | 40 |
| 16 | If given the chance, my baby would always be feeding | 12 | 10 | 8 | 6 | 4 | 40 |
| 17 | Baby enjoyed feeding time | 13 | 9 | 10 | 1 | 2 | 40 |
| 18 | Baby could easily take a feed within 30 minutes of the last one | 10 | 18 | 8 | 4 | 0 | 40 |

Table-2.1.2 showed that highest frequency in the scales are never= 20, rarely=18, sometimes =20 , often =12 and always =12. In percentage form the frequency is given following table :

| Sl. No. of items | 1.Never | 2. Rarely | 3. Sometimes | 4. often | 5. Always | Total |
|---|---|---|---|---|---|---|
| 1 | 5% | 50% | 20% | 25% | 25% | 100% |
| 2 | 38% | 25% | 5% | 10% | 10% | 100% |
| 3 | 10% | 38% | 18% | 28% | 28% | 100% |
| 4 | 18% | 23% | 10% | 0% | 0% | 100% |
| 5 | 28% | 40% | 8% | 5% | 5% | 100% |
| 6 | 23% | 38% | 30% | 10% | 10% | 100% |
| 7 | 15% | 38% | 28% | 20% | 20% | 100% |
| 8 | 28% | 13% | 15% | 0% | 0% | 100% |
| 9 | 35% | 30% | 13% | 13% | 13% | 100% |
| 10 | 10% | 43% | 30% | 10% | 10% | 100% |
| 11 | 13% | 25% | 28% | 30% | 30% | 100% |
| 12 | 25% | 18% | 8% | 0% | 0% | 100% |
| 13 | 38% | 30% | 10% | 18% | 18% | 100% |
| 14 | 10% | 33% | 23% | 5% | 5% | 100% |
| 15 | 33% | 33% | 13% | 8% | 8% | 100% |
| 16 | 25% | 20% | 15% | 10% | 10% | 100% |
| 17 | 23% | 25% | 3% | 5% | 5% | 100% |
| 18 | 45% | 20% | 10% | 0% | 0% | 100% |
| Average | 23% | 30% | 16% | 11% | 11% | 100% |

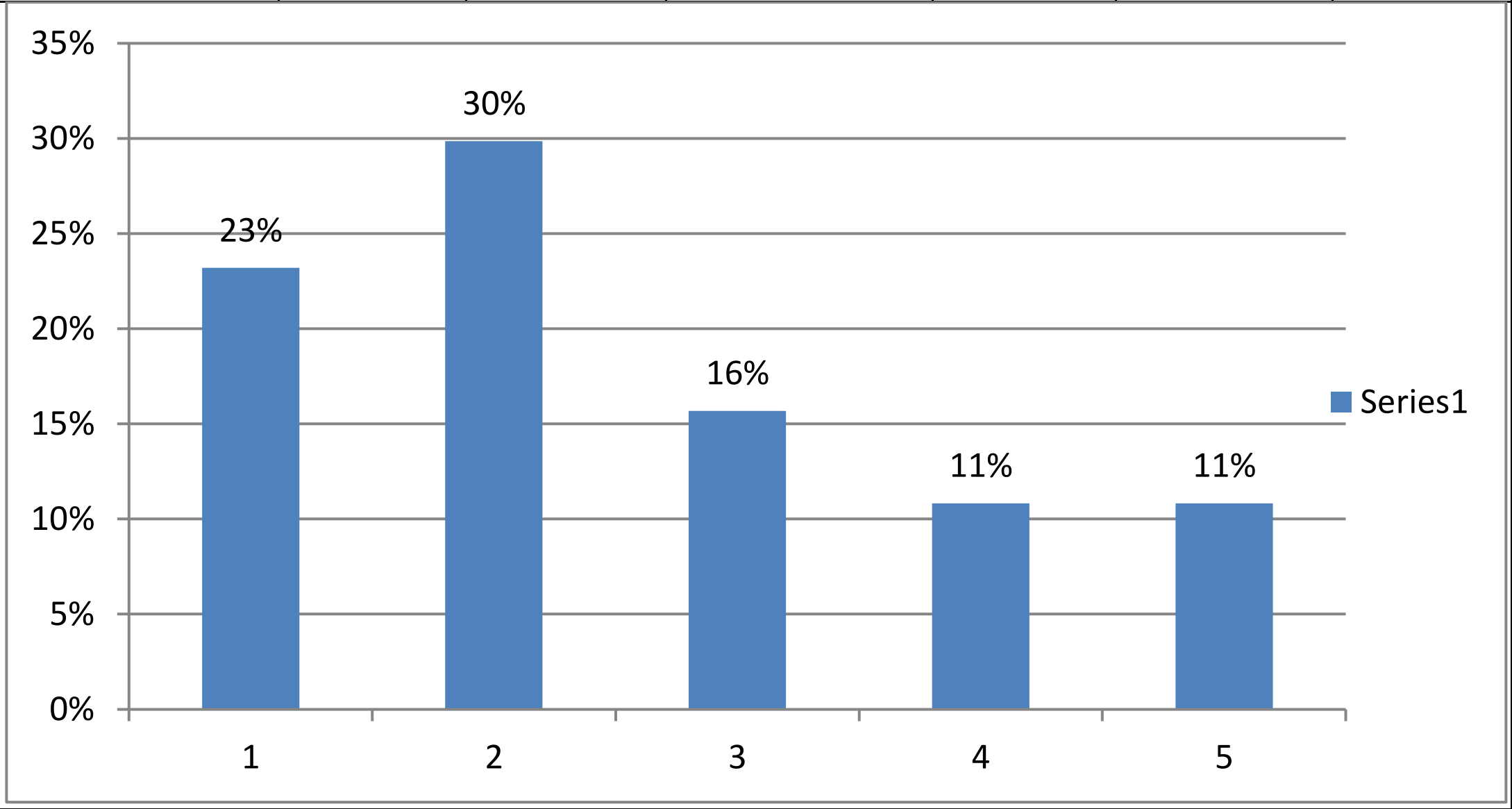

In corresponding of table 2.1.2 we have determined the total point in terms of scaling point of eating behavior as in table 2.1.3.

Table-2.1.3: Total point of baby's eating behavior based on table 2.1.2

| Sl. No. of items | 1.Never | 2. Rarely | 3. Sometimes | 4. often | 5. Always | Total |
|---|---|---|---|---|---|---|
| 1 | 0 | 4 | 60 | 40 | 50 | 154 |
| 2 | 9 | 30 | 30 | 16 | 20 | 105 |
| 3 | 3 | 8 | 45 | 44 | 55 | 155 |
| 4 | 8 | 14 | 27 | 0 | 0 | 49 |
| 5 | 8 | 22 | 48 | 8 | 10 | 96 |
| 6 | 0 | 18 | 45 | 16 | 20 | 99 |
| 7 | | 12 | 45 | 32 | 40 | 129 |
| 8 | 18 | 22 | 15 | 0 | 0 | 55 |
| 9 | 4 | 28 | 36 | 20 | 25 | 113 |
| 10 | 1 | 8 | 51 | 16 | 20 | 96 |
| 11 | 2 | 10 | 30 | 48 | 60 | 150 |
| 12 | 20 | 20 | 21 | 0 | 0 | 61 |
| 13 | 2 | 30 | 36 | 28 | 35 | 131 |
| 14 | 12 | 8 | 39 | 8 | 10 | 77 |
| 15 | 6 | 26 | 39 | 12 | 15 | 98 |
| 16 | 12 | 20 | 24 | 16 | 20 | 92 |
| 17 | 13 | 18 | 30 | 8 | 10 | 79 |
| 18 | 10 | 36 | 24 | 0 | 0 | 70 |
| Total on average of 18 items | | | | | | 100.5 |
| overall Average scale of 40 respondents | | | | | | 2.51 |

Table 2.1.3 showed that the top five items are – my baby loved milk(155) following my baby seemed contented while feeding (154), my baby fed slowly(150), my baby found it difficult to manage a complete feed(131), my baby got full up easily(113) and my baby frequently wanted more milk than I provided(105). The average score of the items is 100.5 and the average score of the respondents is 2.51 . It means the eating behavior lies between rarely and sometimes. Reliability of the questionnaire was checked by Cronbach's Alpha test which is 0.710 that means there is consistency of the items or variables. The item-wise point is presented in the figure 2.1.3.

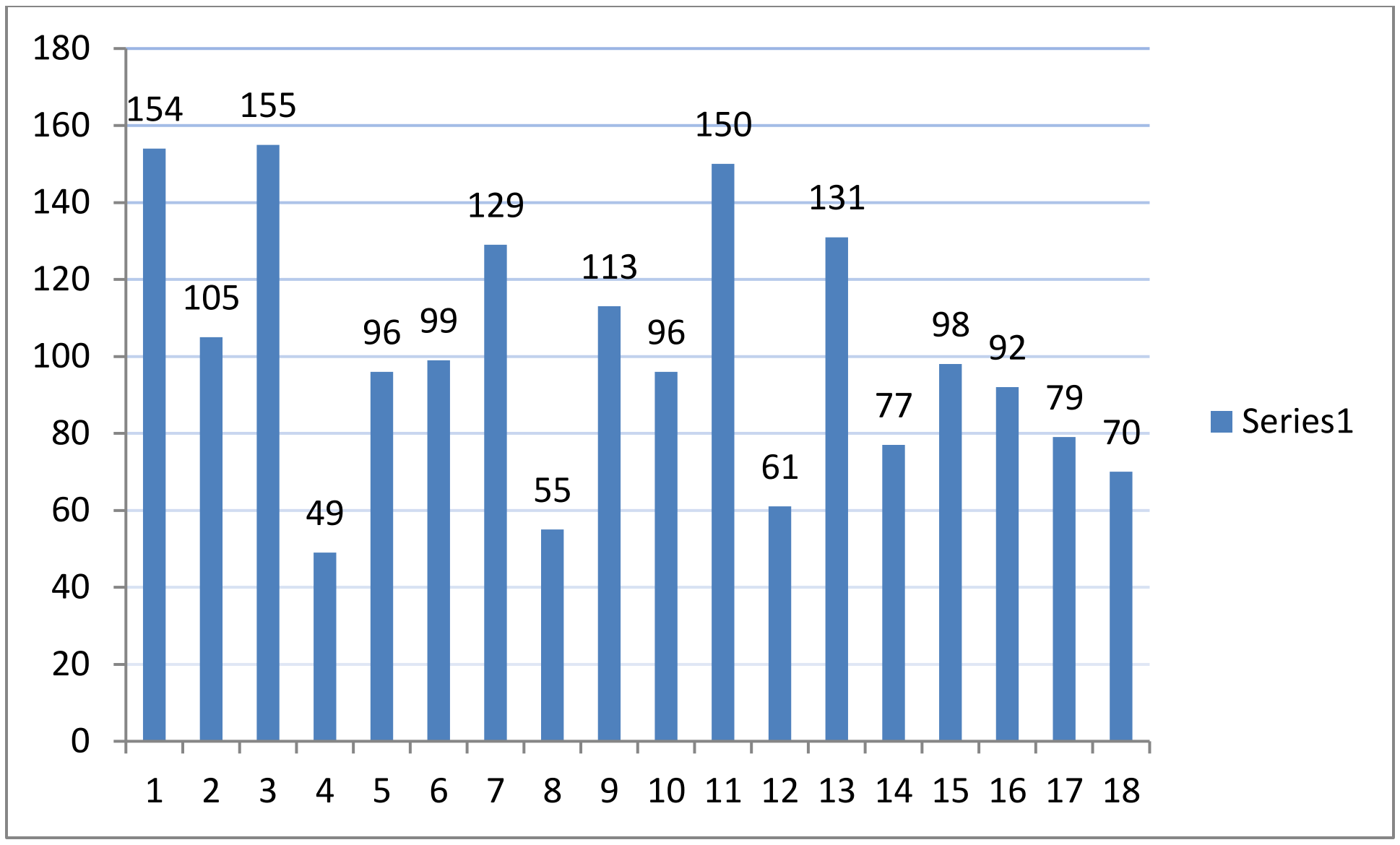

Figure-2.1.3: Item-wise point of baby's eating behavior.

**2.1.3 Factors analysis of baby's eating behavior:**

The researcher has identified some behavioral causes of bay's in eating by asking to the parents . There were eighteen variables or items in the schedule and the respondents were asked to on 5-point scaling as never=1; rarely=2; sometimes =3 often=4 and always=5. The Principal Component Analysis (PCA) with Varimax Orthogonal rotation method explained 75.66% of total variation by the extracted factors with 0.615 Kaiser-Meyer-Olkin (KMO) measure of sampling adequacy (Appendix -1). The Bartlett's Test of Sphericity by Approx. Chi-Square 396.205 and significant level is 0.00. It means the item size is sufficient to explore the factors. Details of the factors or items are given in appendix table 1, 2 and 3 and from the PCA tables we identified six factors as shown in figure 2.1.3.1 based on Eigenvalue.

Figure-2.1.3.1: Scree plot of Factors

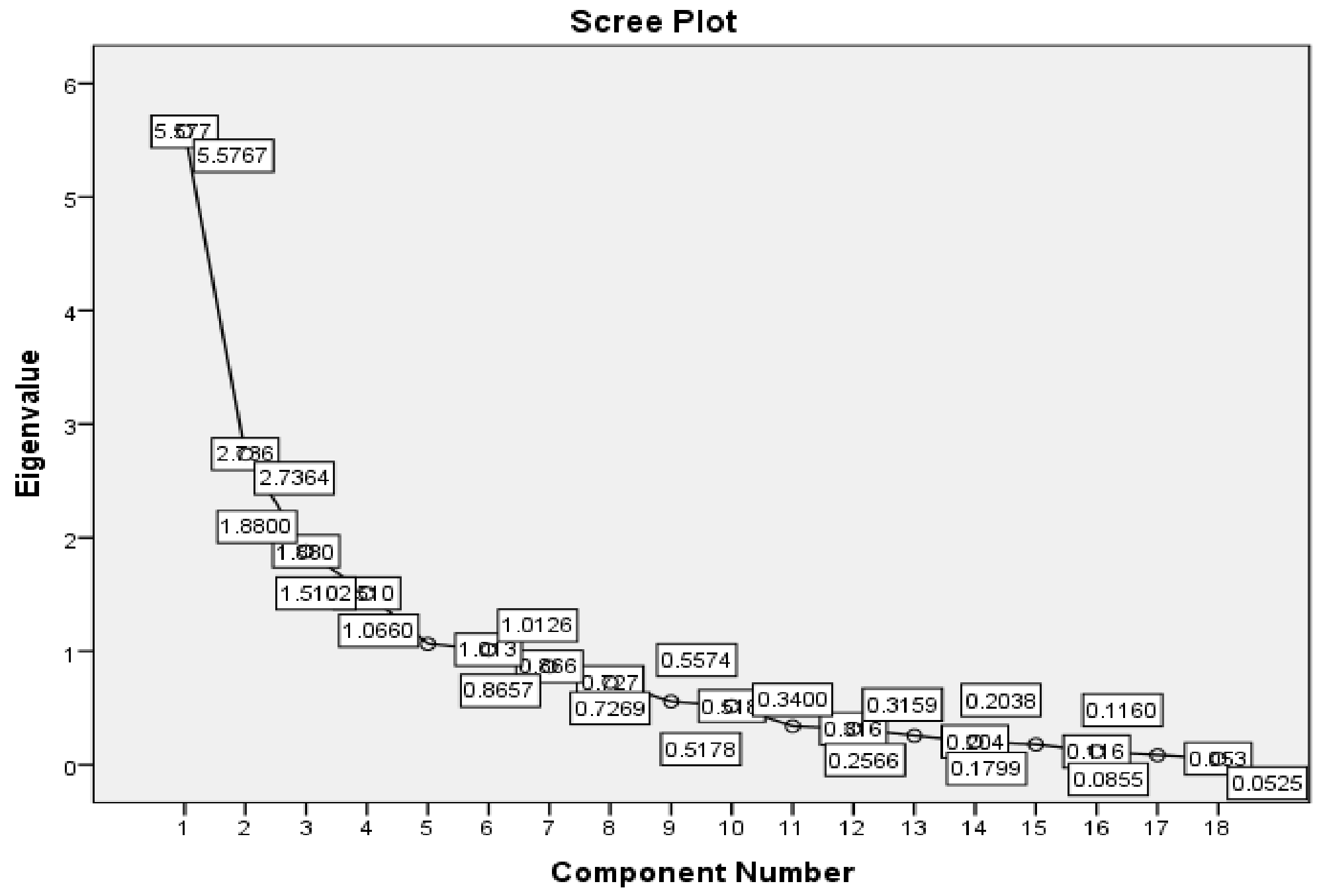

The component **Transformation matrix is given in figure 2.1.3.1.**

| **Table:2.1.3.1:Component Transformation Matrix** | | | | | | |
|---|---|---|---|---|---|---|
| Component | 1 | 2 | 3 | 4 | 5 | 6 |
| 1 | .560 | .390 | -.421 | .147 | -.451 | -.364 |
| 2 | .114 | .477 | .266 | .706 | .244 | .362 |
| 3 | -.357 | .564 | -.475 | -.373 | .018 | .434 |
| 4 | .242 | .423 | .688 | -.524 | -.111 | -.044 |
| 5 | .495 | -.350 | -.040 | -.109 | -.280 | .735 |
| 6 | .493 | -.002 | -.226 | -.234 | .804 | -.073 |
| Extraction Method: Principal Component Analysis.<br>Rotation Method: Varimax with Kaiser Normalization. | | | | | | |

The component score matrix is given in table 2.1.3.2

**Table-2.1.3.2:Component Score Covariance Matrix**

| Component | 1 | 2 | 3 | 4 | 5 | 6 |
|---|---|---|---|---|---|---|
| 1 | 1.000 | .000 | .000 | .000 | .000 | .000 |
| 2 | .000 | 1.000 | .000 | .000 | .000 | .000 |
| 3 | .000 | .000 | 1.000 | .000 | .000 | .000 |
| 4 | .000 | .000 | .000 | 1.000 | .000 | .000 |
| 5 | .000 | .000 | .000 | .000 | 1.000 | .000 |
| 6 | .000 | .000 | .000 | .000 | .000 | 1.000 |
| Extraction Method: Principal Component Analysis. Rotation Method: Varimax with Kaiser Normalization. | | | | | | |

The component Score Covariance Matrix shows that all the six items are significant as the p-value is 0.00 among each other**.** Therefore we have seen in the eating behavior of baby less than 4 years that there are six items of behavior as important statistically significant.

**2.2. Analysis of Child eating behavior between 5-9 years**: In this section an attempt is taken to analyze the child eating behavior between 5-9 years from different perspectives as follows:

**2.2.1. Demographic profile and descriptive statistics of the respondents and the babies:**

Table-2.1: Demographic profile and descriptive statistics of the respondents and the babies

| Variable | Number | Variable | Minimum years | Maximum years | Mean | Std. Deviation |
|---|---|---|---|---|---|---|
| Sex of respondent | | Age of baby | 5 | 9 | 6.98 | 1.527 |
| Father | 20 | Descriptive Statistics of number of Baby | | | | |
| Mother | 20 | Total no. children | 1 | 5 | 2.10 | .955 |
| Total | 40 | 4-9 years | 1.0 | 2.0 | 1.200 | .4051 |
| Descriptive statistics of parents occupation | | | | | | |
| Occupation | | | Frequency | Percent | Valid % | Cumulative % |
| Business | | | 8 | 20 | 20 | 20 |
| Service | | | 18 | 45 | 45 | 65 |
| both service and business | | | 0 | 0 | 0 | 0 |
| Others including housekeeping | | | 14 | 35 | 35 | 100 |
| Total | | | 40 | 100.0 | 100.0 | |
| Descriptive Statistics of Family's annual income and expenditure | | | | | | |
| | | | Minimum | Maximum | Mean | Std. Dev. |
| annual income( in thousand taka) | | | 240 | 1000 | 519.00 | 207.831 |
| annual income( in thousand taka) | | | 180 | 900 | 394.75 | 161.959 |
| Descriptive Statistics of Education of Parents | | | | | | |
| Fathers' Schooling year | | | 10 | 17 | 14.77 | 2.731 |
| Mothers Schooling year | | | 8 | 17 | 12.50 | 2.873 |

Table-2. 2.1 showed that out of 40 respondents 20 mother and 20 fathers. The age of the children varied between 4 year and 9.00 and on average 6.98 with standard deviation 1.57 .The number of total children is fluctuated between 1 and 5 while number of children between 4 -9 years is ranged from 1 to 2. The maximum occupation of the parents are service (18), following others including housekeeping (14), business (8) and both service and business (0). The minimum, maximum and average of annual income of the responding household are 240 thousands, 1000 thousands and 519 thousands respectively and the minimum, maximum and average of education in terms of schooling year of the responding father are 10, 17 and 14.77 respectively while mothers are 8, 17 and 12.50 respectively

**2.2.2 Child's eating behavior**: Childs' eating behavior is analyzed through the following sub sections using the five point scaling of behavior from Scale 1=Never , 2=Rarely, 3=Sometimes , 4= often to 5=Always. Frequency distribution of child's eating behavior: Childs' eating behavior is given in table -2.2.2.1:

Table-:2. 2.2.1: Frequency distribution of child's eating behavior

| Sl. No. | Behavior items or variable | 1=Never | 2=Rarely | 3=Som e-times | 4= often | 5=Alwa ys | Total |
|---|---|---|---|---|---|---|---|
| 1 | Child loves food | 0 | 2 | **26** | 5 | 7 | 40 |
| 2 | Child eats more when worried | 8 | 22 | 9 | 1 | 0 | 40 |
| 3 | Child has big appetite | 9 | 6 | 15 | 6 | 4 | 40 |
| 4 | Child finishes his/her meal quickly | 10 | 11 | 5 | 10 | 4 | 40 |
| 5 | Child is interested in food | 6 | 16 | 11 | 7 | 0 | 40 |
| 6 | Child is always asking for a drink | 3 | 11 | 5 | 9 | 12 | 40 |
| 7 | Child refuses new foods at first | 10 | 8 | 10 | 3 | 9 | 40 |
| 8 | Child eats slowly | 15 | 3 | 2 | 2 | 18 | 40 |
| 9 | Child eats less when angry | 6 | 7 | 4 | 12 | 11 | 40 |
| 10 | Child enjoys tasting new foods | 8 | 17 | 9 | 3 | 3 | 40 |
| 11 | Child eats less when s/he is tired | 5 | 12 | 12 | 7 | 4 | 40 |
| 12 | Child is always asking for food | 3 | 2 | 15 | 10 | 10 | 40 |
| 13 | Child eats more when annoyed | 15 | 21 | 2 | 2 | 0 | 40 |
| 14 | Allowed to, my child would eat too much | **20** | 12 | 4 | 2 | 2 | 40 |
| 15 | Child eats more when anxious | 15 | **19** | 3 | 2 | 1 | 40 |
| 16 | Child enjoys a wide variety of foods | 1 | 3 | 18 | 9 | 9 | 40 |
| 17 | Leaves food on his/her plate at the end of a meal | 7 | 7 | 2 | 5 | 19 | 40 |
| 18 | Child takes more than 30 minutes to finish a meal | 8 | 5 | 7 | 14 | 6 | 40 |
| 19 | The choice, child would eat most of the time | 17 | 6 | 5 | 6 | 5 | 40 |
| 20 | Child looks forward to mealtimes | 10 | 8 | 17 | 2 | 3 | 40 |
| 21 | Child gets full before his/her meal is finished | 1 | 11 | 8 | 10 | 10 | 40 |
| 22 | Child enjoys eating | 0 | 2 | 22 | 8 | 7 | 40 |
| 23 | Child eats more when s/he is happy | 0 | 5 | 21 | 8 | 5 | 40 |
| 24 | Child is difficult to please with meals | 6 | 9 | 4 | **18** | 3 | 40 |
| 25 | Child eats less when upset | 10 | 5 | 8 | 10 | 9 | 40 |
| 26 | Child gets full up easily | 7 | 0 | 4 | 9 | **20** | 40 |
| 27 | Child eats more when s/he has nothing else to do | 14 | 10 | 10 | | 6 | 40 |
| 28 | Child is full finds room to eat his/her favorite food | 7 | 8 | 2 | 4 | 19 | 40 |
| 29 | If given the chance, child drink continuously the day | 8 | 14 | 12 | 5 | 1 | 40 |
| 30 | Child cannot eat a meal if has had a snack just before | 6 | 1 | 6 | 12 | 15 | 40 |
| 31 | The chance my child would always be having a drink | 3 | 22 | 8 | 4 | 3 | 40 |
| 32 | Interested in tasting food hasn't tasted before | 4 | 3 | 13 | 4 | 16 | 40 |
| 33 | Child that s/he doesn't like a food, without tasting | 13 | 4 | 12 | 7 | 4 | 40 |
| 34 | If given the chance, child have food in his mouth | 12 | 10 | 6 | 0 | 12 | 40 |
| 35 | My child eats more and more slowly during meal | 6 | 9 | 10 | 7 | 8 | 40 |

The reliability analysis showed that Cronbach's Alpha is above 0.7 in the table of Reliability Statistics and according to Reliability Statistics if Cronbach's Alpha is above 0.7 then we can say that the data has good level of internal consistency. The table of Item Total statistics (Appendix-4) shows that all the items are important as deleting any of them reduces Cronbach's Alpha. The table of Scale Statistics shows the summary of Likert scale data that there was 35 items and the mean 67.775 and the variance is 158.025 with standard deviation is 12.5708. Table of Reliability Statistics showed that Cronbach's Alpha is 0.769.The table -2.2.2.1 showed that in the different highest frequency are never=20, rarely=19, sometimes =26 ,often= and always=20 . The percentage distribution is given in the following table

| Sl. of Behavior items or variable. | 1=Never | 2=Rarely | 3=Some-times | 4= often | 5=Always | Total |
|---|---|---|---|---|---|---|
| 1 | 0% | 5% | 65% | 13% | 18% | 100 |
| 2 | 20% | 55% | 23% | 3% | 0% | 100 |
| 3 | 23% | 15% | 38% | 15% | 10% | 100 |
| 4 | 25% | 28% | 13% | 25% | 10% | 100 |
| 5 | 15% | 40% | 28% | 18% | 0% | 100 |
| 6 | 8% | 28% | 13% | 23% | 30% | 100 |
| 7 | 25% | 20% | 25% | 8% | 23% | 100 |
| 8 | 38% | 8% | 5% | 5% | 45% | 100 |
| 9 | 15% | 18% | 10% | 30% | 28% | 100 |
| 10 | 20% | 43% | 23% | 8% | 8% | 100 |
| 11 | 13% | 30% | 30% | 18% | 10% | 100 |
| 12 | 8% | 5% | 38% | 25% | 25% | 100 |
| 13 | 38% | 53% | 5% | 5% | 0% | 100 |
| 14 | 50% | 30% | 10% | 5% | 5% | 100 |
| 15 | 38% | 48% | 8% | 5% | 3% | 100 |
| 16 | 3% | 8% | 45% | 23% | 23% | 100 |
| 17 | 18% | 18% | 5% | 13% | 48% | 100 |
| 18 | 20% | 13% | 18% | 35% | 15% | 100 |
| 19 | 43% | 15% | 13% | 15% | 13% | 100 |
| 20 | 25% | 20% | 43% | 5% | 8% | 100 |
| 21 | 3% | 28% | 20% | 25% | 25% | 100 |
| 22 | 0% | 5% | 55% | 20% | 18% | 100 |
| 23 | 0% | 13% | 53% | 20% | 13% | 100 |
| 24 | 15% | 23% | 10% | 45% | 8% | 100 |
| 25 | 25% | 13% | 20% | 25% | 23% | 100 |
| 26 | 18% | 0% | 10% | 23% | 50% | 100 |
| 27 | 35% | 25% | 25% | 0% | 15% | 100 |
| 28 | 18% | 20% | 5% | 10% | 48% | 100 |
| 29 | 20% | 35% | 30% | 13% | 3% | 100 |
| 30 | 15% | 3% | 15% | 30% | 38% | 100 |
| 31 | 8% | 55% | 20% | 10% | 8% | 100 |
| 32 | 10% | 8% | 33% | 10% | 40% | 100 |
| 33 | 33% | 10% | 30% | 18% | 10% | 100 |
| 34 | 30% | 25% | 15% | 0% | 30% | 100 |
| 35 | 15% | 23% | 25% | 18% | 20% | 100 |
| Average | 20% | 22% | 23% | 16% | 19% | 100 |
| | | | | | | |

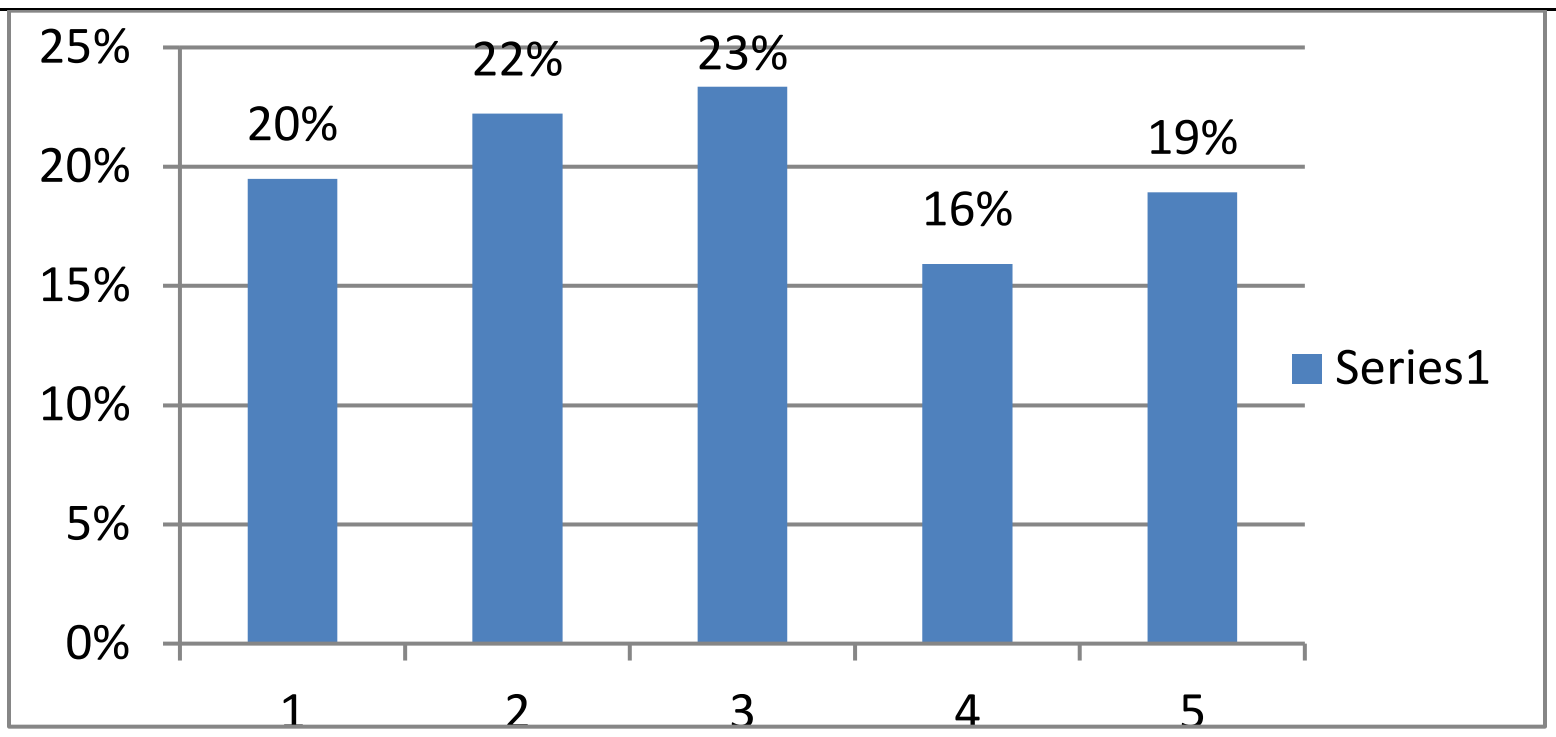

The frequency of table 2.2.1 is converted into five point scale is presented in corresponding to the scale in table 2.2.2.2 below:

Table-2.2.2.2: Point of Eating behavior in in corresponding to the scale in table 2.2.2.1

| Sl.No. | 1=Never | 2=Rarely | 3=Some-times | 4= often | 5=Always | Total | Position |
|---|---|---|---|---|---|---|---|
| 1 | 0 | 4 | 78 | 20 | 35 | 137 | Fifth |
| 2 | 8 | 44 | 27 | 4 | 0 | 83 | |
| 3 | 9 | 12 | 45 | 24 | 20 | 110 | |
| 4 | 10 | 22 | 15 | 40 | 20 | 107 | |
| 5 | 6 | 32 | 33 | 28 | 0 | 99 | |
| 6 | 3 | 22 | 15 | 36 | 60 | 136 | Sixth |
| 7 | 10 | 16 | 30 | 12 | 45 | 113 | |
| 8 | 15 | 6 | 6 | 8 | 90 | 125 | |
| 9 | 6 | 14 | 12 | 48 | 55 | 135 | Seventh |
| 10 | 8 | 34 | 27 | 12 | 15 | 96 | |
| 11 | 5 | 24 | 36 | 28 | 20 | 113 | |
| 12 | 3 | 4 | 45 | 40 | 50 | 142 | Fourth |
| 13 | 15 | 42 | 6 | 8 | 0 | 71 | |
| 14 | 20 | 24 | 12 | 8 | 10 | 74 | |
| 15 | 15 | 38 | 9 | 8 | 5 | 75 | |
| 16 | 1 | 6 | 54 | 36 | 45 | 142 | Fourth |
| 17 | 7 | 14 | 6 | 20 | 95 | 142 | Fourth |
| 18 | 8 | 10 | 21 | 56 | 30 | 125 | |
| 19 | 17 | 12 | 15 | 24 | 25 | 93 | |
| 20 | 10 | 16 | 51 | 8 | 15 | 100 | |
| 21 | 1 | 22 | 24 | 40 | 50 | 137 | Fifth |
| 22 | 0 | 4 | 66 | 32 | 35 | 137 | Fifth |
| 23 | 0 | 10 | 63 | 32 | 25 | 130 | |
| 24 | 6 | 18 | 12 | 72 | 15 | 123 | |
| 25 | 10 | 10 | 24 | 40 | 45 | 129 | |
| 26 | 7 | 0 | 12 | 36 | 110 | 165 | First |
| 27 | 14 | 20 | 30 | 0 | 30 | 94 | |
| 28 | 7 | 16 | 6 | 16 | 95 | 140 | |
| 29 | 8 | 28 | 36 | 20 | 5 | 97 | |
| 30 | 6 | 2 | 18 | 48 | 75 | 149 | Second |
| 31 | 3 | 44 | 24 | 16 | 15 | 102 | |
| 32 | 4 | 6 | 39 | 16 | 80 | 145 | Third |
| 33 | 13 | 8 | 36 | 28 | 20 | 105 | |
| 34 | 12 | 20 | 18 | 0 | 60 | 110 | |
| 35 | 0 | 4 | 78 | 20 | 35 | 137 | |
| Average of items | | | | | | 117.66 | |
| Average of respondents | | | | | | 3.36 | |

The table 2.2.2.2 depicted the total point of each variable and the top seven items are as follows- My child gets full up easily(165) , Child cannot eat a meal if s/he has had a snack just before(149), My child is interested in tasting food s/he hasn't tasted before(145), My child is always asking for food(142), My child enjoys a wide variety of foods(142), My child leaves food on his/her plate at the end of a meal(137), My child loves food(137), My child is always asking for a drink(136), My child eats less when angry(135). The average of items is 117.66 and the average of respondents is 3.36. That means the eating behavior of children varied between sometimes and often. The table 2.2.2.2 depicted the total point of each variable and the top seven items correspondingly based on the total point in figure 2.2.2.2 the total points are presented**.**

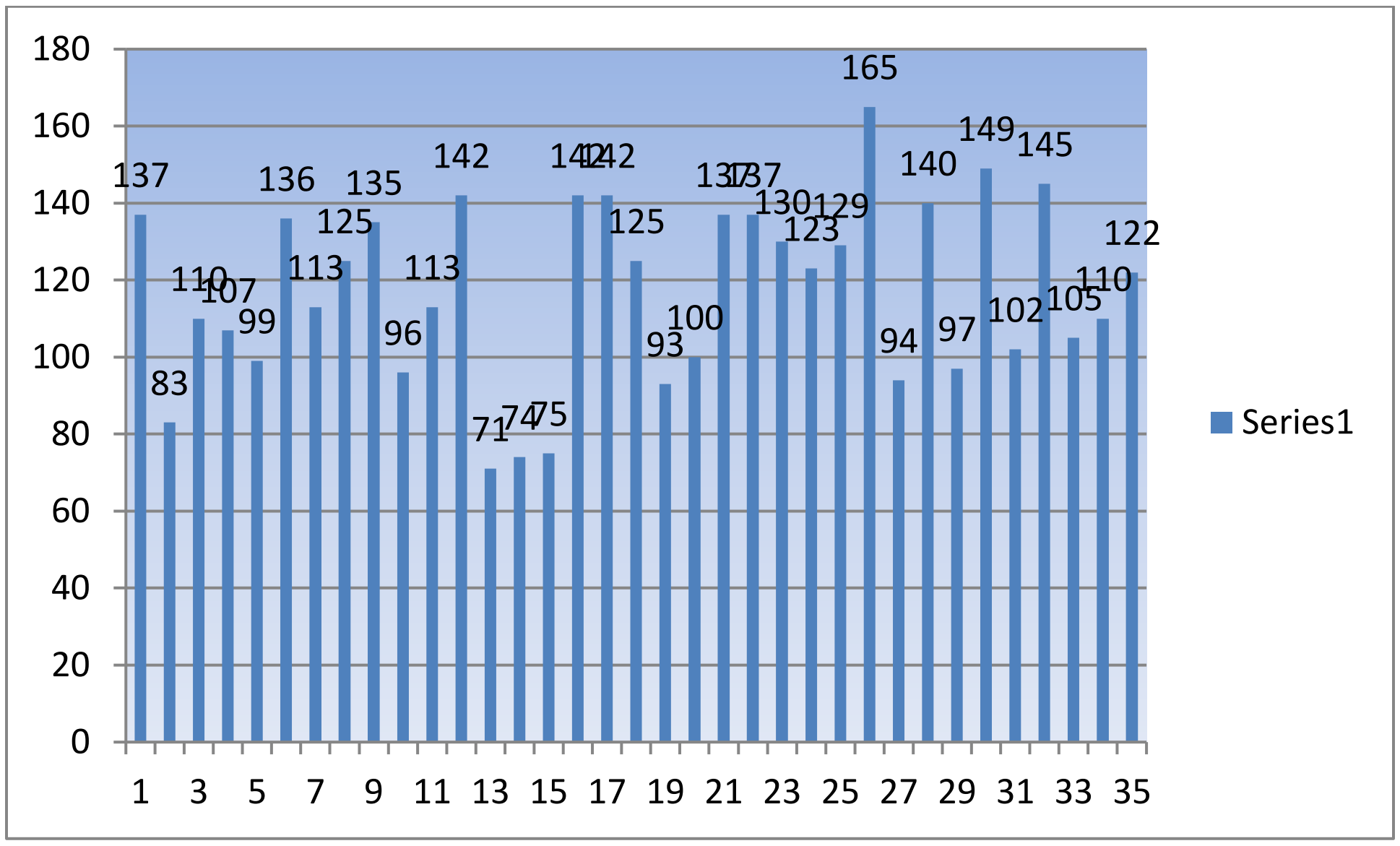

Figure-2.2.2.2: Total point of Child's eating behavior.

**2.2.2.3. Factor Analysis**: The Table of Scale Statistics shows that there are 35 items or variables of Childs 'eating behavior and to reduce the variables factor analysis is done with Principal Component Analysis with Varimax rotation . The communalities of the variables varied between 0.791 and 0.970 (Appendix -5). The total variance explained by the PCA is 88.388% that means the PCA is accepted and based on Eigen values ten items are considered important(Appendix-5). The Scree plot based on Eigen values is given in figure-2.2. 2.3

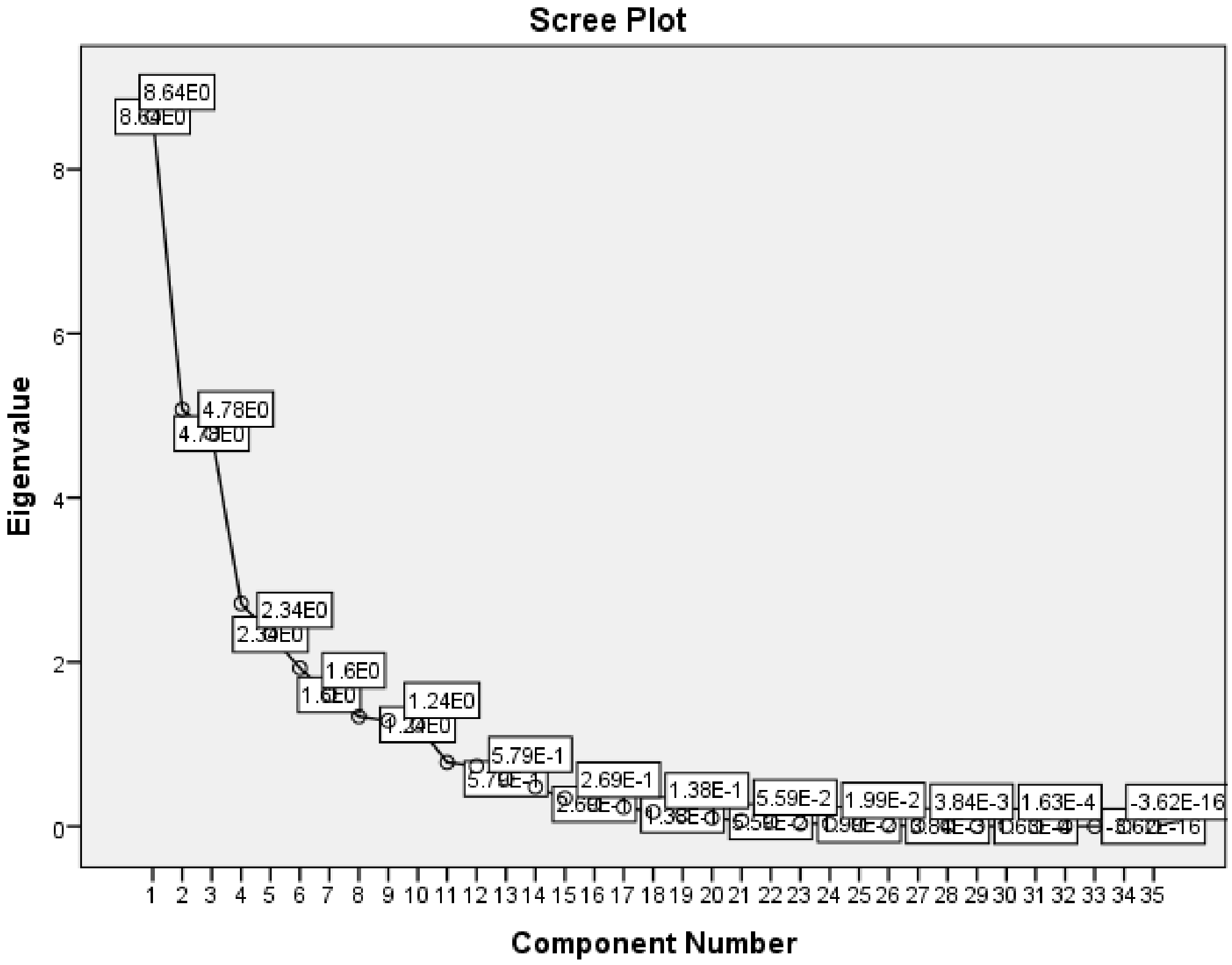

Figure-2.2.2.3: The Scree plot based on Eigen values of Child's Eating Behavior

Therefore the 35 variables are reduced to ten components explaining more than 88% of total variance. As there are more than 25 items the rotted component matrix is impossible to do.

**2.3. Analysis of Adolescent (10-19) Food Habits:** The food habits of adolescent are discussed below:

**2.3.1. Socioeconomic profile of the respondents : The socioeconomic profile of the respondents are discussed through table 2.3.1.1 to 2.3.1.7**:

Table-3.3.1.1: Respondents by Sex

| Sex | Frequency | Valid Percent | Cumulative Percent |
|---|---|---|---|
| Male | 17 | 50.0 | 50.0 |
| Female | 17 | 50.0 | 100.0 |
| Total | 34 | 100.0 | |

Table-2.3.1.1 showed that 50% respondents is male and 50% female i.e. the respondents by sex are equal .

**Table-2.3.1.2: Education status of the respondents**:

| Schooling Year | Frequency | Valid Percent | Cumulative Percent |
|---|---|---|---|
| 5.0 | 2 | 5.9 | 5.9 |
| 7.0 | 1 | 2.9 | 8.8 |
| 8.0 | 3 | 8.8 | 17.6 |
| 9.0 | 5 | 14.7 | 32.4 |
| 10.0 | 2 | 5.9 | 38.2 |
| 11.0 | 5 | 14.7 | 52.9 |
| 12.0 | 11 | 32.4 | 85.3 |
| 13.0 | 4 | 11.8 | 97.1 |
| 15.0 | 1 | 2.9 | 100.0 |
| Total | 34 | 100.0 | |

Table 2.3.1.2 depicted that education status of the respondents in terms of schooling year is classified from 5 years to 15 years and showed that 12 years education is 11(32.4%) following 9 years and 11 years each is 5(14.7%) .

**2.3.1.3. Mother's Education:** Education statuses of the mother of the respondents are given in table 3.1.3:

**Table-2.3.1.3: Mother's Education**

| Schooling Year | Frequency | Valid percent | Cumulative Percent |
|---|---|---|---|
| 5.0 | 1 | 2.9 | 2.9 |
| 8.0 | 5 | 14.7 | 17.6 |
| 10.0 | 7 | 20.6 | 38.2 |
| 11.0 | 3 | 8.8 | 47.1 |
| 12.0 | 3 | 8.8 | 55.9 |
| 13.0 | 1 | 2.9 | 58.8 |
| 14.0 | 1 | 2.9 | 61.8 |
| 16.0 | 6 | 17.6 | 79.4 |
| 18.0 | 7 | 20.6 | 100.0 |
| Total | 34 | 100.0 | |

Table-2.3.1.3 education status of respondent's mother in terms of schooling year varied between 5 years and 18 years and maximum of them is 7 years (20.6) of schooling .

Table-2.3.1.4: Mother's Occupation

| Occupation | Frequency | Valid Percent | Cumulative Percent |
|---|---|---|---|
| service | 1 | 2.9 | 2.9 |
| housekeeping | 24 | 70.6 | 73.5 |
| Both of the above | 8 | 23.5 | 97.1 |
| Business | 1 | 2.9 | 100.0 |
| Total | 34 | 100.0 | |

Table-2. 3.1.4 showed that lion portion (70.6%) of mother's occupation is housekeeping following both service and housekeeping(23.5%).

Table- 2. 3.1.5: Father's Education

| Schooling Year | Frequency | Valid Percent | Cumulative Percent |
|---|---|---|---|
| 1.0 | 1 | 2.9 | 2.9 |
| 10.0 | 5 | 14.7 | 17.6 |
| 11.0 | 1 | 2.9 | 20.6 |
| 12.0 | 8 | 23.5 | 44.1 |
| 13.0 | 1 | 2.9 | 47.1 |
| 15.0 | 1 | 2.9 | 50.0 |
| 16.0 | 5 | 14.7 | 64.7 |
| 17.0 | 2 | 5.9 | 70.6 |
| 18.0 | 10 | 29.4 | 100.0 |
| Total | 34 | 100.0 | |

Table-2.3.1.5 education status of respondent's father in terms of schooling year varied between 1 years and 18 years and maximum of them is 18 years (29.6%) of schooling following 12 years(23.5%), 16 years(14.7%).

Table-2.3.1.6: Father's Occupation

| Occupation | Frequency | Valid Percent | Cumulative Percent |
|---|---|---|---|
| service | 12 | 35.3 | 35.3 |
| Business | 13 | 38.2 | 73.5 |
| Both | 2 | 5.9 | 79.4 |
| others | 7 | 20.6 | 100.0 |
| Total | 34 | 100.0 | |

Table -2.3.1.6 showed that highest number (38.2%) of father's occupation is business following service(35.3%) and others (20.6%).

Table-2.3.1.7: Descriptive statistics annual income and expenditure of the respondent's family

| Variables | Minimum | Maximum | Mean | Std. Deviation |
|---|---|---|---|---|
| annual income( Taka in thousand) | 100 | 1450 | 479.12 | 301.507 |
| annual expenditure( Taka in thousand) | 95 | 1050 | 393.97 | 255.494 |

Table 2.3.1.7 showed that annual income is varied between 100 thousand and 1450 and on average 479.12 while annual expenditure fluctuated between 95 and 1050 with an average 393.97 thousand.

**2.3.2. Descriptive Statistics of Adolescent Food Habits:** The food habits of adolescent is measured in terms of given items or variables with three point scale **as Most** like=2; Neutral=0; and like=1. The descriptive statistics of adolescents' food habits is given in the following tables

**2.3.2.1: Frequency Distribution of food adolescent food habits**: Frequency Distribution of food adolescent food habits is given in table 2.3.2.1

**Table-2.3.2.1: Frequency Distribution of food adolescent food habits**

| sl.no | Items or variables | Most like=2 | like=1 | Neutral=0 | Total |
|---|---|---|---|---|---|
| 1 | If I having lunch away from home, I often choose a low-fat option. | 11 | 20 | 3 | 34 |
| 2 | I usually avoid eating fried foods. | 19 | 15 | 0 | 34 |
| 3 | I usually eat a dessert or pudding if there is one available. | 5 | 29 | 0 | 34 |
| 4 | I make sure I eat at least one serving of fruit a day. | 15 | 19 | 0 | 34 |
| 5 | I try to keep my overall fat intake down. | 11 | 23 | | 34 |
| 6 | If I am buying crisps, I often choose a low-fat brand. | 12 | 11 | 11 | 34 |
| 7 | I avoid eating lots of sausages and burgers. | 13 | 20 | 1 | 34 |
| 8 | I often buy pastries or cakes. | 11 | 20 | 3 | 34 |
| 9 | I try to keep my overall sugar intake down. | 12 | 12 | 10 | 34 |
| 10 | I make sure I eat at least one serving of vegetables or salad a day. | 12 | 20 | 2 | 34 |
| 11 | If I am having a dessert at home, I try to have something low in fat. | 10 | 18 | 6 | 34 |
| 12 | I rarely takeaway meals. | 6 | 28 | 0 | 34 |
| 13 | I try to ensure that I eat plenty of fruit and vegetables. | 9 | 25 | 0 | 34 |
| 14 | I often eat sweet snacks between meals. | 17 | 17 | 0 | 34 |
| 15 | I usually eat at least one serving of vegetables or salad | 13 | 21 | 0 | 34 |
| 16 | When I am buying a soft drink, I usually choose a diet drink. | 26 | 8 | 0 | 34 |
| 17 | When I put butter or margarine on bread, I usually spread it on thinly. | 15 | 13 | 6 | 34 |
| 18 | If I have a packed lunch, I usually include some chocolate and/or biscuits. | 13 | 17 | 4 | 34 |
| 19 | When I have a snack between meals, I often choose fruit. | 9 | 20 | 5 | 34 |
| 20 | If I am having a dessert or pudding in a restaurant, I usually choose the healthiest one. | 12 | 18 | 4 | 34 |
| 21 | I often have cream on desserts. | 12 | 16 | 6 | 34 |
| 22 | I eat at least servings of fruits most days. | 23 | 11 | 0 | 34 |
| 23 | I generally try to have a healthy diet. | 7 | 27 | 0 | 34 |

Table-2.3.2.1 showed that the highest frequency in the scale most like =26, like=29 and netral=11. Reliability of the questionnaire was checked by Cronbach's Alpha test which is 0.8230 that means there is consistency of the items or variables. The percentage distribution of frequency is given below

| sl.no of Items or variables | Most like=2 | like=1 | Neutral=0 | Total | Average figure |
|---|---|---|---|---|---|
| 1 | 32% | 59% | 9% | 100 | |
| 2 | 56% | 44% | 0% | 100 | |
| 3 | 15% | 85% | 0% | 100 | |
| 4 | 44% | 56% | 0% | 100 | |
| 5 | 32% | 68% | 0% | 100 | |
| 6 | 35% | 32% | 31% | 100 | |
| 7 | 38% | 59% | 3% | 100 | |
| 8 | 32% | 59% | 9% | 100 | |
| 9 | 35% | 35% | 29% | 100 | |
| 10 | 35% | 59% | 6% | 100 | |
| 11 | 29% | 53% | 17% | 100 | |
| 12 | 18% | 82% | 0% | 100 | |
| 13 | 26% | 74% | 0% | 100 | |
| 14 | 50% | 50% | 0% | 100 | |
| 15 | 38% | 62% | 0% | 100 | |
| 16 | 76% | 24% | 0% | 100 | |
| 17 | 44% | 38% | 17% | 100 | |
| 18 | 38% | 50% | 11% | 100 | |
| 19 | 26% | 59% | 14% | 100 | |
| 20 | 35% | 53% | 11% | 100 | |
| 21 | 35% | 47% | 17% | 100 | |
| 22 | 68% | 32% | 0% | 100 | |
| 23 | 21% | 79% | 0% | 100 | |
| Average | 37% | 55% | 8% | 100 | |

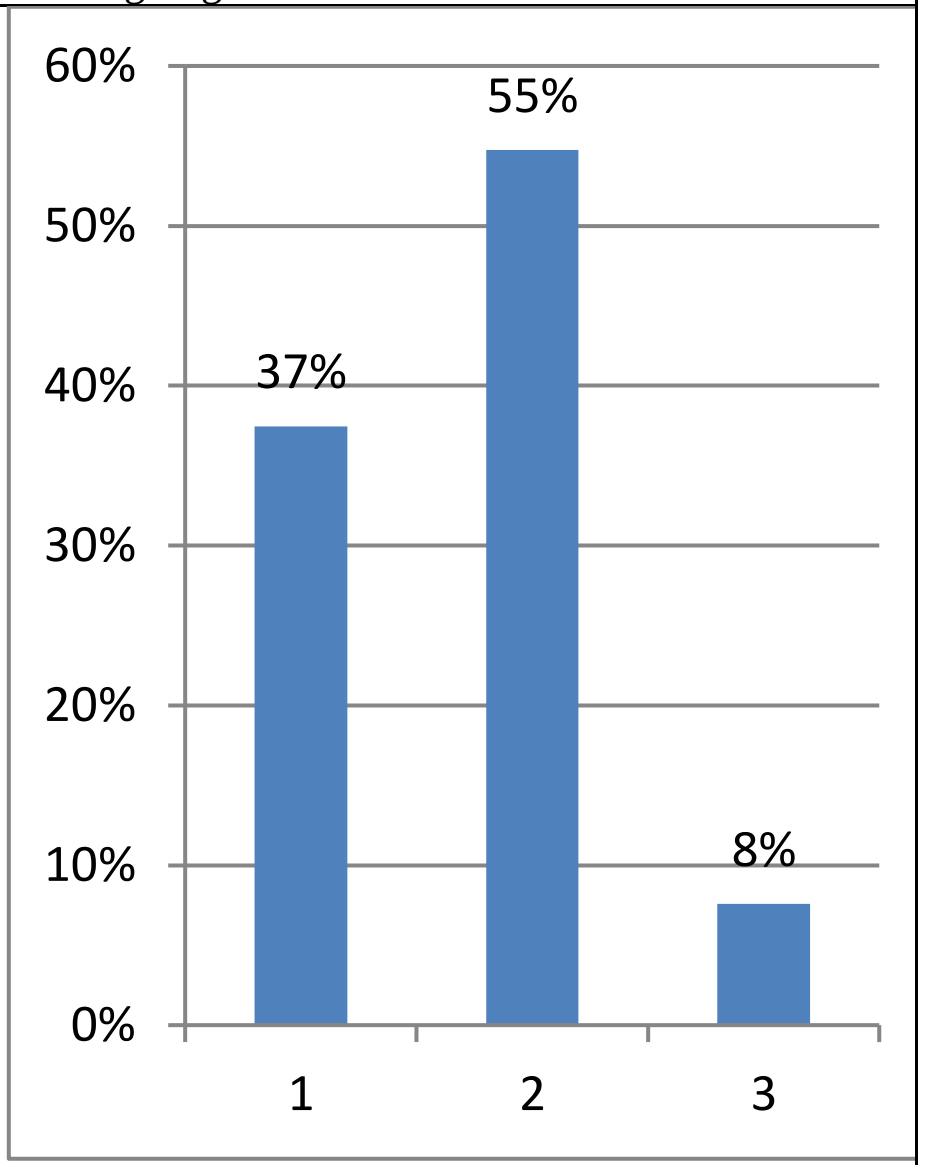

The item-wise point is presented in table 2.3.2.2.

Table-2.3.2.2: Points of food habits of adolescent based on table 2.3.2.1

| Sl. no. of variables/items | Most like=2 | like=1 | Neutral=0 | Total point | position |
|---|---|---|---|---|---|
| 1 | 22 | 20 | 0 | 42 | |
| 2 | 38 | 15 | 0 | 53 | Third |
| 3 | 10 | 29 | 0 | 39 | |
| 4 | 30 | 19 | 0 | 49 | Fifth |
| 5 | 22 | 23 | 0 | 45 | |
| 6 | 24 | 11 | 0 | 35 | |
| 7 | 26 | 20 | 0 | 46 | Sixth |
| 8 | 22 | 20 | 0 | 42 | |
| 9 | 12 | 12 | 0 | 24 | |
| 10 | 24 | 20 | 0 | 44 | |
| 11 | 20 | 18 | 0 | 38 | |
| 12 | 12 | 28 | 0 | 40 | |
| 13 | 18 | 25 | 0 | 43 | |
| 14 | 34 | 17 | 0 | 51 | Fourth |
| 15 | 26 | 21 | 0 | 47 | |
| 16 | 52 | 8 | 0 | 60 | First |
| 17 | 30 | 13 | 0 | 43 | Seventh |
| 18 | 26 | 17 | 0 | 43 | Seventh |
| 19 | 18 | 20 | 0 | 38 | |
| 20 | 24 | 18 | 0 | 42 | |
| 21 | 24 | 16 | 0 | 40 | |
| 22 | 46 | 11 | 0 | 57 | second |
| 23 | 14 | 27 | 0 | 41 | |
| Average of items | | | | 43.56 | |
| Average of respondents | | | | 1.89 | |

The top seven total scoring items are : (i)When I am buying a soft drink, I usually choose a diet drink (60), (ii)I eat at least servings of fruits most days(57),(iii)I usually avoid eating fried foods(53) (iv)I often eat sweet snacks between meals(51),(v) I make sure I eat at least one serving of fruit a day (49)(vi) I avoid eating lots of sausages and burgers.(46) and (vii) When I put butter or margarine on bread and I usually spread it on thinly If I have a packed lunch, I usually include some chocolate and/or biscuits (43). Average of score of the items is 43.56 and the average score of respondents is 1.89.The data of table2. 3.2.2 is presented in figure 2.3.3.2

Figure 2.3.3.2: Points of food habits of adolescent

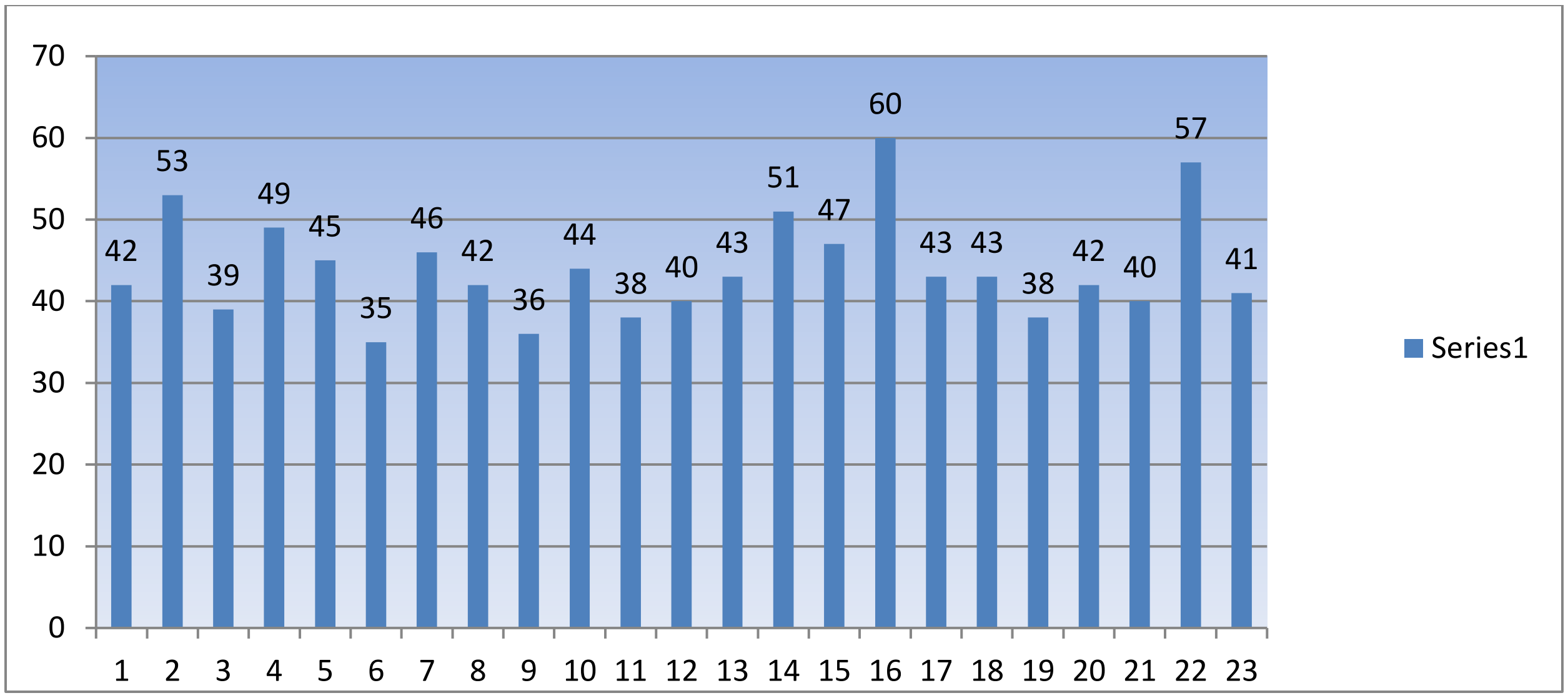

**2.3.3.3 Factor Analysis**: The Table of Scale Statistics shows that there are 23 items or variables of adolescent 'eating behavior and to reduce the variables factor analysis is done with Principal Component Analysis with Varimax rotation. The Bartlett's Test of Sphericity by Approx. Chi-Square 295.205 and significant level is 0.00. It means the item size is sufficient to explore the factors. The communalities of the variables varied between 0.57 and 0.903 (Appendix -6). The total variance explained by the PCA is 79.278% that means the PCA is accepted and based on Eigen values eight items are considered important(Appendix-7). The Scree plot based on Eigen values is given in figure 2.3.3.3.

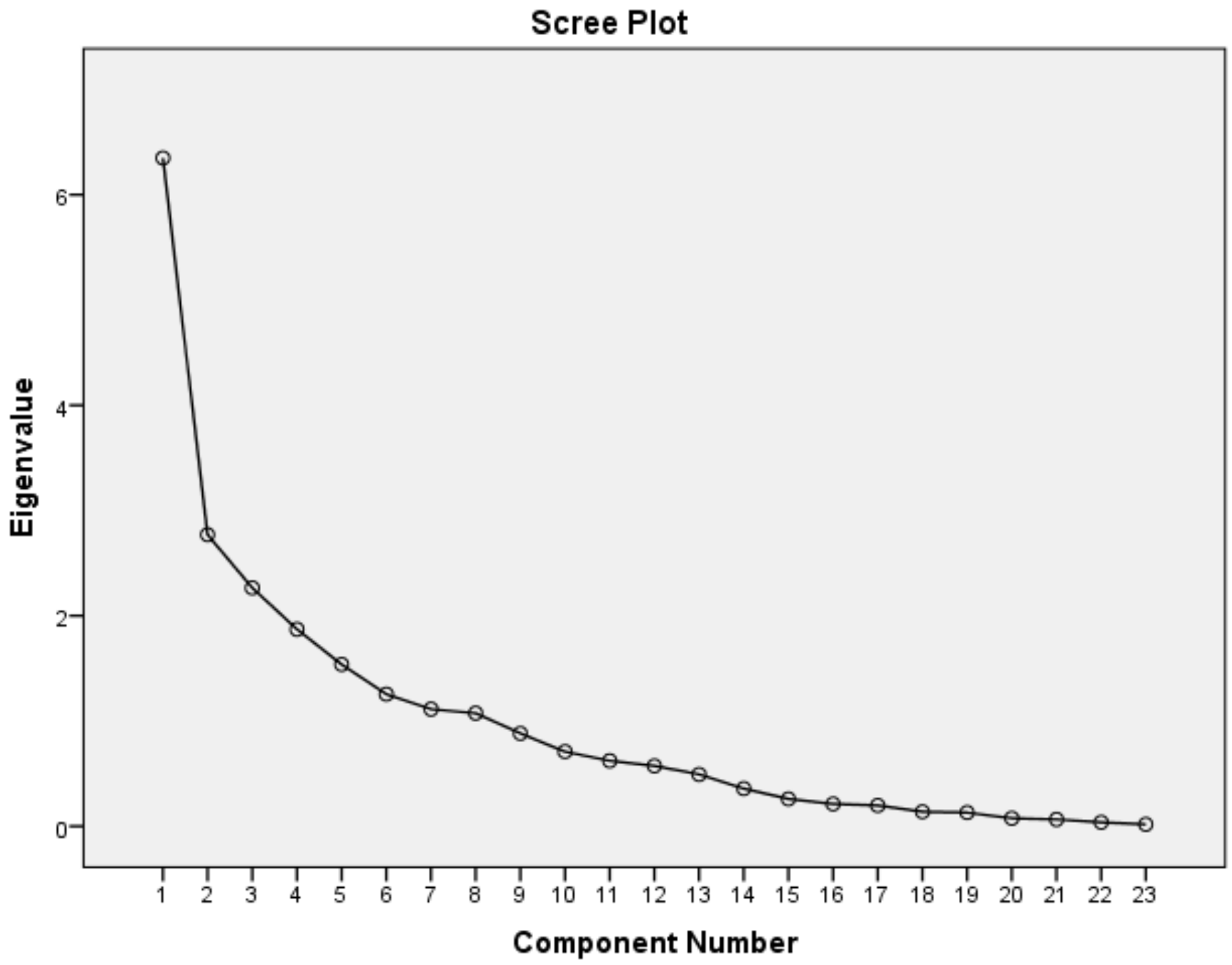

The component transformation matrix is given in the following table

| Component Transformation Matrix | | | | | | | | |
|---|---|---|---|---|---|---|---|---|
| Component | 1 | 2 | 3 | 4 | 5 | 6 | 7 | 8 |
| 1 | .510 | .325 | .488 | .428 | .272 | .358 | -.098 | .025 |
| 2 | .486 | -.788 | .014 | -.007 | .177 | -.043 | .310 | .116 |
| 3 | -.547 | -.303 | -.099 | .217 | .522 | .512 | -.097 | -.088 |
| 4 | -.025 | .150 | .280 | -.644 | .620 | -.219 | -.072 | .218 |
| 5 | .116 | .347 | -.623 | .223 | .313 | -.058 | .391 | .417 |
| 6 | -.213 | .145 | .329 | -.153 | -.115 | .235 | .842 | -.159 |
| 7 | -.327 | -.130 | .371 | .238 | -.217 | -.101 | -.014 | .790 |
| 8 | .195 | .015 | -.198 | -.475 | -.283 | .701 | -.134 | .328 |
| Extraction Method: Principal Component Analysis. Rotation Method: Varimax with Kaiser Normalization. | | | | | | | | |

Therefore the PCA reduces 21 items to eight factors of food habits of adolescent.

**2.4. Parental Feeding Style to his Children up to 9 years-:** The parental feeding style to his children up to 9 years is analyzed below:

### 2.4.1. Demographic profile and descriptive statistics of the respondents and the babies:

Table-2.4.1: Demographic profile and descriptive statistics of the respondents and the babies

| Variable | Number | Variable | Minimum in years | Maximum years | Mean | Std. Deviation |
|---|---|---|---|---|---|---|
| Sex of respondent | | Age of respondents | 28 | 59 | 45.32 | 8.562 |
| Father | 17 | Descriptive Statistics of number of Baby | | | | |
| Mother | 17 | Total no. children | 1 | 5 | 1.82 | 1.141 |
| Total | 34 | 9 years | 1.0 | 3.0 | 1.60 | .731 |
| Descriptive statistics of parents occupation | | | | | | |
| Occupation | | | Frequency | Percent | Valid Percent | Cumulative Percent |
| Business | | | 4 | 12 | 12 | 12 |
| Service | | | 16 | 47 | 47 | 59 |
| both service and business | | | 2 | 5 | 5 | 64 |
| Others including housekeeping | | | 12 | 36 | 36 | 100 |
| Total | | | 40 | 100.0 | 100.0 | |
| **Descriptive Statistics of Family's annual income and expenditure** | | | | | | |
| | | | Minimum | Maximum | Mean | Std. Dev. |
| annual income( in thousand taka) | | | 280 | 1200 | 589.00 | 277.831 |
| annual income( in thousand taka) | | | 246 | 980 | 594.75 | 261.959 |
| | | | | | | |
| **Descriptive Statistics of Education of Parents** | | | | | | |
| Fathers' Schooling year | | | 10 | 19 | 14.50 | 2.631 |
| Mothers Schooling year | | | 8 | 18 | 12 | 2.773 |

Table 2.4.1 showed that out of 34 respondents 17 mother and 17 fathers. The age of the children varied between 4 year and 9.00 and on average 6.98 with standard deviation 1.57 .The number of total children is fluctuated between 1 and 5 while number of children up to 9 years is ranged from 1 to 3. The maximum occupation of the parents are service (16), following others including housekeeping (12), business (4) and both service and business (2). The minimum, maximum and average of annual income of the responding household are 280 thousands, 1200 thousands and 589 thousands respectively and the minimum, maximum and average of annual expenditure are 246, 980 and 594.75 thousands. The minimum, maximum and average education in terms of schooling year of the responding father are 10, 19 and 14.5 respectively while mother's are 8, 18 and 12 respectively.

### 2.4.2. Descriptive statistics of parental feeding style to his/her children up to 9 years

**2.4.2.1: Frequency distribution of feeding style:** The respondents were given a scaling questionnaire on five point scale and the result is given in table 4.2.1

Table-2.4.2.1: Frequency Distribution of parental feeding style to his/her children up to 9 years

| Sl.no. | Items /variable | 1.Never | 2.Rarely | 3.Sometimes | 4.often | 5.Always | Total |
|---|---|---|---|---|---|---|---|
| 1 | I allow child to choose which foods to have for meals | 3 | 1 | 15 | 2 | 13 | 34 |
| 2 | I give child something to eat to make him/her feel better when s/he is feeling upset | 5 | 2 | 9 | 6 | 15 | 34 |
| 3 | I encourage my child to look forward to the meal | 1 | 1 | 11 | 7 | 14 | 34 |
| 4 | I praise my child if s/he eats what I give him/her | 3 | 1 | 7 | 7 | 16 | 34 |
| 5 | I decide how many snacks my child should have | 6 | 4 | 12 | **8** | 4 | 34 |
| 6 | I encourage my child to eat a wide variety of foods | 1 | 2 | 6 | **8** | 17 | 34 |
| 7 | In order to get my child to behave him/herself I promise him/her something to eat | 6 | 4 | 13 | 2 | 9 | 34 |
| 8 | I present food in an attractive way to my child | 7 | 4 | 11 | 4 | 8 | 34 |
| 9 | If child misbehaves I withhold his/her favorite food | 7 | 7 | 10 | 2 | 8 | 34 |
| 10 | I encourage my child to taste each of the foods I serve at mealtimes | 1 | 3 | 11 | 3 | 16 | 34 |
| 11 | I allow my child to wander around during a meal | **14** | 4 | 9 | 3 | 4 | 34 |
| 12 | I encourage my child to try foods that s/he hasn't tasted before | 2 | 0 | 10 | 3 | **19** | 34 |
| 13 | I give my child something to eat to make him/her feel better when s/he has been hurt | 5 | 3 | 11 | 5 | 6 | 34 |
| 14 | I let my child decide when s/he would like to have her meal | 8 | 1 | 13 | 5 | 7 | 34 |
| 15 | I give my child something to eat if s/he is feeling bored | 4 | 3 | 9 | 6 | 12 | 34 |
| 16 | I allow my child to decide when s/he has had enough snacks to eat | 3 | 4 | 13 | 4 | 10 | 34 |
| 17 | I decide when it is time for my child to have a snack | 7 | 1 | 14 | 3 | 9 | 34 |
| 18 | I use puddings as a bribe to get my child to eat his/her main course | 10 | **17** | 10 | 1 | 6 | 34 |
| 19 | I encourage my child to enjoy his/her food | 1 | 3 | 8 | 3 | 19 | 34 |
| 20 | I decide the times when my child eats his/her meals | 6 | 3 | 12 | 5 | 8 | 34 |
| 21 | I give my child something to eat if s/he is feeling worried | 7 | 3 | 12 | 5 | 7 | 34 |
| 22 | I reward my child with something to eat when s/he is well behaved | 6 | 2 | **16** | 4 | 6 | 34 |
| 23 | I let my child eat between meals whenever s/he wants | 3 | 3 | 10 | 6 | 12 | 34 |
| 24 | I insist my child eats meals at the table | 7 | 2 | 10 | 5 | 10 | 34 |
| 25 | I give my child something to eat to make him/her feel better when s/he is feeling angry | 7 | 4 | 13 | 1 | 9 | 34 |
| 26 | I decide what my child eats between meals | 9 | 2 | 12 | 7 | 4 | 34 |
| 27 | I praise my child if s/he eats a new food | 1 | 4 | 12 | 6 | 11 | 34 |

Table-2.4.2.1 depicted that in the five point scale the highest frequency in the never (14), rarely (17) sometimes (16) ,often (8) and always(19). Reliability of the questionnaire was checked by Cronbach's Alpha test which is 0.931 that means there is consistency of the items or variables. The percentage of frequency distribution is given below:

| Sl.no. Items /variable | 1.Never | 2.Rarely | 3.Sometimes | 4.often | 5.Always | Total | Average figure |
|---|---|---|---|---|---|---|---|
| 1 | 9% | 3% | 44% | 6% | 38% | 100 | |
| 2 | 15% | 6% | 26% | 18% | 44% | 100 | |
| 3 | 3% | 3% | 32% | 21% | 41% | 100 | |
| 4 | 9% | 3% | 21% | 21% | 47% | 100 | |
| 5 | 18% | 12% | 35% | 24% | 12% | 100 | |
| 6 | 3% | 6% | 18% | 24% | 50% | 100 | |
| 7 | 18% | 12% | 38% | 6% | 26% | 100 | |
| 8 | 21% | 12% | 32% | 12% | 24% | 100 | |
| 9 | 21% | 21% | 29% | 6% | 24% | 100 | |
| 10 | 3% | 9% | 32% | 9% | 47% | 100 | |
| 11 | 41% | 12% | 26% | 9% | 12% | 100 | |
| 12 | 6% | 0% | 29% | 9% | 56% | 100 | |
| 13 | 15% | 9% | 32% | 15% | 18% | 100 | |
| 14 | 24% | 3% | 38% | 15% | 21% | 100 | |
| 15 | 12% | 9% | 26% | 18% | 35% | 100 | |
| 16 | 9% | 12% | 38% | 12% | 29% | 100 | |
| 17 | 21% | 3% | 41% | 9% | 26% | 100 | |
| 18 | 29% | 50% | 29% | 3% | 18% | 100 | |
| 19 | 3% | 9% | 24% | 9% | 56% | 100 | |
| 20 | 18% | 9% | 35% | 15% | 24% | 100 | |
| 21 | 21% | 9% | 35% | 15% | 21% | 100 | |
| 22 | 18% | 6% | 47% | 12% | 18% | 100 | |
| 23 | 9% | 9% | 29% | 18% | 35% | 100 | |
| 24 | 21% | 6% | 29% | 15% | 29% | 100 | |
| 25 | 21% | 12% | 38% | 3% | 26% | 100 | |
| 26 | 26% | 6% | 35% | 21% | 12% | 100 | |
| 27 | 3% | 12% | 35% | 18% | 32% | 100 | |
| average | 15% | 10% | 33% | 13% | 30% | | |

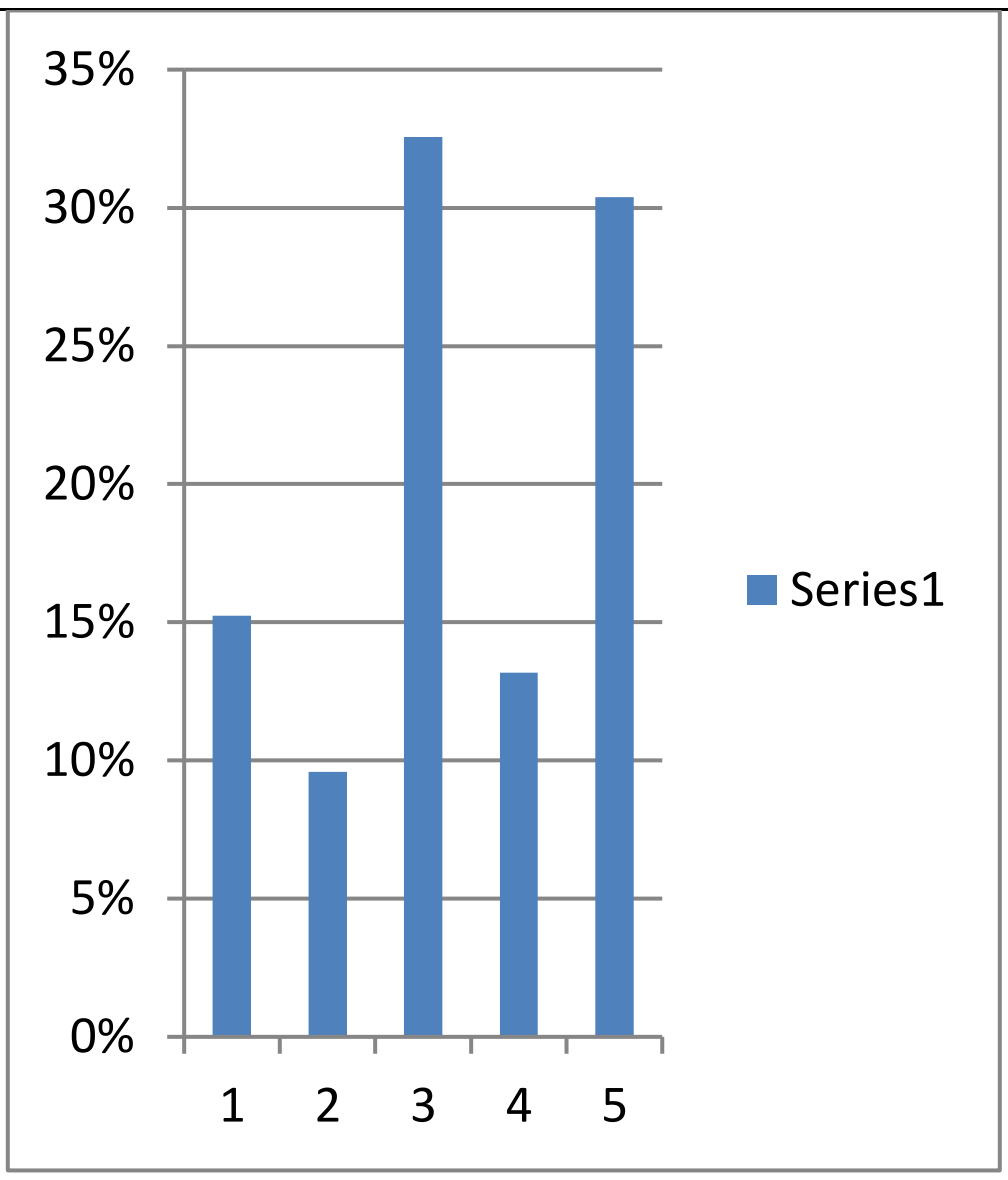

The total point corresponding to the frequency table 2.4.2.1 is given in table 2.4.2.2.

Table-2.4.2.2: Total Scaling Point of parental feeding style to his/her children up to 9 years based on table-2.4.2.1

| Variables no. | 1.Never | 2.Rarely | 3.Sometimes | 4.Often | 5.Always | Total | position |
|---|---|---|---|---|---|---|---|
| 1 | 3 | 2 | 45 | 8 | 65 | 123 | Eight |
| 2 | 5 | 4 | 27 | 24 | 75 | 135 | Fourth |
| 3 | 1 | 2 | 33 | 28 | 70 | 134 | Fifth |
| 4 | 3 | 2 | 21 | 28 | 80 | 134 | Fifth |
| 5 | 6 | 8 | 36 | 32 | 20 | 102 | |
| 6 | 1 | 4 | 18 | 32 | 85 | 140 | First |
| 7 | 6 | 8 | 39 | 8 | 45 | 106 | |
| 8 | 7 | 8 | 33 | 16 | 40 | 104 | |
| 9 | 7 | 14 | 30 | 8 | 40 | 99 | |
| 10 | 1 | 6 | 33 | 12 | 80 | 132 | sixth |
| 11 | 14 | 8 | 27 | 12 | 20 | 81 | |
| 12 | 2 | 0 | 30 | 12 | 95 | 139 | Second |
| 13 | 5 | 6 | 33 | 20 | 30 | 94 | |
| 14 | 8 | 2 | 39 | 20 | 35 | 104 | |
| 15 | 4 | 6 | 27 | 24 | 60 | 121 | |
| 16 | 3 | 8 | 39 | 16 | 50 | 116 | |
| 17 | 7 | 2 | 42 | 12 | 45 | 108 | |
| 18 | 10 | 34 | 30 | 4 | 30 | 108 | |
| 19 | 1 | 6 | 24 | 12 | 95 | 138 | Second |
| 20 | 6 | 6 | 36 | 20 | 40 | 108 | |
| 21 | 7 | 6 | 36 | 20 | 35 | 104 | |
| 22 | 6 | 4 | 48 | 16 | 30 | 104 | |
| 23 | 3 | 6 | 30 | 24 | 60 | 123 | Eight |
| 24 | 7 | 2 | 10 | 5 | 10 | 111 | |
| 25 | 7 | 4 | 13 | 1 | 9 | 103 | |
| 26 | 9 | 2 | 12 | 7 | 4 | 97 | |
| 27 | 1 | 4 | 12 | 6 | 11 | 124 | Seventh |
| Average of items | | | | | | 114.51 | |
| Average of respondents | | | | | | 4.24 | |

The total point score of table 2.4.2.2. is given in figure 2. 4.2.2 for easy understanding

Figure-2. 4.2.2 : The total point score of table 2.4.2.2.

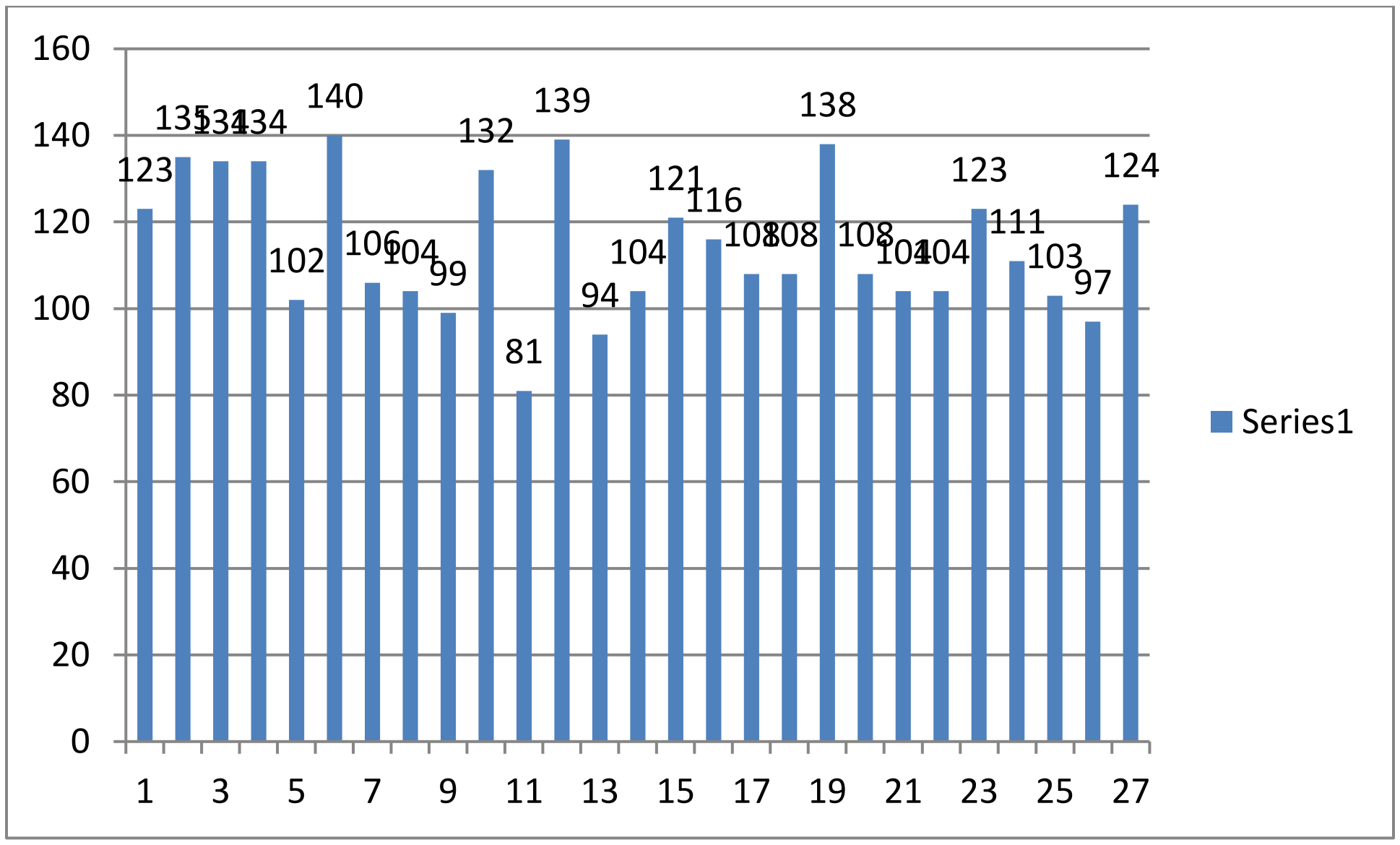

**2.4.3. Factor Analysis:** The Table of Scale Statistics shows that there are 27 items or variables of parent's 'eating behavior to their children and to reduce the variables factor analysis is done with Principal Component Analysis with Varimax rotation . Kaiser-Meyer-Olkin Measure of Sampling Adequacy is .670. The Bartlett's Test of Sphericity by Approx. Chi-Square 806.205 and significant level is 0.00. It means the item size is sufficient to explore the factors. The communalities of the variables varied between 0.51 and 0.893 (Appendix -9). The total variance explained by the PCA is 78.96% that means the PCA is accepted and based on Eigen values seven items are considered important(Appendix-10). The Scree plot based on Eigen values is given in figure-2. 4.3.1.

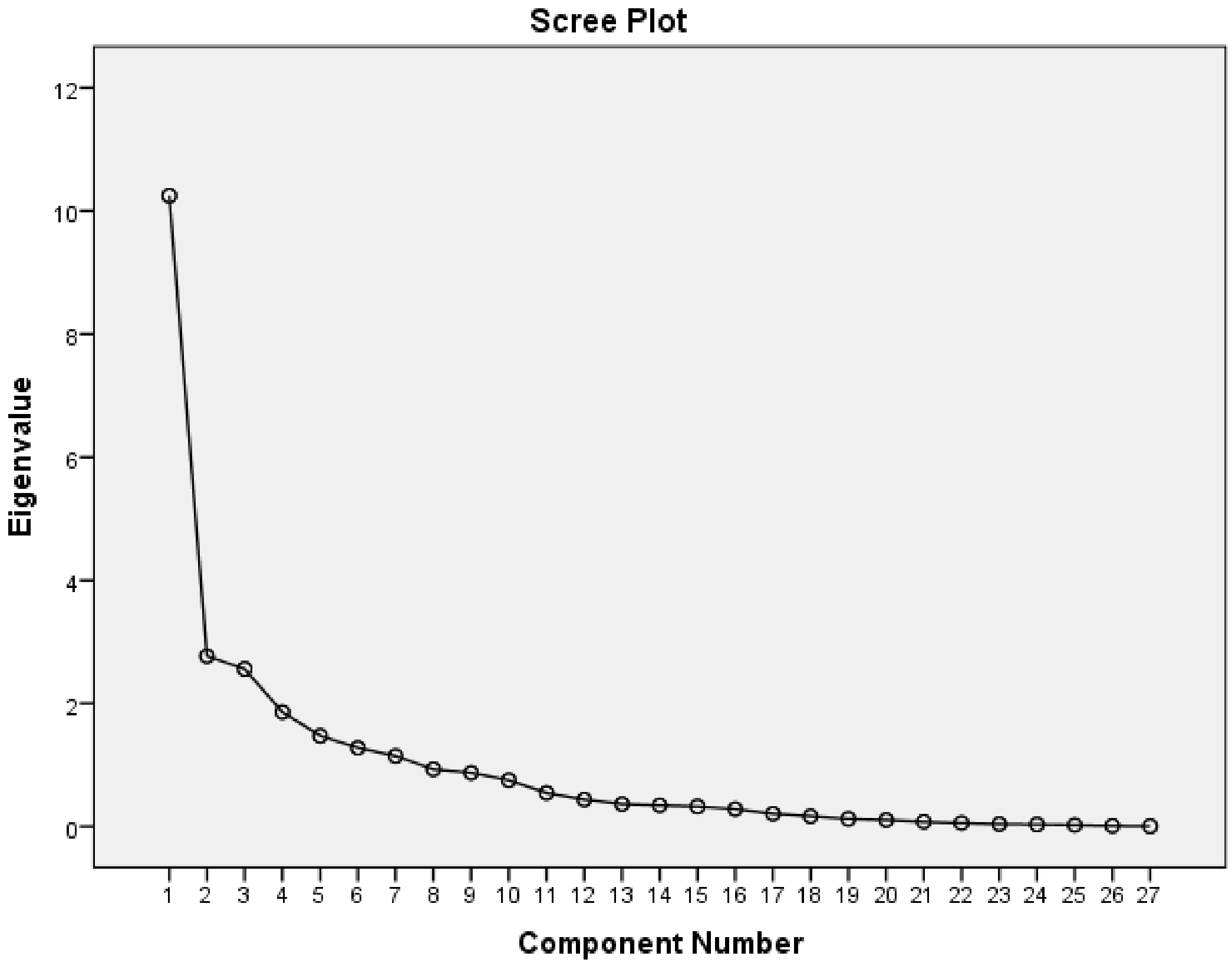

Figure-2.4.3.1

The component matrix is given in appendix-11 and 12. The component score Coefficient Matrix is given in appendix 13. The Component Score Covariance Matrix is given in table2. 4.3.3

Table - 2. 4.3.3: Component Score Covariance Matrix

| Component | 1 | 2 | 3 | 4 | 5 | 6 | 7 |
|---|---|---|---|---|---|---|---|
| 1 | 1.000 | .000 | .000 | .000 | .000 | .000 | .000 |
| 2 | .000 | 1.000 | .000 | .000 | .000 | .000 | .000 |
| 3 | .000 | .000 | 1.000 | .000 | .000 | .000 | .000 |
| 4 | .000 | .000 | .000 | 1.000 | .000 | .000 | .000 |
| 5 | .000 | .000 | .000 | .000 | 1.000 | .000 | .000 |
| 6 | .000 | .000 | .000 | .000 | .000 | 1.000 | .000 |
| 7 | .000 | .000 | .000 | .000 | .000 | .000 | 1.000 |
| Extraction Method: Principal Component Analysis. Rotation Method: Varimax with Kaiser Normalization. | | | | | | | |

Therefore PCA produced seven factors from 27 through extraction of parent's eating behavior to their children .

### 2.5. Food Preference For Adults (20-29) Unmarried Male And Female:

**2.5.1: Socioeconomic status of the respondents:** The socioeconomic status of the respondents food preferences for adults (20-29) unmarried male and female are discussed through table 2.5.1 to 2.5.6 as given below:

Table-2.5.1: Demographic profile and descriptive statistics of the respondents and the babies

| Variable | Number |
|---|---|
| Sex of respondent | |
| Male | 20 |
| Female | 20 |
| Total | 40 |
| Marital status | |
| Unmarried | 40 |
| Total | 40 |

Table-2.5.1 depicted that there were 20 male and 20 female respondents and all them were unmarried of 20-29 aged.

Table -2. 5.2: Age Structure of the respondents:

| Age | Frequency | Valid Percent | Cumulative Percent |
|---|---|---|---|
| 21 | 13 | 31.7 | 31.7 |
| 22 | 12 | 29.3 | 61.0 |
| 23 | 7 | 17.1 | 78.0 |
| 24 | 4 | 9.8 | 87.8 |
| 25 | 3 | 7.3 | 95.1 |
| 26 | 1 | 2.4 | 97.6 |
| Total | 40 | 100.0 | 100.00 |

Table-2.5.2 showed that age structure of the respondents varied between 21 and 26 and 13(31.7%) of respondents have 21 age following 12 (29.3%), 7(17.1%), 4(9.8%) and 3(7.3%) and only 1(2.4) .Table-2.5.3: Occupation of the respondents

| occupation | Frequency | Percent | Valid Percent | Cumulative Percent |
|---|---|---|---|---|
| Service | 3 | 7.3 | 7.3 | 7.3 |
| Business | 1 | 2.4 | 2.4 | 9.8 |
| Unemployed | 4 | 9.8 | 9.8 | 19.5 |
| Others( including Students) | 32 | 78.0 | 78.0 | 97.6 |
| Total | 40 | 100.0 | 100.0 | 100.00 |

Table-2.5.3 showed that occupation of most of the respondents is others including students32(78%) following unemployed 4(9.8%) service3(7.3%) and business 1(2.4%).

Table-2.5.4.Education status of the respondents

| Schooling year | Frequency | Valid Percent | Cumulative Percent |
|---|---|---|---|
| 12 | 1 | 2.4 | 2.4 |
| 14 | 2 | 4.9 | 7.3 |
| 15 | 22 | 53.7 | 61.0 |
| 16 | 4 | 9.8 | 70.7 |
| 17 | 4 | 9.8 | 80.5 |
| 18 | 1 | 2.4 | 82.9 |
| 19 | 2 | 4.9 | 87.8 |
| 20 | 3 | 7.3 | 95.1 |
| 9 | 1 | 2.4 | 97.6 |
| Total | 40 | 100.0 | |

Table-2.5.4 showed that occupation of the respondents varied between 9 and 20 and 22(53.7%) of respondents have 15 years education following 16 years of 4(9.8%), 17 years of 4(9.8%) and 20 years 3(7.3%).

Table 2.5.5.Annual income of respondents

| income (in thousand) | Frequency | Valid Percent | Cumulative Percent |
|---|---|---|---|
| 100 | 34 | 82.9 | 82.9 |
| 200 | 1 | 2.4 | 85.4 |
| 250 | 1 | 2.4 | 87.8 |
| 36 | 1 | 2.4 | 90.2 |
| 48 | 1 | 2.4 | 92.7 |
| 60 | 2 | 4.9 | 97.6 |
| Total | 40 | 100.0 | |

Table-2.5.5 showed that annual income of most of the respondents 34( 82.9%) is 100 thousand following 60 thousands 2(4.9%) and 200 thousand , 250 thousand 36 thousand and 48 thousand 1(2.4%) each. In table 2.5.6 annual expenditure of respondents is presented and it showed that annual expenditure of the respondents varied between 96 thousands and 200 thousands.

Table-2.5.6: Annual expenditure of respondents

| income (in thousand ) | Frequency | Valid Percent | Cumulative Percent |
|---|---|---|---|
| 100 | 3 | 7.3 | 7.3 |
| 120 | 1 | 2.4 | 9.8 |
| 150 | 1 | 2.4 | 12.2 |
| 200 | 1 | 2.4 | 14.6 |
| 48 | 1 | 2.4 | 17.1 |
| 60 | 3 | 7.3 | 24.4 |
| 70 | 2 | 4.9 | 29.3 |
| 72 | 6 | 14.6 | 43.9 |
| 78 | 2 | 4.9 | 48.8 |
| 80 | 6 | 14.6 | 63.4 |
| 82 | 2 | 4.9 | 68.3 |
| 84 | 5 | 12.2 | 80.5 |
| 85 | 2 | 4.9 | 85.4 |
| 90 | 1 | 2.4 | 87.8 |
| 94 | 2 | 4.9 | 92.7 |
| 95 | 1 | 2.4 | 95.1 |
| 96 | 1 | 2.4 | 97.6 |
| Total | 40 | 100.0 | 100 |

**2.5.2. Descriptive Statistics of food preference:** Table-2.5.2.1 depicted that in the five point scale the highest frequency in the dislike a lot (39), dislike a little (12) neither like or dislike (26) , like a little (31) and like a lot (34). Reliability of the questionnaire was checked by Cronbach's Alpha test which is 0.851 that means there is consistency of the items or variables. The total point corresponding to the frequency table 2.5.2.1 is given in table 2.5.2.2

Table-2.5.2.1: Frequency distribution in percentage of food preference of unmarried

| Sl.No. | Food Item | 1- Dislike a lot | 2- Dislike a little | 3- Neither like or dislike | 4- Like a little | 5- Like a lot | Total |
|---|---|---|---|---|---|---|---|
| 1 | Beef | 0% | 0% | 10% | 78% | 13% | 100 |
| 2 | Beef burgers | 3% | 0% | 18% | 58% | 23% | 100 |
| 3 | Lamb | 8% | 8% | 10% | 5% | 70% | 100 |
| 4 | Chicken | 3% | 0% | 10% | 20% | 68% | 100 |
| 5 | Bacon | 15% | 20% | 65% | 0% | 0% | 100 |
| 6 | Ham | 70% | 25% | 5% | 0% | 0% | 100 |
| 7 | Sausage(Meat) | 33% | 5% | 5% | 23% | 35% | 100 |
| 8 | White fish | 40% | 5% | 10% | 23% | 23% | 100 |
| 9 | Oily fish | 35% | 8% | 3% | 23% | 33% | 100 |
| 10 | Smoked Salmon | 70% | 8% | 23% | 0% | 0% | 100 |
| 11 | Tinned Tuna | 33% | 8% | 18% | 43% | 0% | 100 |
| 12 | Eggs | 5% | 8% | 28% | 60% | 0% | 100 |
| 13 | Baked beans | 5% | 18% | 13% | 63% | 5% | 100 |
| 14 | Bread or Bread rolls | 3% | 5% | 33% | 60% | 0% | 100 |
| 15 | Bran cereal | 25% | 10% | 23% | 20% | 23% | 100 |
| 16 | Porridge | 30% | 10% | 15% | 43% | 3% | 100 |
| 17 | Plain boiled rice | 3% | 5% | 13% | 78% | 3% | 100 |
| 18 | Sugared cereal | 40% | 30% | 20% | 10% | 0% | 100 |
| 19 | Hummus | 98% | 0% | 3% | 0% | 0% | 100 |
| 20 | Wheat cereal | 13% | 8% | 35% | 20% | 25% | 100 |
| 21 | Potatoes (boiled or mashed) | 0% | 8% | 93% | 0% | 0% | 100 |
| 22 | Chips | 3% | 0% | 3% | 15% | 80% | 100 |
| 23 | Rice or corn cereal | 8% | 0% | 20% | 30% | 43% | 100 |
| 24 | Soft cheese | 10% | 8% | 35% | 48% | 0% | 100 |
| 25 | Hard cheese | 18% | 5% | 35% | 43% | 0% | 100 |
| 26 | Cottage Cheese | 13% | 3% | 8% | 30% | 40% | 100 |
| 27 | Plain, Low-fat yoghurt | 15% | 3% | 13% | 70% | 0% | 100 |
| 28 | Oranges | 0% | 3% | 3% | 3% | 93% | 100 |
| 29 | Grapes | 0% | 0% | 10% | 5% | 85% | 100 |

| | | | | Table continued | | | |
|---|---|---|---|---|---|---|---|
| 30 | Apples | 0% | 13% | 8% | 20% | 60% | 100 |
| 31 | Melon | 8% | 3% | 13% | 78% | 0% | 100 |
| 32 | Peaches | 70% | 13% | 8% | 8% | 0% | 100 |
| 33 | Apricots | 78% | 3% | 3% | 5% | 13% | 100 |
| 34 | Strawberries | 28% | 3% | 30% | 10% | 30% | 100 |
| 35 | Avocadoes | 70% | 3% | 13% | 15% | 0% | 100 |
| 36 | Spinach | 20% | 3% | 15% | 8% | 55% | 100 |
| 37 | Carrots | 3% | 13% | 15% | 70% | 0% | 100 |
| 38 | Green beans | 25% | 30% | 30% | 15% | 0% | 100 |
| 39 | Cucumber | 5% | 8% | 10% | 13% | 65% | 100 |
| 40 | Celery | 3% | 5% | 8% | 10% | 75% | 100 |
| 41 | Mushrooms | 18% | 3% | 8% | 23% | 50% | 100 |
| 42 | Parsnips | 30% | 8% | 15% | 20% | 28% | 100 |
| 43 | Peas | 23% | 18% | 15% | 25% | 20% | 100 |
| 44 | Sweet corn | 3% | 15% | 15% | 68% | 0% | 100 |
| 45 | Broccoli | 30% | 23% | 43% | 5% | 0% | 100 |
| 46 | Salad leaves | 38% | 3% | 25% | 18% | 18% | 100 |
| 47 | Red peppers | 8% | 10% | 20% | 63% | 0% | 100 |
| 48 | Raw tomatoes | 15% | 10% | 13% | 23% | 53% | 100 |
| 49 | Beetroot | 20% | 18% | 10% | 53% | 0% | 100 |
| 50 | Brussels sprouts | 58% | 5% | 23% | 15% | 0% | 100 |
| 51 | Butter | 45% | 15% | 13% | 28% | 0% | 100 |
| 52 | Butter-like spreads | 8% | 3% | 13% | 78% | 0% | 100 |
| 53 | Cream | 70% | 5% | 13% | 13% | 0% | 100 |
| 54 | Mayonnaise | 25% | 5% | 5% | 65% | 0% | 100 |
| 55 | Plain biscuits | 3% | 5% | 35% | 58% | 0% | 100 |
| 56 | Chocolate biscuits | 5% | 5% | 8% | 83% | 0% | 100 |
| 57 | Cake | 0% | 0% | 8% | 20% | 73% | 100 |
| 58 | Ice cream | 0% | 3% | 3% | 10% | 85% | 100 |
| 59 | Custard | 13% | 3% | 3% | 83% | 0% | 100 |
| 60 | Chocolate | 0% | 0% | 3% | 10% | 88% | 100 |
| 61 | Crisps | 0% | 0% | 0% | 5% | 95% | 100 |
| 62 | Chewy gummy sweets | 10% | 23% | 20% | 48% | 0% | 100 |
| | Average | 21% | 8% | 17% | 31% | 24% | 100 |

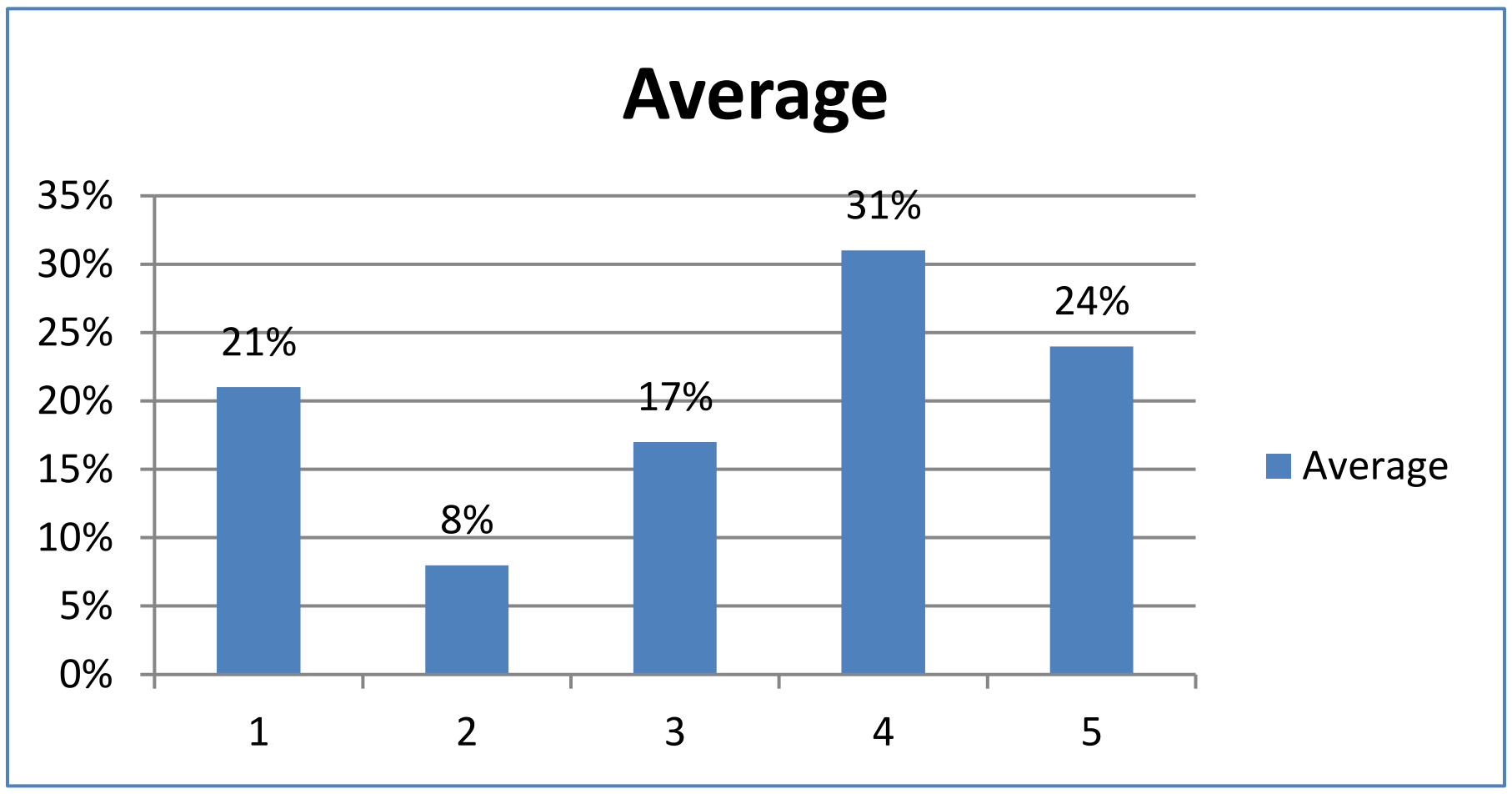

Table-2.5.2.2: Total Scaling Point of food preference un married between 20-29 years based on table-2.5.2.2

| Variable No. | 1- Dislike a lot | 2- Dislike a little | 3- Neither like or dislike | 4- Like a little | 5- Like a lot | Total | Position |
|---|---|---|---|---|---|---|---|
| 1 | 0 | 0 | 12 | 124 | 25 | 161 | |
| 2 | 1 | 0 | 21 | 92 | 45 | 159 | |
| 3 | 3 | 6 | 12 | 8 | 140 | 169 | |
| 4 | 1 | 0 | 12 | 32 | 135 | 180 | seventh |
| 5 | 6 | 16 | 78 | 0 | 0 | 100 | |
| 6 | 28 | 20 | 6 | 0 | 0 | 54 | |
| 7 | 13 | 4 | 6 | 36 | 70 | 129 | |
| 8 | 16 | 4 | 12 | 36 | 45 | 113 | |
| 9 | 14 | 6 | 3 | 36 | 65 | 124 | |
| 10 | 28 | 6 | 27 | 0 | 0 | 61 | |
| 11 | 13 | 6 | 21 | 68 | 0 | 108 | |
| 12 | 2 | 6 | 33 | 96 | 0 | 137 | |
| 13 | 2 | 14 | 15 | 100 | 10 | 141 | |
| 14 | 1 | 4 | 39 | 96 | 0 | 140 | |
| 15 | 10 | 8 | 27 | 32 | 45 | 122 | |
| 16 | 12 | 8 | 18 | 68 | 5 | 111 | |
| 17 | 1 | 4 | 15 | 124 | 5 | 149 | |
| 18 | 16 | 24 | 24 | 16 | 0 | 80 | |
| 19 | 39 | 0 | 3 | 0 | 0 | 42 | |
| 20 | 5 | 6 | 42 | 32 | 50 | 135 | |
| 21 | 0 | 6 | 111 | 0 | 0 | 117 | |
| 22 | 1 | 0 | 3 | 24 | 160 | 188 | Fifth |
| 23 | 3 | 0 | 24 | 48 | 85 | 160 | |
| 24 | 4 | 6 | 42 | 76 | 0 | 128 | |
| 25 | 7 | 4 | 42 | 68 | 0 | 121 | |
| 26 | 5 | 2 | 9 | 48 | 80 | 144 | |
| 27 | 6 | 2 | 15 | 112 | 0 | 135 | |
| 28 | 0 | 2 | 3 | 4 | 185 | 194 | second |
| 29 | 0 | 0 | 12 | 8 | 170 | 190 | fourth |
| 30 | 0 | 10 | 9 | 32 | 120 | 171 | Eighth |
| 31 | 3 | 2 | 15 | 124 | 0 | 144 | |
| 32 | 28 | 10 | 9 | 12 | 0 | 59 | |
| 33 | 31 | 2 | 3 | 8 | 25 | 69 | |
| 34 | 11 | 2 | 36 | 16 | 60 | 125 | |
| 35 | 28 | 2 | 15 | 24 | 0 | 69 | |
| 36 | 8 | 2 | 18 | 12 | 110 | 150 | |
| 37 | 1 | 10 | 18 | 112 | 0 | 141 | |
| 38 | 10 | 24 | 36 | 24 | 0 | 94 | |
| 39 | 2 | 6 | 12 | 20 | 130 | 170 | |
| 40 | 1 | 4 | 9 | 16 | 150 | 180 | seventh |
| 41 | 7 | 2 | 9 | 36 | 100 | 154 | |
| 42 | 12 | 6 | 18 | 32 | 55 | 123 | |
| 43 | 9 | 14 | 18 | 40 | 40 | 121 | |
| 44 | 1 | 12 | 18 | 108 | 0 | 139 | |
| 45 | 12 | 18 | 51 | 8 | 0 | 89 | |
| 46 | 15 | 2 | 30 | 28 | 35 | 110 | |
| 47 | 3 | 8 | 24 | 100 | 0 | 135 | |
| 48 | 6 | 8 | 15 | 36 | 105 | 170 | |
| 49 | 8 | 14 | 12 | 84 | 0 | 118 | |
| 50 | 23 | 4 | 27 | 24 | 0 | 78 | |
| 51 | 18 | 12 | 15 | 44 | 0 | 89 | |
| 52 | 3 | 2 | 15 | 124 | 0 | 144 | |
| 53 | 28 | 4 | 15 | 20 | 0 | 67 | |
| 54 | 10 | 4 | 6 | 104 | 0 | 124 | |
| 55 | 1 | 4 | 42 | 92 | 0 | 139 | |
| 56 | 2 | 4 | 9 | 132 | 0 | 147 | |
| 57 | 0 | 0 | 9 | 32 | 145 | 186 | sixth |
| 58 | 0 | 2 | 3 | 16 | 170 | 191 | Third |
| 59 | 5 | 2 | 3 | 132 | 0 | 142 | |
| 60 | 0 | 0 | 3 | 16 | 175 | 194 | Second |
| 61 | 0 | 0 | 0 | 8 | 190 | 198 | First |
| 62 | 4 | 18 | 24 | 76 | 0 | 122 | |
| Average score of items | | | | | | 131.35 | |
| Average score of respondents | | | | | | 2.11 | |

The total point of each variable or item of food preference is given in figure -2.5.2 .1

Figure-2.5.2.1: total point of each variable or item of food preference

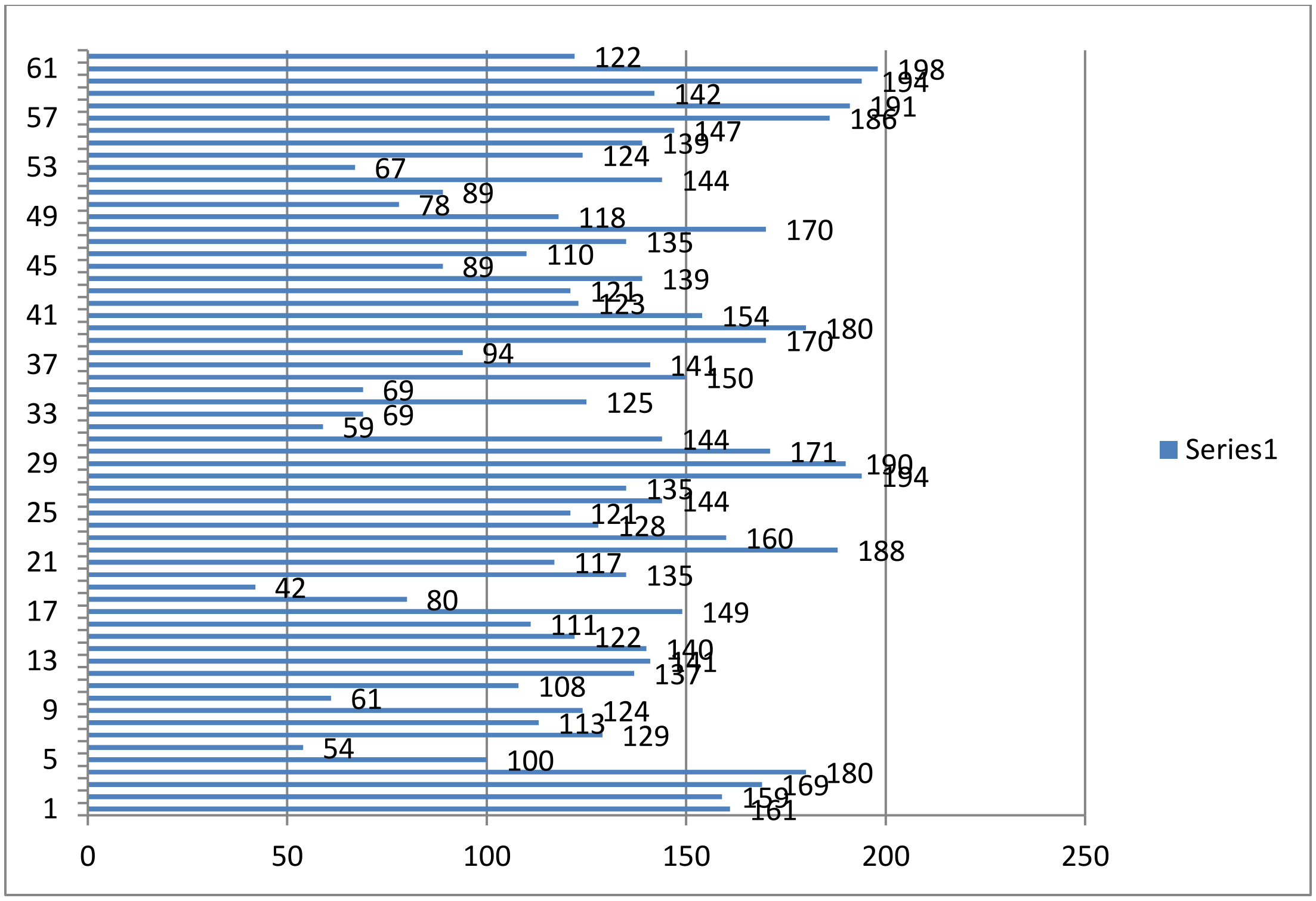

From the above analysis we can finalize the top ten preferred food items as given in table 2.5.2.3

**Table-2.5.2.3: top ten preferred food items**

| Serial No. | Food items | Total point | Position |
|---|---|---|---|
| 1 | Crisps | 198 | First |
| 2 | Chocolate | 194 | Second |
| 3 | Oranges | 194 | second |
| 4 | Ice cream | 191 | Third |
| 5 | Grapes | 190 | fourth |
| 6 | Chips | 188 | Fifth |
| 7 | cake | 186 | sixth |
| 8 | Celery | 180 | seventh |
| 9 | Chicken | 180 | seventh |
| 10 | Apples | 171 | Eighth |

## 2.6. Food Preference for Adults (29-64) Married Male and Female:

**2.6.1: Socioeconomic Profile of Respondents:** Socioeconomic profiles of respondents are discussed through table 2.6.1.1 to 2.6.1.5 below:

Table-2.6.1.1: Respondents by sex

| Variable | Number |
|---|---|
| Sex of respondent | |
| Male | 20 |
| Female | 20 |
| Total | 40 |
| Marital status | |
| married | 40 |
| Total | 40 |

Table-2.6.1.1 depicted that there were 20 male and 20 female respondents and all them were married of 29 -64aged.

Table: 2.6.1.2: age, annual income and expenditure of the respondents

| Variables | Minimum | Maximum | Mean | Std. Deviation |
|---|---|---|---|---|
| age | 24.0 | 59.0 | 42.425 | 9.2733 |
| annual income(000) | 120 | 6000 | 574.15 | 902.991 |
| annual expenditure(000) | 120 | 1200 | 419.20 | 221.168 |

Table-2.6.1.2 showed that age ranged from 24 to 59 with an average 42 and standard deviation 9. The annual income varied between 120 thousands and 6000 thousand with an average 574 and standard deviation 902 while annual expenditure fluctuated between 120 thousands and 1200 thousands with an average 419 and standard deviation 221.

Table-2.6.1.3: Occupation of respondents

| Occupation | Frequency | Valid Percent | Cumulative Percent |
|---|---|---|---|
| Service | 8 | 20.0 | 20.0 |
| Business | 26 | 65.0 | 85.0 |
| Both of them | 4 | 10.0 | 95.0 |
| Others | 2 | 5.0 | 100.0 |
| Total | 40 | 100.0 | |

Table -2.6.1.3 showed that the occupation of lion portion of respondents is business 26(65%) following service 8(20%) and both business and service 4(10%).

Table-2.6.1.4: Occupation of spouse

| Occupation | Frequency | Valid Percent | Cumulative Percent |
|---|---|---|---|
| Household | 23 | 57.5 | 57.5 |
| Service | 12 | 30.0 | 87.5 |
| Both service and Household | 4 | 10.0 | 97.5 |
| Business | 1 | 2.5 | 100.0 |
| Total | 40 | 100.0 | |

Table-2. 6.1.4 showed that the occupation of lion portion of spouse is household 23( 57.5%) following service 12(30%) and both business and service 4(10%) and business 1(2.5%).

Table-2.6.1.5.Education status of the respondents

| Year of Schooling | Frequency | Valid Percent | Cumulative Percent |
|---|---|---|---|
| 5.0 | 4 | 10.0 | 10.0 |
| 6.0 | 1 | 2.5 | 12.5 |
| 8.0 | 2 | 5.0 | 17.5 |
| 9.0 | 2 | 5.0 | 22.5 |
| 10.0 | 8 | 20.0 | 42.5 |
| 11.0 | 2 | 5.0 | 47.5 |
| 12.0 | 7 | 17.5 | 65.0 |
| 14.0 | 3 | 7.5 | 72.5 |
| 16.0 | 5 | 12.5 | 85.0 |
| 17.0 | 1 | 2.5 | 87.5 |
| 18.0 | 5 | 12.5 | 100.0 |
| Total | 40 | 100.0 | |

Table-2.6.1.5 showed that education of the respondents in terms of schooling year varied between 5 and 18 years and 10 years schooling is highest number of respondents 8(20%) following 12 years 7(17.5%) .

**2.6.2. Descriptive Statistics of food preference married adults :** Table-2.6.2.1 depicted that in the five point scale the highest frequency in the dislike a lot (35), dislike a little (15) neither like or dislike (15) , like a little (32) and like a lot (26). Reliability of the questionnaire was checked by Cronbach's Alpha test which is 0.876 that means there is consistency of the items or variables. The total point corresponding to the frequency table 2.6.2.1 is given in table 2.6.2.2.

Table-2.6.2.1: Frequency distribution in percentage of food preference of married adults

| Serial No. | Food Item/Variable | 1- Dislike a lot | 2- Dislike a little | 3- Neither like or dislike | 4- Like a little | 5- Like a lot | Total |
|---|---|---|---|---|---|---|---|
| 1 | Beef | 3% | 5% | 5% | 80% | 8% | 100 |
| 2 | Beef burgers | 40% | 18% | 10% | 3% | 30% | 100 |
| 3 | Lamb | 70% | 13% | 8% | 10% | 0% | 100 |
| 4 | Chicken | 0% | 3% | 5% | 38% | 55% | 100 |
| 5 | Bacon | 73% | 15% | 8% | 5% | 0% | 100 |
| 6 | Ham | 70% | 15% | 10% | 5% | 0% | 100 |
| 7 | Sausage(Meat) | 33% | 5% | 5% | 23% | 35% | 100 |
| 8 | White fish | 18% | 20% | 30% | 20% | 18% | 100 |
| 9 | Oily fish | 25% | 8% | 20% | 23% | 25% | 100 |
| 10 | Smoked Salmon | 30% | 20% | 18% | 18% | 15% | 100 |
| 11 | Tinned Tuna | 18% | 25% | 33% | 18% | 8% | 100 |
| 12 | Eggs | 30% | 33% | 25% | 10% | 3% | 100 |
| 13 | Baked beans | 43% | 20% | 23% | 10% | 5% | 100 |
| 14 | Bread or Bread rolls | 38% | 30% | 33% | 0% | 0% | 100 |
| 15 | Bran cereal | 50% | 23% | 15% | 8% | 5% | 100 |
| 16 | Porridge | 3% | 3% | 8% | 38% | 50% | 100 |
| 17 | Plain boiled rice | 3% | 5% | 13% | 38% | 43% | 100 |
| 18 | Sugared cereal | 0% | 5% | 15% | 30% | 50% | 100 |
| 19 | Hummus | 0% | 13% | 13% | 25% | 50% | 100 |
| 20 | Wheat cereal | 0% | 0% | 5% | 30% | 65% | 100 |
| 21 | Potatoes | 58% | 25% | 18% | 0% | 0% | 100 |
| 22 | Chips | 88% | 13% | 0% | 0% | 0% | 100 |

| | | | Table continued | | | | |
|---|---|---|---|---|---|---|---|
| 23 | Rice or corn cereal | 0% | 0% | 3% | 30% | 68% | 100 |
| 24 | Soft cheese | 23% | 20% | 18% | 25% | 15% | 100 |
| 25 | Hard cheese | 43% | 30% | 20% | 8% | 0% | 100 |
| 26 | Cottage Cheese | 38% | 18% | 23% | 13% | 10% | 100 |
| 27 | Plain, Low-fat yoghurt | 5% | 8% | 15% | 30% | 43% | 100 |
| 28 | Oranges | 8% | 15% | 15% | 15% | 48% | 100 |
| 29 | Grapes | 13% | 13% | 33% | 30% | 13% | 100 |
| 30 | Apples | 8% | 18% | 13% | 45% | 18% | 100 |
| 31 | Melon | 5% | 18% | 20% | 43% | 15% | 100 |
| 32 | Peaches | 63% | 13% | 13% | 13% | 0% | 100 |
| 33 | Apricots | 40% | 20% | 20% | 15% | 5% | 100 |
| 34 | Strawberries | 23% | 15% | 38% | 15% | 10% | 100 |
| 35 | Avocadoes | 43% | 18% | 20% | 15% | 5% | 100 |
| 36 | Spinach | 45% | 25% | 18% | 13% | 0% | 100 |
| 37 | Carrots | 30% | 20% | 23% | 28% | 0% | 100 |
| 38 | Green beans | 23% | 25% | 20% | 18% | 15% | 100 |
| 39 | Cucumber | 25% | 30% | 35% | 10% | 0% | 100 |
| 40 | Celery | 30% | 23% | 23% | 18% | 8% | 100 |
| 41 | Mushrooms | 3% | 8% | 10% | 20% | 60% | 100 |
| 42 | Parsnips | 25% | 15% | 20% | 18% | 23% | 100 |
| 43 | Peas | 23% | 18% | 40% | 13% | 8% | 100 |
| 44 | Sweet corn | 45% | 23% | 15% | 18% | 0% | 100 |
| 45 | Broccoli | 20% | 20% | 15% | 30% | 20% | 100 |
| 46 | Salad leaves | 30% | 20% | 25% | 18% | 8% | 100 |
| 47 | Red peppers | 25% | 20% | 25% | 20% | 10% | 100 |
| 48 | Raw tomatoes | 38% | 25% | 25% | 13% | 0% | 100 |
| 49 | Beetroot | 45% | 38% | 10% | 8% | 0% | 100 |
| 50 | Brussels sprouts | 58% | 5% | 23% | 15% | 0% | 100 |
| 51 | Butter | 50% | 20% | 15% | 15% | 0% | 100 |
| 52 | Butter-like spreads | 73% | 15% | 8% | 5% | 0% | 100 |
| 53 | Cream | 78% | 10% | 8% | 5% | 0% | 100 |
| 54 | Mayonnaise | 93% | 8% | 0% | 0% | 0% | 100 |
| 55 | Plain biscuits | 50% | 25% | 15% | 8% | 3% | 100 |
| 56 | Chocolate biscuits | 70% | 10% | 10% | 10% | 0% | 100 |
| 57 | Cake | 80% | 10% | 8% | 3% | 0% | 100 |
| 58 | Ice cream | 95% | 3% | 3% | 0% | 0% | 100 |
| 59 | Custard | 13% | 3% | 3% | 83% | 0% | 100 |
| 60 | Chocolate | 75% | 13% | 10% | 3% | 0% | 100 |
| 61 | Crisps | 63% | 18% | 15% | 5% | 0% | 100 |
| 62 | Chewy gummy sweets | 75% | 23% | 3% | 0% | 0% | 100 |
| | Average | 37% | 16% | 16% | 18% | 14% | 100 |

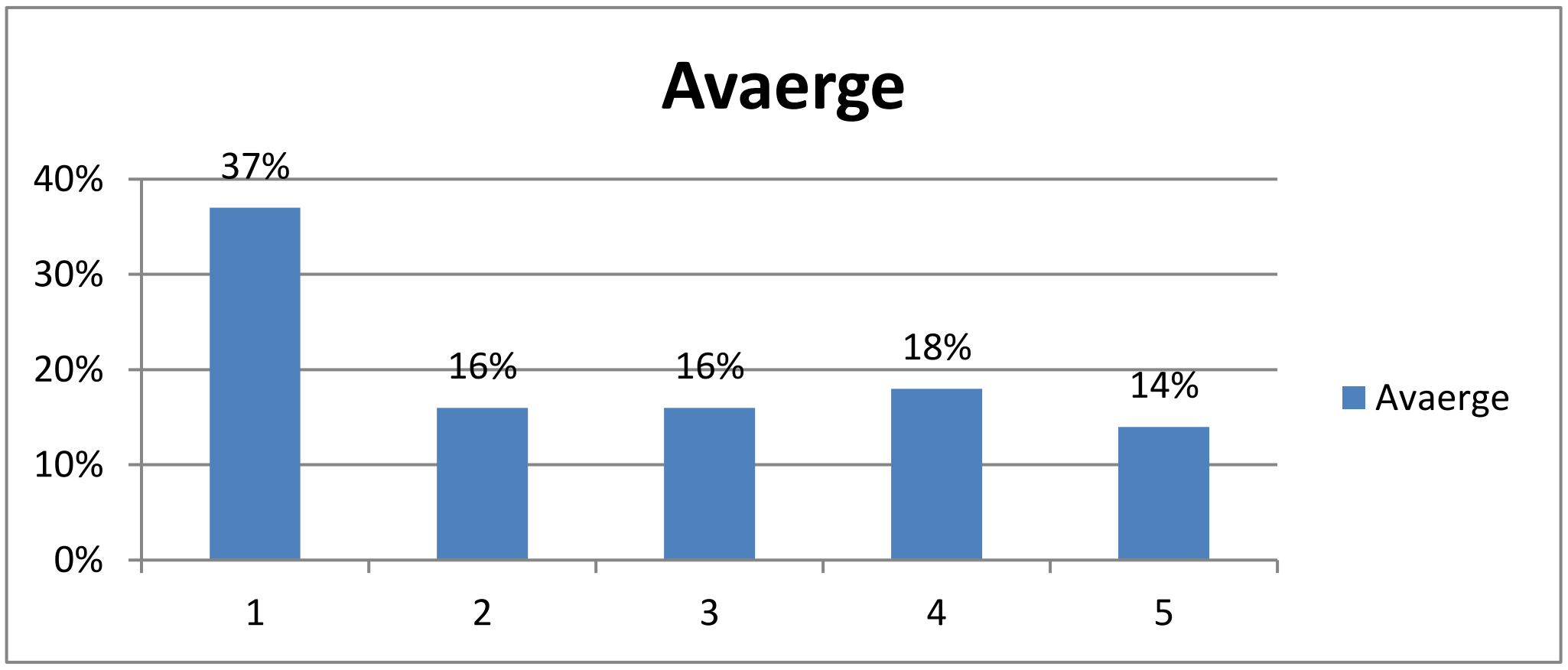

Table-2.6.2.2: Total Scaling Point of food preference married between 29-64 years based on table-2.6.2.1

| Variable No. | 1- Dislike a lot | 2- Dislike a little | 3- Neither like or dislike | 4- Like a little | 5- Like a lot | Total | Position |
|---|---|---|---|---|---|---|---|
| 1 | 1 | 4 | 6 | 128 | 15 | 154 | Ninth |
| 2 | 16 | 14 | 12 | 4 | 60 | 106 | |
| 3 | 28 | 10 | 9 | 16 | 0 | 63 | |
| 4 | 0 | 2 | 6 | 60 | 110 | 178 | third |
| 5 | 29 | 12 | 9 | 8 | 0 | 58 | |
| 6 | 28 | 12 | 12 | 8 | 0 | 60 | |
| 7 | 13 | 4 | 6 | 36 | 70 | 129 | |
| 8 | 7 | 16 | 36 | 32 | 35 | 126 | |
| 9 | 10 | 6 | 24 | 36 | 50 | 126 | |
| 10 | 12 | 16 | 21 | 28 | 30 | 107 | |
| 11 | 7 | 20 | 39 | 28 | 15 | 109 | |
| 12 | 12 | 26 | 30 | 16 | 5 | 89 | |
| 13 | 17 | 16 | 27 | 16 | 10 | 86 | |
| 14 | 15 | 24 | 39 | 0 | 0 | 78 | |
| 15 | 20 | 18 | 18 | 12 | 10 | 78 | |
| 16 | 1 | 2 | 9 | 60 | 100 | 172 | Fourth |
| 17 | 1 | 4 | 15 | 60 | 85 | 165 | seventh |
| 18 | 0 | 4 | 18 | 48 | 100 | 170 | sixth |
| 19 | 0 | 10 | 15 | 40 | 100 | 165 | seventh |
| 20 | 0 | 0 | 6 | 48 | 130 | 184 | Second |
| 21 | 23 | 20 | 21 | 0 | 0 | 64 | |
| 22 | 35 | 10 | 0 | 0 | 0 | 45 | |
| 23 | 0 | 0 | 3 | 48 | 135 | 186 | First |
| 24 | 9 | 16 | 21 | 40 | 30 | 116 | |
| 25 | 17 | 24 | 24 | 12 | 0 | 77 | |
| 26 | 15 | 14 | 27 | 20 | 20 | 96 | |
| 27 | 2 | 6 | 18 | 48 | 85 | 159 | Eight |
| 28 | 3 | 12 | 18 | 24 | 95 | 152 | |
| 29 | 5 | 10 | 39 | 48 | 25 | 127 | |
| 30 | 3 | 14 | 15 | 72 | 35 | 139 | |
| 31 | 2 | 14 | 24 | 68 | 30 | 138 | |
| 32 | 25 | 10 | 15 | 20 | 0 | 70 | |
| 33 | 16 | 16 | 24 | 24 | 10 | 90 | |
| 34 | 9 | 12 | 45 | 24 | 20 | 110 | |
| 35 | 17 | 14 | 24 | 24 | 10 | 89 | |
| 36 | 18 | 20 | 21 | 20 | 0 | 79 | |
| 37 | 12 | 16 | 27 | 44 | 0 | 99 | |
| 38 | 9 | 20 | 24 | 28 | 30 | 111 | |
| 39 | 10 | 24 | 42 | 16 | 0 | 92 | |
| 40 | 12 | 18 | 27 | 28 | 15 | 100 | |
| 41 | 1 | 6 | 12 | 32 | 120 | 171 | Fifth |
| 42 | 10 | 12 | 24 | 28 | 45 | 119 | |
| 43 | 9 | 14 | 48 | 20 | 15 | 106 | |
| 44 | 18 | 18 | 18 | 28 | 0 | 82 | |
| 45 | 8 | 16 | 18 | 48 | 40 | 130 | |
| 46 | 12 | 16 | 30 | 28 | 15 | 101 | |
| 47 | 10 | 16 | 30 | 32 | 20 | 108 | |
| 48 | 15 | 20 | 30 | 20 | 0 | 85 | |
| 49 | 18 | 30 | 12 | 12 | 0 | 72 | |
| 50 | 23 | 4 | 27 | 24 | 0 | 78 | |
| 51 | 20 | 16 | 18 | 24 | 0 | 78 | |
| 52 | 29 | 12 | 9 | 8 | 0 | 58 | |
| 53 | 31 | 8 | 9 | 8 | 0 | 56 | |
| 54 | 37 | 6 | 0 | 0 | 0 | 43 | |
| 55 | 20 | 20 | 18 | 12 | 5 | 75 | |
| 56 | 28 | 8 | 12 | 16 | 0 | 64 | |
| 57 | 32 | 8 | 9 | 4 | 0 | 53 | |
| 58 | 38 | 2 | 3 | 0 | 0 | 43 | |
| 59 | 5 | 2 | 3 | 132 | 0 | 142 | |
| 60 | 30 | 10 | 12 | 4 | 0 | 56 | |
| 61 | 25 | 14 | 18 | 8 | 0 | 65 | |
| 62 | 30 | 18 | 3 | 0 | 0 | 51 | |
| Average score of items | | | | | | 102.87 | |
| Average score of respondents | | | | | | 1.66 | |

Figure: 2.6.2.2.The total point od food preference of adults married

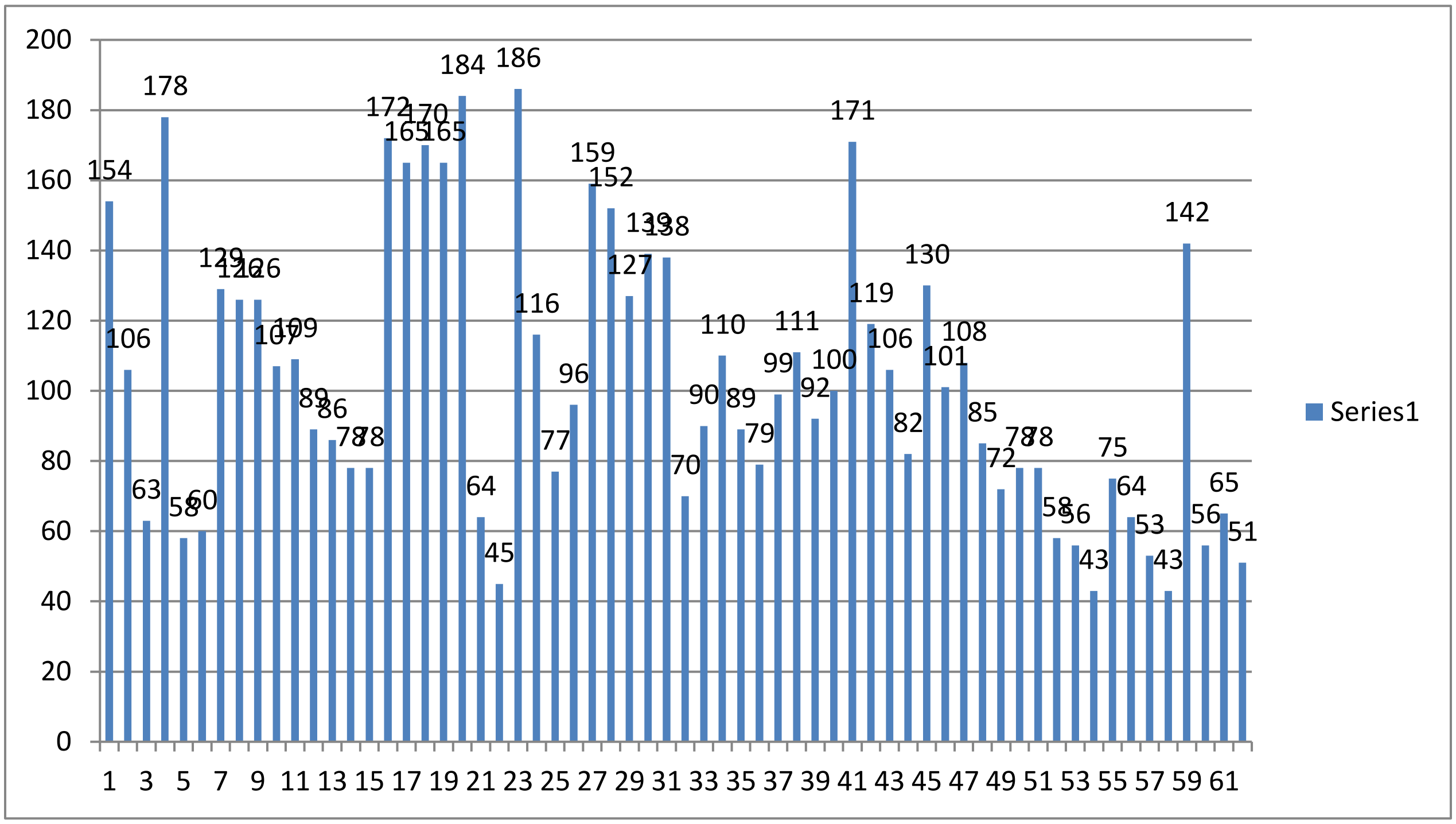

From the above analysis the top ten preferred food items of married adults are listed below

| Serial No | Food Item | Total | Position |
|---|---|---|---|
| 1 | Rice or corn cereal | 186 | First |
| 2 | Wheat cereal | 184 | Second |
| 3 | Chicken | 178 | third |
| 4 | Porridge | 172 | Fourth |
| 5 | Mushrooms | 171 | Fifth |
| 6 | Sugared cereal | 170 | sixth |
| 7 | Plain boiled rice | 165 | seventh |
| 8 | Hummus | 165 | seventh |
| 9 | Plain, Low-fat yoghurt | 159 | Eight |
| 10 | Beef | 154 | Ninth |

**2.7. A Comparison of food preference between adults of 20-29 unmarried and 29-64 married:** Being the two groups were given same checklist of food preference so we can compare the preferences of them.

**2.7.1: Comparison of total point score by food items of the group**s: From table-2.5.2.2 and table-2.6.2.2 a comparative table is made as below in table 2.7.1

Table-2.7.1: Comparative food preference score of unmarried adults of (20-29) married adults (29-64 ) based on table 5.2.2.2 and 6.2.2.2

| Variable No. | unmarried adults (20-29) | married adults(29-64 ) |
|---|---|---|
| 1 | 161 | 154 |
| 2 | 159 | 106 |
| 3 | 169 | 63 |
| 4 | 180 | 178 |
| 5 | 100 | 58 |
| 6 | 54 | 60 |
| 7 | 129 | 129 |
| 8 | 113 | 126 |
| 9 | 124 | 126 |
| 10 | 61 | 107 |
| 11 | 108 | 109 |
| 12 | 137 | 89 |
| 13 | 141 | 86 |
| 14 | 140 | 78 |
| 15 | 122 | 78 |
| 16 | 111 | 172 |
| 17 | 149 | 165 |
| 18 | 80 | 170 |
| 19 | 42 | 165 |
| 20 | 135 | 184 |
| 21 | 117 | 64 |
| 22 | 188 | 45 |
| 23 | 160 | 186 |
| 24 | 128 | 116 |
| 25 | 121 | 77 |
| 26 | 144 | 96 |
| 27 | 135 | 159 |
| 28 | 194 | 152 |
| 29 | 190 | 127 |
| 30 | 171 | 139 |
| 31 | 144 | 138 |
| 32 | 59 | 70 |
| 33 | 69 | 90 |
| 34 | 125 | 110 |
| 35 | 69 | 89 |
| 36 | 150 | 79 |
| 37 | 141 | 99 |
| 38 | 94 | 111 |
| 39 | 170 | 92 |
| 40 | 180 | 100 |
| 41 | 154 | 171 |
| 42 | 123 | 119 |
| 43 | 121 | 106 |
| 44 | 139 | 82 |
| 45 | 89 | 130 |
| 46 | 110 | 101 |
| 47 | 135 | 108 |
| 48 | 170 | 85 |
| 49 | 118 | 72 |
| 50 | 78 | 78 |
| 51 | 89 | 78 |
| 52 | 144 | 58 |
| 53 | 67 | 56 |
| 54 | 124 | 43 |
| 55 | 139 | 75 |
| 56 | 147 | 64 |
| 57 | 186 | 53 |
| 58 | 191 | 43 |
| 59 | 142 | 142 |
| 60 | 194 | 56 |
| 61 | 198 | 65 |
| 62 | 122 | 51 |

Figure-2.7.1: Comparative food preference of unmarried adults and married adults

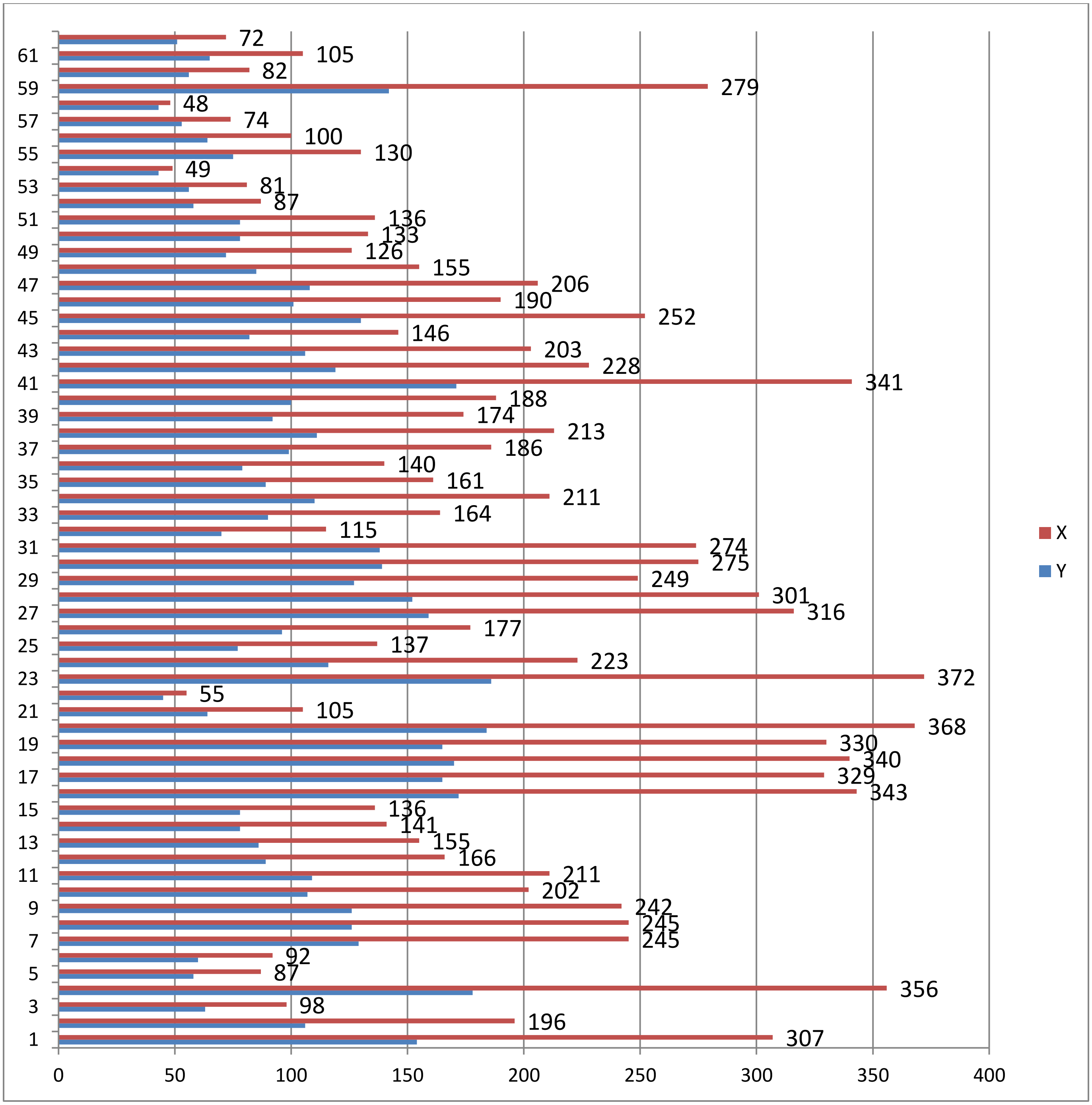

Note: X= Food preference of unmarried adults and Y= Food preference of married adults

**2.7.2: Two –sample t- test**: The food preference of un married adults and food preference of married are compared and test whether there is any significant difference among their choices. And for this two –sample t- test is applied.

**Table-2.7.2.1Group Statistics**

| Group | N | Mean | Std. Deviation | Std. Error Mean |
|---|---|---|---|---|
| Unmarried | 62 | 191.10 | 89.873 | 11.414 |
| Married | 62 | 102.87 | 39.889 | 5.066 |

**Table-2.7.2.2:Independent Samples Test**

| | Levene's Test for Equality of Variances | | t-test for Equality of Means | | | | | | |
|---|---|---|---|---|---|---|---|---|---|
| | | | | | | | | 95% Confidence Interval of the Difference | |
| | F | Sig. | t | df | Sig. (2-tailed) | Mean Difference | Std. Error Difference | Lower | Upper |
| Equal variances assumed | 36.142 | .000 | 7.065 | 122 | .000 | 88.226 | 12.488 | 63.505 | 112.946 |
| Equal variances not assumed | | | 7.065 | 84.135 | .000 | 88.226 | 12.488 | 63.393 | 113.058 |

In the table of group statistics we can find that mean score of unmarried and married are 191.10 and 102.87 correspondingly with Std. deviation 89.873 and 39.889 for unmarried and married respectively. Therefore the average score of unmarried adult is higher than that of married adults. The table of independent Samples Test depicted two test - Levene's Test for Equality of Variances and t-test for Levene's Test for Equality of means. In the first test i.e. Levene's Test for Equality of Variances here our null hypothesis is that the both the variances are equal and alternative is that the variances are different for the two groups. The Levene's Test for Equality of Variances showed the p- value is 0.00 that means there are significant differences of variances of the two groups. The second test is t-test for Equality of Means which is presented in the second row and showed that the observed test statistics is 7.065 with p- value 0 .000. In this case we can also reject the null hypothesis and can conclude that there is significant difference between the means of the two groups.

Finally we can conclude that there is significant difference between the food preference of unmarried adults and married adults and their preferences s are not same.

**8. Overall awareness and behavior of food hygiene and nutrition**: There are six kinds of respondents and after analysis the overall average score each group is given in table 8.

Table 8: Overall average score of the respondents in different scale

| Serial No. | Respondents | Number of respondents | Scale of questionnaire | Item-wise Average score | Average score by respondents |
|---|---|---|---|---|---|
| 1. | Parents of baby up to 4 years | 40 | 5.00 | 100.5 | 2.51 |
| 2. | Parents of Child eating behavior between 5-9 years | 40 | 5.00 | 117.66 | 3.36 |
| 3. | Adolescent (10-19 years ) Food Habits | 34 | 3.00 | 43.56 | 1.89 |
| 4. | Parental Feeding Style to his Children up to 9 years | 34 | 5.00 | 114.52 | 4.24 |
| 5. | Food preference of adults unmarried (20-29) | 40 | 5.00 | 131.35 | 2.12 |
| 6. | Food Preference for Adults (29-64) Married Male and Female | 40 | 5.00 | 102.87 | 1.66 |

**9. Conclusion:** This chapter has analyzed collected data through different descriptive and inferential statistical techniques and methods. The major findings produced from the analyses are given in the next chapter.

# CHAPTER THREE

# FINDINGS AND CONCLUSION

**Introduction**: In the previous chapter the data has been analyzed using different statistical techniques of six groups of our respondents in this chapter the major findings and suggestions are given below:

**3.1.Findings on f Baby's Eating Behavior up to 4(Four Year) Responded by Parents:** Table-2.1.2: Frequency of distribution baby's eating behavior showed the highest frequency as follows: (i) in the scale never =1 "Even when my baby had just eaten well he/she was happy to feed again if offered" showed highest frequency 50% (20); (ii) in the scale rarely=2 "Baby could easily take a feed within 30 minutes of the last one showed 45% (18), (iii) in the scale sometimes =3 "Baby seemed contented while feeding 50% (20), in the scale often=4 "Baby became distressed while feeding and Baby got full before taking all the milk I think he/she should have" 30% (12) each and in the scale 5=Always "Always Baby fed slowly" 30%(12). The average frequency in percentage in the different scales are 23%, 30%, 16%,11%and 11% respectively. Table 2.1.3 showed that the top five score items are – my baby loved milk(155) following my baby seemed contented while feeding (154), my baby fed slowly(150), my baby found it difficult to manage a complete feed(131), my baby got full up easily(113) and my baby frequently wanted more milk than I provided (105). The average score of the items is 100.5 and the average score of the respondents is 2.51. It means the eating behavior of baby lies between rarely and sometimes. Reliability of the questionnaire was checked by Cronbach's Alpha test which is 0.710 that means there is consistency of the items or variables. The PCA reduces 21 items into six factors significant for eating behavior of parents.

**3.2**. **Child eating behavior between 5-9 years:** The table -2.2.2.1 depicted the highest frequency among 35 items that in the scale never=1 " If allowed to, my child would eat too much item is highest 50% (20), in the scale rarely=2 " my child eats more when anxious item is highest" 47.5%(19) in the scale sometimes = 3, " my child loves food" 65%(26), in the scale often =4, " my child is difficult to please with meals item is highest" 45% 18 and in the scale always=5, "my child gets full up easily item is highest" 55%(22). The average percentage of frequency distribution in the five point scales are 20%,27%, 23%, 16% and 19% respectively.

The table 2.2.2.2 depicted the total point of each variable and the top seven items are as follows- My child gets full up easily(165) , Child cannot eat a meal if s/he has had a snack just before(149), My child is interested in tasting food s/he hasn't tasted before(145), My child is always asking for food(142), My child enjoys a wide variety of foods(142), My child leaves food on his/her plate at the end of a meal(137), My child loves food(137), My child is always asking for a drink(136), My child eats less when angry(135). The average of items is 117.66 and the average of respondents is 3.36. That means the eating behavior of children varied between sometimes and often. Through the PCA 35 items are reduced to 10 factors.

**3.3. Adolescent (10-19) Food Habits**: The food habits of adolescent is measured in terms of given items or variables with three point scale as Most like=2; Neutral=0; and like=1. Table 2.3.2.1 showed that the highest frequency in the scale Most like=2 is " When I am buying a soft drink, I usually choose a diet drink" 76%(26); in the scale like=1 the item "I rarely takeaway meals 82% (28)" and in the scale Neutral = 0,If I am buying crisps, I often choose a low-fat brand 32% (11) among twenty three items . The average percentages in the scales are 37%, 55% and 8% respectively. The top seven total scoring items are : (i)When I am buying a soft drink, I usually choose a diet drink (60), (ii)I eat at least servings of fruits most days(57),(iii)I usually avoid eating fried foods(53) (iv)I often eat sweet snacks between meals(51),(v) I make sure I eat at least one serving of fruit a day (49)(vi) I avoid eating lots of sausages and burgers.(46) and (vii) When I put butter or margarine on bread and I usually spread it on thinly If I have a packed lunch, I usually include some chocolate and/or biscuits (43). Average of score of the items is 43.56 and the average score of respondents is 1.89.

The PCA reduces 21 items to eight factors of food habits of adolescent.

**3.4. Parental Feeding Style to his Children up to 9 years-:** Table-2.4.2.1 provided that in the five point scale the highest frequency showing items are in the scales never=1 is " I allow my child to wander around during a meal(14)", rarely =2 is " I use puddings as a bribe to get my child to eat his/her main course rarely (17)" sometimes=3 is " I reward my child with something to eat when s/he is well behaved (16)" often =4 is " I decide how many snacks my child should have(8) and I encourage my child to eat a wide variety of foods (8)" and always=5 is " I encourage my child to try foods that s/he hasn't tasted before (19). The average percentage of frequency distribution in the five point scales are 15%, 10%, 33%, 13% and 30% respectively.

The top seven score point items in the scales are I encourage my child to eat a wide variety of foods (140), I encourage my child to try foods that s/he hasn't tasted before(139), I encourage my child to enjoy his/her food(138), I give my child something to eat to make him/her feel better when s/he is feeling upset(135), I encourage my child to look forward to the meal and I praise my child if s/he eats what I give him/her(134), I encourage my child to taste each of the foods I serve at mealtimes(132), I praise my child if s/he eats a new food(124)( Table-2.4.2.2). The total variance explained by the PCA is 78.96% that means the PCA is accepted and based on Eigen values seven items are considered important. The Average of items is 114.52 and the average of respondents is 4.24 that means the parental Feeding Style to his Children up to 9 years varies between often and always.

**3.5. Food preference of adults unmarried (20-29):** From table-2.5.2.1 we can find that in the five point scale the highest frequency in the scales are dislike a lot=1 is hummus 97.5% (39), dislike a little=2 is Green beans 30% (12) ,neither like or dislike=3 is Bacon 65% (26) , like a little=4 is Oranges 92.5% (37). The average percentages of frequency distribution in the different scales are 21%, 8%, 17% 31% and 24% respectively. The top ten preferred food items are Crisps(198), Chocolate(194), Oranges(194), Ice cream(191), Grapes(190), Chips(188) , cake(186), Celery(180), Chicken(180) and apples(171)(Table -2.5.2.3). The Average of items is 131.35 and the average of respondents is 2.12 that mean the food preference of adults (20-29 years) varies between neither like or dislike and like a little.

**3.6. Food Preference for Adults (29-64) Married Male and Female:** From table-2.6.2.1 we can find that in the five point scale the highest frequency in the scales are dislike a lot =1 is Chips 87.5% (35), dislike a little=2 is Beetroot 37.5% (15) neither like or dislike=3 is Strawberries 37.5% (15) , like a little=4 is Custard 82.5% (33) and like a lot =5 is Rice or corn cereal 67.5% (27). The average percentages of frequency distribution in the different scales are 37%, 16%, 16% 18% and 14% respectively. The top ten preferred food items are Rice or corn cereal(186), Wheat cereal(184), Chicken(178), Porridge(172) Mushrooms(171) , Sugared cereal(170) Plain boiled rice(165), Hummus(165), Plain, Low-fat yoghurt(159) and Beef(154)(Table--2.6.2.3). The Average of items is 102.87 and the average of respondents is 1.66 that means the food preference of adults (29-64 years ) varies between dislike a little and neither like nor dislike .

**3.7. A Comparison of food preference between adults of 20-29 unmarried and 29-64 married:** Being the two groups were given same checklist of food preference so we can compare the preferences of them. The average score of adults of 20-29 unmarried is 131.35 while it is 102.87 for married 102.87 so the average score of adults of 20-29 unmarried is higher than that of married 29-64 . Two sample t-test was applied to confirm whether there is any similarity between these two groups. In the table 2.7.1 of group statistics we can find that mean score of unmarried and married are 191.10 and 102.87 correspondingly with Std. deviation 89.873 and 39.889 for unmarried and married respectively. Therefore the average score of unmarried adult is higher than that of married adults. The table of independent Samples Test (table-2.7.2) depicted two test **-** Levene's Test for Equality of Variances and t-test for Levene's Test for Equality of means. In the first test i.e. Levene's Test for Equality of Variances here our null hypothesis is that the both the variances are equal and alternative is that the variances are different for the two groups. The Levene's Test for Equality of Variances showed the p- value is 0.00 that means there are significant differences of variances of the two groups. The second test is t-test for Equality of Means which is presented in the second row and showed that the observed test statistics is 7.065 with p- value 0 .000. In this case we can also reject the null hypothesis and can conclude that there is significant difference between the means of the two groups. Finally we can conclude that there is significant difference between the food preference of unmarried adults and married adults and their preferences s are not same.

**3.8 Suggestions for policy implications:** Following the analysis and findings the suggestions are prescribed below for better awareness of food safety hygiene and nutrition:

(i)The average awareness of parents for their baby up to four years is 2.51 out of 5 around 50%. Therefore, it is needed to increase awareness of the parents in feeding the babies.

(ii) The average awareness of parents to their child’s eating behavior between 5-9 years is 3.36 out of 5. The awareness is around 67% so we should be more careful in this regard.

(iii) The average awareness adolescent (10-19 years) food habit is 1.89 on three point scale which about 63% only. Therefore the consciousness of adolescent (10-19 years) is be to increase in taking food.

(iv) The average feeding styles of parents is 4.24 out of 5 to their children up to 9 years and in percentage it is 84%. Therefore it is about to satisfactory .

**REFERNCES**

ALAM, M. M., & HOSSAIN, M. K. (2018). Policy options on sustainable resource utilization and food security in Haor areas of Bangladesh: A theoretical approach. *International Journal of Social, Political and Economic Research*, 5(1), 11-28.

Alam, M. M., & Zakaria, A. F. M. (2021). A Probit Estimation of Urban Bases of Environmental Awareness: Evidence from Sylhet City, Bangladesh. *arXiv preprint* arXiv:2107.08342.

Bhuvneswari, R. and Muthukumar Dr. E.(2015), Food Safety Awareness Among Customers To Avoid The Risk Of Food Borne Illness, *Journal of Business Management & Social Sciences Research* (JBM&SSR) ISSN No: 2319-5614,Volume 4, No.8

Bruhn Christine M. (1997), Consumer Concerns: Motivating to Action, *Emerging Infectious Diseases,Vol.3.No.4*

Bruhn Christine M and Schutz Howard G. (1999), Consumer Food Safety Knowledge And Practices, Journal of Food Safety, Vol. 19,pp.73-87

Buzby JC, Roberts T 1995), ERS estimates US foodborne disease costs, Food Review, vol 18, pp37–42)

Cameron S, Walker W, Beers M et al 1995, Enterohaemorrhagic Escherichia coli outbreak in South Australia associated with the consumption of mettwurst, Communicable Diseases Intelligence, vol 19, pp70–100)

Emmanuel K. Yiridoe, Samuel Bonti-Ankomah and Ralph C. Martin(2005), Comparison of consumer perceptions and preference toward organic versus conventionally produced foods: A review and update of the literature, *Renewable Agriculture and Food Systems*,Vol. 20,Isuue.4 pp. 193–205 DOI: 10. 1079 / RAF 2005113

Daily star Newspaper(2105), Daily star, Bangladesh April 8, 2015

Doaa M., Nasr-Eldein, Sahar M. Soliman, Samar El Hossiny(2017), Effect of Applying Different Educational Strategies on Dietary Habits of Preparatory Schools' Pupils, *IOSR Journal of Nursing and Health Science* (IOSR-JNHS) e-ISSN: 2320–1959.p- ISSN: 2320–1940 Volume 6, Issue 4 Ver. IV. PP. 12-23

Government of Bangladesh (2018) Second Country Investment Plan (CIP) for Nutrition-sensitive Food Systems.

Finance Division. "Climate Fiscal Framework." (2014).

Finance Division. (2014). Bangladesh Climate Fiscal Framework, Ministry of Finance, GPRB Finance Division, Ministry of Finance, Government of the People's Republic of Bangladesh, Dhaka

FAO (2014) Guidelines for assessing nutrition-related Knowledge, Attitudes and Practices, KAP Manual.

Field, A. (2009). Discovering Statistics using SPSS. Sage: London

Fung Fred, Wang Huei Shyong &Menon Suresh(2018), Food Safety in 21st Century, *Biomedical Journal,* Science Direct ,Vol. 41, pp-88- 95

Hoddinott Jhon, Ahmed Akhter,Karachiwall Naureen I and Roy Shalini(2017), Nutrition behavior change communication causes sustained effects on IYCN knowledge in two cluster randomized trials in Bangladesh, *Maternal and Child Nutrition* ,Wiley.

Islam, M. R., Alam, M., & Afzal, M. N. İ. (2021). Nighttime Light Intensity and Child Health Outcomes in Bangladesh. *arXiv preprint* arXiv:2108.00926.

Karmakar P, Jahan N, Banik S, Das A, Rahman KA, Kundu SK and Sattar MM(2016), Food

Habits, Obesity and Introduction, Journal of Food & Nutritional Disorders , Nutritional Knowledge among the University Students in Noakhali Region of Bangladesh:ACross Sectional Study Journal of Food & Nutritional Disorders, Vol. 5, Issu.4 http://dx.doi.org/10.4172/2324-9323.1000201

Karmakar P., Jahan N., Banik S., Das A., Rahman K.A., Kundu S.K. and Sattar M.M. (2017) Food Habits, Obesity and Nutritional Knowledge among the University Students in Noakhali Region of Bangladesh: A Cross Sectional Study, Journal of Food and Nutritional Disorders 5(4)

Nutrition Iqbal Kabir, A.K.M, S. K. Roy, S. Khatoon (2013) Development of a Complementary Feeding Manual for Bangladesh, BBF for NFPCSP

Institute of Public Health and WHO(1994, 2003

Manikam Logan , Raghu Lingam , Isabel Lever , Emma C. Alexander , Chidi Amadi , Yasmin Milner , Taimur Shafi , Lucy Stephenson , Sonia Ahmed and Monica Lakhanpaul(2018), Complementary Feeding Practices for South Asian Young Children Living in High-Income Countries: A Systematic Review, *Nutrients* , 10, 1676; doi:10.3390/nu10111676 www.mdpi.com/journal/nutrients

Mahmud, I. and N. Mbuya (2016) Water, Sanitation, Hygiene, and Nutrition in Bangladesh - Can Building Toilets Affect Children’s Growth? World Bank Study

Nevin Sanlier, Ece Konaklioglu, (2012) "Food safety knowledge, attitude and food handling practices of students", British Food Journal, Vol. 114 Issue: 4, pp.469-480, https:// doi.org/ 10.1108/ 0007070 12 112 19504

Nesbitt Andrea, Thomas M.Kate,Marshal Barbara, Snedeker Kate,Meleta Kathryn,Watson Brenda and Bienefeld Monika(2014), Baseline for consumer food safety knowledge and behavior in Canada, Food Control, Vol.38, pp.157-173.

Nina Michaelidou and Hassan Louise M(n.d), The Role of Health Consciousness, Food Safety Concern and Ethical Identity on Attitudes and Intentions towards Organic Food, University of Birmingham Birmingham Business School University House Edgbaston and University of Stirling and the Open University Institute for Social Marketing University of Stirling

Nunnally JC (1978). *Psychometric Theory, 2nd ed*. New York: McGraw-Hill.

Nurul Alam, Swapan Kumar Roy, Tahmeed Ahmed, and A.M. Shamsir Ahmed(2010), Nutritional Status, Dietary Intake, and Relevant Knowledge of Adolescent Girls in Rural Bangladesh, J HEALTH POPUL NUTR 2010 Feb;28(1):86-94 ISSN 1606-0997 | $ 5.00+0.20

Quinlan Jennifer J. (2013), Foodborne Illness Incidence Rates and Food Safety Risks for Populations of Low Socioeconomic Status and Minority Race/Ethnicity: A Review of the Literature, *Int. J. Environ. Res. Public Health* Vol. 10, 3634-3652; doi:10.3390/ijerph10083634

Scallan, Elaine & Hoekstra, Mike & J Angulo, Frederick & V Tauxe, Robert & Widdowson, Marc-Alain & Roy, Sharon & L Jones, Jeffery & M Griffin, Patricia. (2011). Foodborne Illness Acquired in the United States—Major Pathogens. Emerging infectious diseases. 17. 7-15. 10.3201/eid1701.091101p1

Sedef Akgüngör, Bülent Miran , Canan Abay (2007), Consumer Willingness to Pay for Organic Products in Urban Turk, Contributed Paper prepared for presentation at the 105th EAAE Seminar ‘International Marketing and International Trade of Quality Food Products’, Bologna, Italy, March 8-10, 2007.

Saha ML, T Bari, MR Khan and S Hoque 2011. Bacteriological and physico-chemical properties of the Gulshan lake, Dhaka. Bangladesh J. Bot. Vol.40,Iss.2, pp.105-111

Saha, S., M.K. Zahid, S. Rasheed (2011) The Study of the Level of Knowledge, Attitude, Practices (KAP) as well as the Effects of School Environment on the Nutritional Status of Children (7-12) Coming from Affluent Families in the Dhaka City in Bangladesh, Bangladesh Journal of Nutrition Vol. 24-25

SPRING (2017) Bangladesh: Farmer Nutrition School Cohort Study. Sustainability of Improved Practices Following Graduation. Strengthening Partnerships, Results, and Innovations in Nutrition Globally (SPRING) project

Sienny, T. and Serli, W.(2010), The concern and awareness of consumers and food service operators towards food safety and food hygiene in small and medium restaurants in Surabaya, Indonesia, *International Food Research Journal* ,Vol.17, pp.641-650.

Subba rao G. M, Sudershan R. V., Rao Pratima , Vishnu Rao M., Polasa Kalpagam, "Food safety knowledge, attitudes and practices of mothers – findings from focus group studies in south India", Appetite 49(2007),

WHO(2000), The world Health Report, 2000. WHO, Geneva.

WFP (2015) Safety Nets: Cash With Nutrition Education Has Greatest Impact On Child Nutrition, https://www.wfp.org/news/news-release/safety-nets-cash-nutrition-education-has-greatest-impactchild-nutrition.

Van Dillen Sonja ME, Hiddink Gerrit J, Koelen Maria A, Graaf Cees de, and Woerkum Cees MJ van(2003), Understanding nutrition communication between health professionals and consumers: development of a model for nutrition awareness based on qualitative consumer research, *American Society for Clinical Nutrition* ,Vol. 77(suppl), pp.1065S–72s

Appendix -1

| | **Appendix -1Communalities of Baby's eating behavior** | | |
|---|---|---|---|
| S.No. | Variables/ items of behavior | Initial | Extraction |
| 1 | My baby seemed contented while feeding | 1.000 | .769 |
| 2 | My baby frequently wanted more milk than I provided | 1.000 | .844 |
| 3 | My baby loved milk | 1.000 | .905 |
| 4 | My baby had a big appetite | 1.000 | .782 |
| 5 | My baby finished feeding quickly | 1.000 | .745 |
| 6 | My baby became distressed while feeding | 1.000 | .859 |
| 7 | 13.7 My baby got full up easily | 1.000 | .654 |
| 8 | If allowed to, my baby would take too much milk | 1.000 | .813 |
| 9 | My baby took more than 30 minutes to finish feeding | 1.000 | .781 |
| 10 | My baby got full before taking all the milk I think he/she should have | 1.000 | .733 |
| 11 | My baby fed slowly | 1.000 | .710 |
| 12 | Even when my baby had just eaten well he/she was happy to feed again if offered | 1.000 | .734 |
| 13 | My baby found it difficult to manage a complete feed | 1.000 | .855 |
| 14 | My baby was always demanding a feed | 1.000 | .722 |
| 15 | My baby sucked more and more slowly during the course of a feed | 1.000 | .717 |
| 16 | If given the chance, my baby would always be feeding | 1.000 | .830 |
| 17 | My baby enjoyed feeding time | 1.000 | .634 |
| 18 | My baby could easily take a feed within 30 minutes of the last one | 1.000 | .694 |
| | Extraction Method: Principal Component Analysis. | | |

Appendix-2:Total value explained by PCA of Baby's Eating Behavior

| Component | Initial Eigenvalues | | | Rotation Sums of Squared Loadings | | |
|---|---|---|---|---|---|---|
| | Total | % of Variance | Cumulative % | Total | % of Variance | Cumulative % |
| 1 | 5.577 | 30.982 | 30.982 | 2.617 | 14.540 | 14.540 |
| 2 | 2.736 | 15.202 | 46.184 | 2.473 | 13.737 | 28.277 |
| 3 | 1.880 | 10.444 | 56.628 | 2.374 | 13.191 | 41.469 |
| 4 | 1.510 | 8.390 | 65.018 | 2.228 | 12.376 | 53.845 |
| 5 | 1.066 | 5.922 | 70.940 | 2.053 | 11.407 | 65.252 |
| 6 | 1.013 | 5.626 | 76.566 | 2.037 | 11.314 | 76.566 |
| 7 | .866 | 4.809 | 81.375 | | | |
| 8 | .727 | 4.038 | 85.414 | | | |
| 9 | .557 | 3.097 | 88.511 | | | |
| 10 | .518 | 2.876 | 91.387 | | | |
| 11 | .340 | 1.889 | 93.276 | | | |
| 12 | .316 | 1.755 | 95.031 | | | |
| 13 | .257 | 1.426 | 96.457 | | | |
| 14 | .204 | 1.132 | 97.589 | | | |
| 15 | .180 | 1.000 | 98.589 | | | |
| 16 | .116 | .644 | 99.233 | | | |
| 17 | .086 | .475 | 99.708 | | | |
| 18 | .053 | .292 | 100.000 | | | |

**Appendix-3: Rotated Component Matrix** of by PCA of Baby's Eating Behavior

| Items/ Variable | Component | | | | | |
|---|---|---|---|---|---|---|
| | 1 | 2 | 3 | 4 | 5 | 6 |
| 1 | -.103 | -.513 | .029 | .070 | .189 | .674 |
| 2 | -.167 | .028 | .122 | .891 | -.082 | -.004 |
| 3 | .328 | .094 | -.002 | .847 | .264 | -.040 |
| 4 | .280 | .768 | -.204 | .248 | .096 | -.022 |
| 5 | .815 | .197 | -.027 | .023 | -.102 | -.174 |
| 6 | -.354 | .177 | .059 | .117 | -.031 | .827 |
| 7 | .010 | -.031 | .793 | -.040 | .126 | .084 |
| 8 | .527 | .263 | -.135 | .634 | -.136 | .164 |
| 9 | -.404 | .164 | .010 | .127 | .730 | .202 |
| 10 | -.056 | -.385 | .589 | .090 | .435 | .195 |
| 11 | -.725 | -.006 | .222 | -.108 | .279 | .213 |
| 12 | .471 | .475 | .018 | .115 | -.410 | -.325 |
| 13 | -.009 | .149 | .334 | -.301 | .418 | .675 |
| 14 | .120 | .159 | -.824 | .014 | .000 | .063 |
| 15 | -.161 | -.208 | .300 | .012 | .746 | .035 |
| 16 | .021 | .893 | -.053 | .019 | -.076 | .152 |
| 17 | .520 | .409 | -.142 | .131 | -.393 | -.071 |
| 18 | .316 | -.117 | -.606 | -.273 | -.160 | -.336 |

Extraction Method: Principal Component Analysis.
Rotation Method: Varimax with Kaiser Normalization.

a. Rotation converged in 8 iterations.

**Appendix-4 Item-Total Statistics of Childs' eating behavior**

| Variable or Items of behavior | Scale Mean if Item Deleted | Scale Variance if Item Deleted | Corrected Item-Total Correlation | Cronbach's Alpha if Item Deleted |
|---|---|---|---|---|
| My child loves food | 65.350 | 153.105 | .201 | .762 |
| My child eats more when worried | 66.700 | 150.267 | .404 | .754 |
| My child has big appetite | 66.025 | 154.999 | .046 | .773 |
| My child finishes his/her meal quickly | 66.100 | 165.836 | -.275 | .798 |
| My child is interested in food | 65.300 | 161.395 | -.176 | .783 |
| My child is always asking for a drink | 65.375 | 145.266 | .328 | .751 |
| My child refuses new foods at first | 65.950 | 156.356 | -.014 | .780 |
| My child eats slowly | 65.650 | 143.259 | .252 | .757 |
| My child eats less when angry | 65.400 | 158.810 | -.079 | .784 |
| My child enjoys tasting new foods | 64.925 | 150.584 | .254 | .758 |
| My child eats less when s/he is tired | 65.500 | 156.462 | -.005 | .777 |
| My child is always asking for food | 66.375 | 150.292 | .234 | .759 |
| My child eats more when annoyed | 67.000 | 153.026 | .232 | .761 |
| If allowed to, my child would eat too much | 66.925 | 150.892 | .213 | .761 |
| My child eats more when anxious | 66.900 | 155.785 | .058 | .770 |
| My child enjoys a wide variety of foods | 65.225 | 153.666 | .133 | .766 |
| My child leaves food on his/her plate at the end of a meal | 65.225 | 144.281 | .283 | .754 |
| My child takes more than 30 minutes to finish a meal | 65.650 | 146.387 | .291 | .754 |
| Given the choice, my child would eat most of the time | 66.400 | 137.579 | .526 | .732 |
| My child looks forward to mealtimes | 66.275 | 149.230 | .264 | .757 |
| My child gets full before his/her meal is finished | 65.350 | 145.105 | .390 | .747 |
| My child enjoys eating | 65.250 | 149.269 | .388 | .753 |
| My child eats more when s/he is happy | 65.400 | 148.503 | .399 | .752 |
| My child is difficult to please with meals | 65.700 | 149.703 | .216 | .760 |
| My child eats less when upset | 65.800 | 147.446 | .239 | .758 |
| My child gets full up easily | 64.900 | 163.990 | -.215 | .796 |
| My child eats more when s/he has nothing else to do | 66.425 | 139.533 | .514 | .735 |
| Even if child is full ups/he finds room to eat his/her favorite food | 65.400 | 158.656 | -.081 | .788 |
| If given the chance, child drink continuously throughout the day | 66.350 | 143.362 | .548 | .740 |
| Child cannot eat a meal if s/he has had a snack just before | 65.050 | 140.818 | .461 | .739 |
| if given the chance my child would always be having a drink | 66.250 | 144.910 | .469 | .744 |
| My child is interested in tasting food s/he hasn't tasted before | 65.150 | 158.900 | -.079 | .783 |
| Child decides that s/he doesn't like a food, even without tasting it | 66.150 | 145.669 | .316 | .752 |
| If given the chance, child would always have food in his/her mouth | 67.025 | 155.307 | .061 | .770 |
| My child eats more and more slowly during the course of a meal | 65.900 | 141.579 | .484 | .739 |

Appendix-5: **Communalities of variables of Childs' eating behavior**

| **Communalities** | | |
|---|---|---|
| | Initial | Extraction |
| My child loves food | 1.000 | .798 |
| My child eats more when worried | 1.000 | .857 |
| My child has big appetite | 1.000 | .807 |
| My child finishes his/her meal quickly | 1.000 | .901 |
| My child is interested in food | 1.000 | .890 |
| My child is always asking for a drink | 1.000 | .945 |
| My child refuses new foods at first | 1.000 | .847 |
| My child eats slowly | 1.000 | .960 |
| My child eats less when angry | 1.000 | .775 |
| My child enjoys tasting new foods | 1.000 | .905 |
| My child eats less when s/he is tired | 1.000 | .851 |
| My child is always asking for food | 1.000 | .868 |
| My child eats more when annoyed | 1.000 | .856 |
| If allowed to, my child would eat too much | 1.000 | .807 |
| My child eats more when anxious | 1.000 | .911 |
| My child enjoys a wide variety of foods | 1.000 | .831 |
| My child leaves food on his/her plate at the end of a meal | 1.000 | .940 |
| My child takes more than 30 minutes to finish a meal | 1.000 | .906 |
| Given the choice, my child would eat most of the time | 1.000 | .866 |
| My child looks forward to mealtimes | 1.000 | .884 |
| My child gets full before his/her meal is finished | 1.000 | .931 |
| My child enjoys eating | 1.000 | .925 |
| My child eats more when s/he is happy | 1.000 | .898 |
| My child is difficult to please with meals | 1.000 | .900 |
| My child eats less when upset | 1.000 | .947 |
| My child gets full up easily | 1.000 | .970 |
| My child eats more when s/he has nothing else to do | 1.000 | .923 |
| Even if child is full ups/he finds room to eat his/her favorite food | 1.000 | .912 |
| If given the chance, child drink continuously throughout the day | 1.000 | .922 |
| Child cannot eat a meal if s/he has had a snack just before | 1.000 | .791 |
| if given the chance my child would always be having a drink | 1.000 | .888 |
| My child is interested in tasting food s/he hasn't tasted before | 1.000 | .842 |
| Child decides that s/he doesn't like a food, even without tasting it | 1.000 | .931 |
| If given the chance, child would always have food in his/her mouth | 1.000 | .900 |
| My child eats more and more slowly during the course of a meal | 1.000 | .851 |
| Extraction Method: Principal Component Analysis. | | |

**Appendix-5: Total Variance Explained by PCA of Childs' Eating Behavior**

| Component | Initial Eigenvalues | | | Extraction Sums of Squared Loadings | | |
|---|---|---|---|---|---|---|
| | Total | % of Variance | Cumulative % | Total | % of Variance | Cumulative % |
| 1 | 8.639 | 24.683 | 24.683 | 8.639 | 24.683 | 24.683 |
| 2 | 5.075 | 14.499 | 39.183 | 5.075 | 14.499 | 39.183 |
| 3 | 4.781 | 13.661 | 52.844 | 4.781 | 13.661 | 52.844 |
| 4 | 2.713 | 7.752 | 60.596 | 2.713 | 7.752 | 60.596 |
| 5 | 2.337 | 6.676 | 67.272 | 2.337 | 6.676 | 67.272 |
| 6 | 1.927 | 5.506 | 72.777 | 1.927 | 5.506 | 72.777 |
| 7 | 1.602 | 4.577 | 77.355 | 1.602 | 4.577 | 77.355 |
| 8 | 1.337 | 3.821 | 81.176 | 1.337 | 3.821 | 81.176 |
| 9 | 1.286 | 3.675 | 84.850 | 1.286 | 3.675 | 84.850 |
| 10 | 1.238 | 3.538 | 88.388 | 1.238 | 3.538 | 88.388 |
| 11 | .779 | 2.226 | 90.615 | | | |
| 12 | .733 | 2.094 | 92.709 | | | |
| 13 | .579 | 1.653 | 94.362 | | | |
| 14 | .479 | 1.368 | 95.730 | | | |
| 15 | .341 | .975 | 96.706 | | | |
| 16 | .269 | .768 | 97.474 | | | |
| 17 | .233 | .664 | 98.138 | | | |
| 18 | .177 | .505 | 98.643 | | | |
| 19 | .138 | .395 | 99.038 | | | |
| 20 | .102 | .292 | 99.330 | | | |
| 21 | .069 | .196 | 99.526 | | | |
| 22 | .056 | .160 | 99.686 | | | |
| 23 | .036 | .104 | 99.790 | | | |
| 24 | .028 | .079 | 99.870 | | | |
| 25 | .020 | .057 | 99.926 | | | |
| 26 | .012 | .033 | 99.959 | | | |
| 27 | .008 | .024 | 99.983 | | | |
| 28 | .004 | .011 | 99.994 | | | |
| 29 | .001 | .004 | 99.998 | | | |
| 30 | .001 | .002 | 100.000 | | | |
| 31 | .000 | .000 | 100.000 | | | |
| 32 | 5.788E-16 | 1.654E-15 | 100.000 | | | |
| 33 | 2.478E-16 | 7.081E-16 | 100.000 | | | |
| 34 | 1.658E-16 | 4.737E-16 | 100.000 | | | |
| 35 | -3.625E-16 | -1.036E-15 | 100.000 | | | |
| Extraction Method: Principal Component Analysis. | | | | | | |

Appendix 6: **Communalities of Variables of Adolescent behavior**

| Sl. No. of Variable | Initial | Extraction |
|---|---|---|
| 1 | 1.000 | .864 |
| 2 | 1.000 | .574 |
| 3 | 1.000 | .903 |
| 4 | 1.000 | .797 |
| 5 | 1.000 | .808 |
| 6 | 1.000 | .736 |
| 7 | 1.000 | .838 |
| 8 | 1.000 | .815 |
| 9 | 1.000 | .800 |
| 10 | 1.000 | .859 |
| 11 | 1.000 | .854 |
| 12 | 1.000 | .631 |
| 13 | 1.000 | .787 |
| 14 | 1.000 | .780 |
| 15 | 1.000 | .747 |
| 16 | 1.000 | .811 |
| 17 | 1.000 | .880 |
| 18 | 1.000 | .701 |
| 19 | 1.000 | .778 |
| 20 | 1.000 | .802 |
| 21 | 1.000 | .760 |
| 22 | 1.000 | .897 |
| 23 | 1.000 | .812 |

| Extraction Method: Principal Component Analysis. |
|---|

Appendix-7: **Total Variance Explained**

| Component | Initial Eigenvalues | | | Extraction Sums of Squared Loadings | | |
|---|---|---|---|---|---|---|
| | Total | % of Variance | Cumulative % | Total | % of Variance | Cumulative % |
| 1 | 6.350 | 27.607 | 27.607 | 6.350 | 27.607 | 27.607 |
| 2 | 2.770 | 12.042 | 39.649 | 2.770 | 12.042 | 39.649 |
| 3 | 2.265 | 9.846 | 49.495 | 2.265 | 9.846 | 49.495 |
| 4 | 1.871 | 8.135 | 57.631 | 1.871 | 8.135 | 57.631 |
| 5 | 1.538 | 6.685 | 64.316 | 1.538 | 6.685 | 64.316 |
| 6 | 1.255 | 5.456 | 69.772 | 1.255 | 5.456 | 69.772 |
| 7 | 1.112 | 4.836 | 74.608 | 1.112 | 4.836 | 74.608 |
| 8 | 1.074 | 4.670 | 79.278 | 1.074 | 4.670 | 79.278 |
| 9 | .882 | 3.836 | 83.114 | | | |
| 10 | .709 | 3.081 | 86.195 | | | |
| 11 | .622 | 2.705 | 88.900 | | | |
| 12 | .573 | 2.492 | 91.392 | | | |
| 13 | .492 | 2.140 | 93.532 | | | |
| 14 | .357 | 1.554 | 95.086 | | | |
| 15 | .259 | 1.127 | 96.213 | | | |
| 16 | .211 | .918 | 97.131 | | | |
| 17 | .197 | .858 | 97.989 | | | |
| 18 | .137 | .596 | 98.585 | | | |
| 19 | .130 | .564 | 99.149 | | | |
| 20 | .076 | .331 | 99.480 | | | |
| 21 | .065 | .281 | 99.760 | | | |
| 22 | .037 | .162 | 99.923 | | | |
| 23 | .018 | .077 | 100.000 | | | |

Appendix -8 - **Component Matrix**

| **Component Matrix[a]** | | | | | | | | |
|---|---|---|---|---|---|---|---|---|
| | Component | | | | | | | |
| | 1 | 2 | 3 | 4 | 5 | 6 | 7 | 8 |
| 1_lowfat | .672 | .091 | .136 | -.457 | -.078 | .261 | .212 | .241 |
| 2_eating | .507 | .101 | .079 | -.249 | -.258 | -.355 | -.158 | .146 |
| 3_dessert | -.299 | .786 | .118 | .054 | -.333 | -.254 | -.052 | .046 |
| 4_least | .782 | .168 | -.061 | .354 | -.107 | -.071 | -.076 | -.079 |
| 5_fat | .681 | -.150 | -.166 | -.185 | -.124 | -.129 | .264 | -.398 |
| 6_crisps | .513 | .046 | .482 | -.100 | .171 | .177 | -.137 | .387 |
| 7_sausages | .534 | -.082 | -.089 | .220 | -.262 | .432 | .476 | .086 |
| 8_pastries | -.268 | .777 | -.037 | -.222 | .008 | .093 | .279 | -.034 |
| 9_sugar | .568 | .135 | .071 | .201 | -.483 | .201 | .101 | -.360 |
| 10_vegetables | .590 | -.279 | -.375 | .262 | -.168 | -.049 | -.435 | -.067 |
| 11_home | .697 | -.356 | -.054 | -.273 | .258 | -.094 | .131 | -.266 |
| 12_meals | .380 | -.294 | .404 | .021 | -.263 | -.015 | -.050 | .405 |
| 13_fruit | .766 | -.015 | .006 | .129 | -.373 | -.036 | -.030 | .206 |
| 14_sweet | .012 | .419 | -.169 | -.018 | .223 | .679 | -.231 | -.106 |
| 15_salad | .419 | .109 | -.577 | -.325 | .219 | -.233 | .007 | .136 |
| 16_softdrink | .516 | .583 | -.352 | .074 | .178 | -.073 | -.197 | .020 |
| 17_butter | .180 | .180 | -.053 | .445 | .391 | -.321 | .544 | .249 |
| 18_chocolate | .412 | .177 | .293 | .581 | .258 | -.045 | -.035 | -.081 |
| 19_snack | .385 | -.051 | .623 | -.286 | .352 | .074 | -.111 | -.126 |
| 20_healthiest | .385 | .193 | .585 | .378 | .267 | -.081 | -.061 | -.225 |
| 21_cream | .457 | -.458 | -.406 | .126 | .359 | .122 | -.003 | .133 |
| 22_servings | .671 | .528 | -.318 | -.044 | .129 | .081 | -.124 | .160 |
| 23_diet | .608 | .205 | .270 | -.489 | .057 | -.188 | .018 | -.222 |
| Extraction Method: Principal Component Analysis. | | | | | | | | |
| a. 8 components extracted. | | | | | | | | |

**pendix**-9 **Communalities of variables of parents eating behavior**

| Variable No. | Initial | Extraction |
|---|---|---|
| 1 | 1.000 | .713 |
| 2 | 1.000 | .806 |
| 3 | 1.000 | .740 |
| 4 | 1.000 | .819 |
| 5 | 1.000 | .820 |
| 6 | 1.000 | .843 |
| 7 | 1.000 | .514 |
| 8 | 1.000 | .882 |
| 9 | 1.000 | .822 |
| 10 | 1.000 | .818 |
| 11 | 1.000 | .659 |
| 12 | 1.000 | .783 |
| 13 | 1.000 | .763 |
| 14 | 1.000 | .862 |
| 15 | 1.000 | .893 |
| 16 | 1.000 | .859 |
| 17 | 1.000 | .871 |
| 18 | 1.000 | .845 |
| 19 | 1.000 | .694 |
| 20 | 1.000 | .822 |
| 21 | 1.000 | .805 |
| 22 | 1.000 | .831 |
| 23 | 1.000 | .793 |
| 24 | 1.000 | .629 |
| 25 | 1.000 | .810 |
| 26 | 1.000 | .747 |
| 27 | 1.000 | .880 |
| Extraction Method: Principal Component Analysis. | | |

Appendix-10 : The total variance explained

| Component | Initial Eigenvalues | | | Extraction Sums of Squared Loadings | | |
|---|---|---|---|---|---|---|
| | Total | % of Variance | Cumulative % | Total | % of Variance | Cumulative % |
| 1 | 10.247 | 37.952 | 37.952 | 10.247 | 37.952 | 37.952 |
| 2 | 2.767 | 10.248 | 48.200 | 2.767 | 10.248 | 48.200 |
| 3 | 2.559 | 9.479 | 57.679 | 2.559 | 9.479 | 57.679 |
| 4 | 1.856 | 6.874 | 64.553 | 1.856 | 6.874 | 64.553 |
| 5 | 1.473 | 5.456 | 70.009 | 1.473 | 5.456 | 70.009 |
| 6 | 1.279 | 4.736 | 74.745 | 1.279 | 4.736 | 74.745 |
| 7 | 1.143 | 4.231 | 78.976 | 1.143 | 4.231 | 78.976 |
| 8 | .930 | 3.444 | 82.420 | | | |
| 9 | .873 | 3.233 | 85.653 | | | |
| 10 | .750 | 2.778 | 88.431 | | | |
| 11 | .544 | 2.013 | 90.444 | | | |
| 12 | .436 | 1.615 | 92.060 | | | |
| 13 | .363 | 1.344 | 93.403 | | | |
| 14 | .342 | 1.266 | 94.669 | | | |
| 15 | .324 | 1.199 | 95.868 | | | |
| 16 | .282 | 1.045 | 96.913 | | | |
| 17 | .207 | .765 | 97.678 | | | |
| 18 | .165 | .612 | 98.290 | | | |
| 19 | .123 | .457 | 98.748 | | | |
| 20 | .104 | .384 | 99.132 | | | |
| 21 | .075 | .278 | 99.410 | | | |
| 22 | .054 | .202 | 99.612 | | | |
| 23 | .038 | .142 | 99.754 | | | |
| 24 | .033 | .120 | 99.874 | | | |
| 25 | .021 | .079 | 99.953 | | | |
| 26 | .009 | .033 | 99.986 | | | |
| 27 | .004 | .014 | 100.000 | | | |

Appendix-11: Component Matrix

| **Component Matrix[a]** | | | | | | | |
|---|---|---|---|---|---|---|---|
| | Component | | | | | | |
| Variable number | 1 | 2 | 3 | 4 | 5 | 6 | 7 |
| 1 | .433 | -.208 | -.412 | .197 | .498 | .062 | .146 |
| 2 | .636 | -.240 | -.359 | -.038 | -.294 | .058 | .352 |
| 3 | .604 | .343 | -.244 | .059 | -.373 | .055 | -.226 |
| 4 | .623 | .364 | -.304 | .367 | -.192 | -.171 | .076 |
| 5 | .357 | .434 | .235 | -.530 | .178 | .348 | .123 |
| 6 | .624 | .132 | -.367 | -.321 | -.200 | -.394 | -.047 |
| 7 | .531 | -.031 | .298 | -.058 | -.027 | -.266 | -.260 |
| 8 | .770 | .016 | .286 | -.101 | -.202 | .392 | .052 |
| 9 | .072 | .199 | .787 | .207 | -.166 | .252 | -.157 |
| 10 | .650 | .506 | -.277 | .242 | -.020 | -.033 | .049 |
| 11 | .568 | -.414 | .073 | .169 | .026 | .162 | .324 |
| 12 | .552 | .397 | -.075 | .453 | .126 | .254 | -.171 |
| 13 | .757 | -.203 | -.009 | .258 | -.216 | .163 | -.096 |
| 14 | .693 | -.454 | -.152 | -.046 | .374 | .101 | .028 |
| 15 | .720 | -.332 | .109 | -.278 | -.267 | .168 | -.273 |
| 16 | .702 | -.423 | -.307 | .047 | .079 | .237 | -.168 |
| 17 | .653 | .441 | .220 | -.213 | .320 | .223 | .069 |
| 18 | .571 | -.541 | .346 | -.067 | -.020 | -.148 | .282 |
| 19 | .627 | .335 | -.297 | -.217 | -.208 | -.082 | .058 |
| 20 | .635 | .331 | .040 | -.456 | -.059 | .041 | .308 |
| 21 | .817 | -.246 | .169 | .185 | -.065 | -.069 | -.074 |
| 22 | .777 | .031 | .222 | .286 | -.053 | -.257 | .166 |
| 23 | .552 | .301 | .250 | .138 | .445 | -.253 | -.234 |
| 24 | .445 | .078 | .407 | .455 | .048 | -.106 | .198 |
| 25 | .573 | -.404 | .240 | -.257 | -.057 | -.226 | -.375 |
| 26 | .530 | .130 | .375 | -.231 | .192 | -.428 | .189 |
| 27 | .678 | -.014 | -.388 | -.178 | .380 | .035 | -.304 |
| Extraction Method: Principal Component Analysis. | | | | | | | |
| a. 7 components extracted. | | | | | | | |

**Appendix-12: Rotated component matrix**

| Rotated Component Matrix[a] | | | | | | | |
|---|---|---|---|---|---|---|---|
| | Component | | | | | | |
| Variable number | 1 | 2 | 3 | 4 | 5 | 6 | 7 |
| 1 | .148 | .260 | -.113 | -.018 | .767 | .101 | -.109 |
| 2 | .405 | .652 | .164 | .115 | .184 | -.138 | -.350 |
| 3 | .776 | .030 | .322 | .173 | .010 | -.048 | -.015 |
| 4 | .844 | .193 | -.025 | -.007 | .114 | .228 | -.072 |
| 5 | .042 | -.047 | .058 | .896 | .029 | .061 | .075 |
| 6 | .516 | .059 | .373 | .195 | .091 | .194 | -.591 |
| 7 | .160 | .088 | .517 | .093 | .010 | .449 | .047 |
| 8 | .314 | .499 | .421 | .526 | .042 | .028 | .282 |
| 9 | -.028 | .062 | .173 | .183 | -.409 | .209 | .737 |
| 10 | .811 | .073 | -.060 | .227 | .220 | .226 | -.004 |
| 11 | .058 | .732 | .136 | .062 | .290 | .093 | .077 |
| 12 | .656 | .023 | -.007 | .139 | .342 | .145 | .442 |
| 13 | .465 | .509 | .441 | .005 | .233 | .042 | .194 |
| 14 | .013 | .453 | .335 | .136 | .706 | .142 | -.084 |
| 15 | .170 | .371 | .802 | .252 | .129 | -.055 | .006 |
| 16 | .256 | .406 | .471 | .020 | .624 | -.128 | -.018 |
| 17 | .276 | .056 | .083 | .755 | .257 | .335 | .194 |
| 18 | -.173 | .729 | .375 | .075 | .081 | .355 | -.075 |
| 19 | .619 | .102 | .190 | .382 | .064 | .071 | -.329 |
| 20 | .316 | .247 | .123 | .740 | -.015 | .185 | -.255 |
| 21 | .323 | .511 | .498 | .029 | .234 | .344 | .130 |
| 22 | .446 | .499 | .202 | .071 | .073 | .571 | .079 |
| 23 | .261 | -.138 | .176 | .202 | .320 | .697 | .214 |
| 24 | .241 | .404 | -.041 | -.005 | -.018 | .522 | .365 |
| 25 | -.046 | .191 | .813 | .040 | .126 | .292 | -.085 |
| 26 | .033 | .188 | .180 | .370 | -.024 | .715 | -.172 |
| 27 | .321 | -.057 | .380 | .256 | .722 | .109 | -.177 |
| Extraction Method: Principal Component Analysis. Rotation Method: Varimax with Kaiser Normalization. | | | | | | | |
| a. Rotation converged in 14 iterations. | | | | | | | |

Appendix 13: **Appendix 13:Component Score Coefficient Matrix**

| | Component | | | | | | |
|---|---|---|---|---|---|---|---|
| | 1 | 2 | 3 | 4 | 5 | 6 | 7 |
| 1 | -.055 | .061 | -.201 | -.016 | .357 | .048 | -.016 |
| 2 | .078 | .298 | -.109 | .025 | -.076 | -.150 | -.180 |
| 3 | .259 | -.100 | .150 | -.037 | -.111 | -.141 | .031 |
| 4 | .268 | .036 | -.109 | -.130 | -.069 | .063 | -.051 |
| 5 | -.102 | -.029 | -.055 | .435 | .027 | -.093 | .051 |
| 6 | .116 | -.090 | .123 | -.033 | -.102 | .091 | -.340 |
| 7 | -.002 | -.131 | .227 | -.083 | -.059 | .183 | -.008 |
| 8 | .015 | .144 | .060 | .196 | -.077 | -.190 | .176 |
| 9 | .008 | .002 | .108 | .055 | -.166 | -.013 | .374 |
| 10 | .224 | -.028 | -.127 | -.005 | .018 | .038 | .003 |
| 11 | -.068 | .317 | -.143 | .022 | .052 | -.040 | .039 |
| 12 | .182 | -.091 | -.054 | -.023 | .147 | -.048 | .290 |
| 13 | .110 | .099 | .104 | -.091 | -.008 | -.119 | .141 |
| 14 | -.138 | .066 | -.004 | .032 | .287 | .007 | -.003 |
| 15 | -.013 | -.018 | .334 | .040 | -.052 | -.179 | .049 |
| 16 | -.001 | .018 | .128 | -.040 | .219 | -.174 | .076 |
| 17 | -.050 | -.052 | -.091 | .302 | .115 | .040 | .108 |
| 18 | -.158 | .278 | -.015 | -.005 | -.067 | .133 | -.101 |

| | |
|---|---|
| 1. Res. No. | Sex: Father =0 Mother=1 |
| 2. Age | |
| 3. Number of total children | |
| 4. Number of baby up to 4 year | |
| 5. No. Son baby up to 4 year | |
| 6. No. Daughter baby up to 4 year | |
| 7. Name | |
| 8. Contact no. | |
| 9. Res. Occupation | 1. Business |
| | 2. Service |
| | 3. both B&S |
| | 4. Others |
| 10. Annual Income of Family | |
| 11. Annual Expenditure of Family | |
| 12. Education (in years) | |

| | | | | | | | |
|---|---|---|---|---|---|---|---|
| 19 | .150 | -.016 | .008 | .088 | -.093 | -.037 | -.177 |
| 20 | -.011 | .105 | -.103 | .304 | -.106 | -.006 | -.169 |
| 21 | .028 | .072 | .108 | -.106 | -.003 | .071 | .061 |
| 22 | .072 | .141 | -.078 | -.099 | -.088 | .225 | -.024 |
| 23 | -.013 | -.236 | .038 | -.038 | .164 | .331 | .085 |
| 24 | .037 | .164 | -.159 | -.081 | -.056 | .212 | .132 |
| 25 | -.086 | -.138 | .381 | -.093 | -.019 | .097 | -.051 |
| 26 | -.106 | .018 | -.051 | .077 | -.073 | .366 | -.197 |
| 27 | -.014 | -.234 | .134 | .038 | .315 | -.004 | -.021 |

Extraction Method: Principal Component Analysis. Rotation Method: Varimax with Kaiser Normalization.

**Questionnaire Set: 1:**

**Baby Eating Behavior up to 4 years**

13. How would you describe your baby's feeding style at a day time feed?

| Items /variables | 1 Never | 2 Rarely | 3. Some-times | 4. often | 5. Always |
|---|---|---|---|---|---|
| 13.1 My baby seemed contented while feeding | | | | | |
| 13.2 My baby frequently wanted more milk than I provided | | | | | |
| 13.3 My baby loved milk | | | | | |
| 13.4 My baby had a big appetite | | | | | |
| 13.5 My baby finished feeding quickly | | | | | |
| 13.6 My baby became distressed while feeding | | | | | |
| 13.7 My baby got full up easily | | | | | |
| 13.8 If allowed to, my baby would take too much milk | | | | | |
| 13.9 My baby took more than 30 minutes to finish feeding | | | | | |
| 13.10 My baby got full before taking all the milk I think he/she should have | | | | | |
| 13.11 My baby fed slowly | | | | | |
| 13.12 Even when my baby had just eaten well he/she was happy to feed again if offered | | | | | |
| 13.13 My baby found it difficult to manage a complete feed | | | | | |
| 13.14 My baby was always demanding a feed | | | | | |
| 13.15 My baby sucked more and more slowly during the course of a feed | | | | | |
| 13.16 If given the chance, my baby would always be feeding | | | | | |
| 13.17 My baby enjoyed feeding time | | | | | |
| 13.18 My baby could easily take a feed within 30 minutes of the last one | | | | | |

**Set-2:Child Eating Behavior Questionnaire (5-9 Years)**

13. Please read the following statements and tick the boxes most appropriate to your child's eating behavior.

| | 1. Never | 2. Rarely | 3. Some-times | 4. often | 5. Always |
|---|---|---|---|---|---|
| 13.1 My child loves food | | | | | |
| 13.2 My child eats more when worried | | | | | |
| 13.3 My child has big appetite | | | | | |
| 13.4 My child finishes his/her meal quickly | | | | | |
| 13.5 My child is interested in food | | | | | |
| 13.6 My child is always asking for a drink | | | | | |
| 13.7 My child refuses new foods at first | | | | | |
| 13.8 My child eats slowly | | | | | |
| 13.9 My child eats less when angry | | | | | |
| 13.10 My child enjoys tasting new foods | | | | | |
| 13.11 My child eats less when s/he is tired | | | | | |
| 13.12 My child is always asking for food | | | | | |
| 13.13 My child eats more when annoyed | | | | | |
| 13.14 If allowed to, my child would eat too much | | | | | |
| 13.15 My child eats more when anxious | | | | | |
| 13.16 My child enjoys a wide variety of foods | | | | | |
| 13.17 My child leaves food on his/her plate at the end of a meal | | | | | |
| 13.18 My child takes more than 30 minutes to finish a meal | | | | | |
| 13.19 Given the choice, my child would eat most of the time | | | | | |
| 13.20 My child looks forward to mealtimes | | | | | |
| 13.21 My child gets full before his/her meal is finished | | | | | |
| 13.22 My child enjoys eating | | | | | |
| 13.23 My child eats more when s/he is happy | | | | | |
| 13.24 My child is difficult to please with meals | | | | | |
| 13.25 My child eats less when upset | | | | | |
| 13.26 My child gets full up easily | | | | | |
| 13.27 My child eats more when s/he has nothing else to do | | | | | |
| 13.28 Even if child is full finds room to eat his/her favorite food | | | | | |
| 13.29 If given the chance, drink continuously throughout the day | | | | | |
| 13.30 Child cannot eat a meal if s/he has had a snack just before | | | | | |
| 13.31 if given the chance my child would always be having a drink | | | | | |
| 13.32 My child is interested in tasting food s/he hasn't tasted before | | | | | |
| 13.33 Child decides that s/he doesn't like a food, without tasting | | | | | |
| 13.34 If given the chance, child always have food in his/her mouth | | | | | |
| 13.35 My child eats more and more slowly the course of a meal | | | | | |

Adolescent Food Habits (10-19 years)

| | | |
|---|---|---|
| Res. No. | | |
| 1. Sex | M | 0 |
| | F | 1 |
| 2. Res. Age | | |
| 3. Schooling in years | | |
| 4. Father's Schooling in years | | |
| 5. Mother's Schooling in years | | |
| 6. Res. Name | | |
| 7. Father's Profession | 1.Service | |
| | 2.Business | |
| | 3.Both | |
| | 4.Others | |
| 8. Mother's Profession | 1.House | |
| | 2.Service | |
| | 3.Both | |
| | 4.Business | |
| | 5.Others | |
| 9. Annual Income of Family | | |
| 10. Annual Expenditure of Family | | |
| 11. Res. Contact no. | | |

| | | Most like=2 | Neutral=0 | like=1 |
|---|---|---|---|---|
| 12.1_lowfat | If I am having lunch away from home, I often choose a low-fat option. | | | |
| 12.2_eating | I usually avoid eating fried foods. | | | |
| 12.3_dessert | I usually eat a dessert or pudding if there is one available. | | | |
| 12.4_least | I make sure I eat at least one serving of fruit a day. | | | |
| 12.5_fat | I try to keep my overall fat intake down. | | | |
| 12.6_crisps | If I am buying crisps, I often choose a low-fat brand. | | | |
| 12.7_sausages | I avoid eating lots of sausages and burgers. | | | |
| 12.8_pastries | I often buy pastries or cakes. | | | |
| 12.9_sugar | I try to keep my overall sugar intake down. | | | |
| 12.10_vegetables | I make sure I eat at least one serving of vegetables or salad a day. | | | |
| 12.11_home | If I am having a dessert at home, I try to have something low in fat. | | | |
| 12.12_meals | I rarely takeaway meals. | | | |
| 12.13_fruit | I try to ensure that I eat plenty of fruit and vegetables. | | | |
| 12.14_sweet | I often eat sweet snacks between meals. | | | |

| | | | |
|---|---|---|---|
| Res. No. | 1. Sex | M | 0 |
| | | F | 1 |
| 2. Occ_sp | 1.Household | | |
| | 2. Service | | |
| | 3. both service & household | | |
| | 4. Business | | |
| | 5. Others | | |
| 3. Res. Age | | | |
| 4. no_tc | | | |
| 5. occ. respondance | 1. Service | | |
| | 2. Business | | |
| | 3. both Service & Business | | |
| | 4. Others | | |
| 6. No. of total child up to 9 | | | |
| 7. No. of Son child up to 9 | | | |
| 8. No. of daughter child up to 9 | | | |
| 9. Res. Education years | | | |

| | | | | |
|---|---|---|---|---|
| 12.15_salad | I usually eat at least one serving of vegetables (excluding potatoes) or salad with my evening meal. | | | |
| 12.16_softdrink | When I am buying a soft drink, I usually choose a diet drink. | | | |
| 12.17_butter | When I put butter or margarine on bread, I usually spread it on thinly. | | | |
| 12.18_chocolate | If I have a packed lunch, I usually include some chocolate and/or biscuits. | | | |
| 12.19_snack | When I have a snack between meals, I often choose fruit. | | | |
| 12.20_healthiest | If I am having a dessert or pudding in a restaurant, I usually choose the healthiest one. | | | |
| 12.21_cream | I often have cream on desserts. | | | |
| 12.22_servings | I eat at least 3 servings of fruits most days. | | | |
| 12.23_diet | I generally try to have a healthy diet. | | | |

Parental Feeding Style to his Children up to 9 years-

| | |
|---|---|
| 10. Annual Income | |
| 11. Annual Expenditure | |
| 12. Education of spouse | |
| 13. Conjugal life | |

| Items/variable | | 1.Never | 2.Rarely | 3.Sometimes | 4.Often | 5.Always |
|---|---|---|---|---|---|---|
| I allow my child to choose which foods to have for meals | 14.1 choose | | | | | |
| I give my child something to eat to make him/her feel better when s/he is feeling upset | 14.2 upset | | | | | |
| I encourage my child to look forward to the meal | 14.3 encourage | | | | | |
| I praise my child if s/he eats what I give him/her | 14.4 praise | | | | | |
| I decide how many snacks my child should have | 14.5 decide | | | | | |
| I encourage my child to eat a wide variety of foods | 14.6 variety | | | | | |
| In order to get my child to behave him/herself I promise him/her something to eat | 14.7 order | | | | | |
| I present food in an attractive way to my child | 14.8 present | | | | | |
| If my child misbehaves I withhold his/her favorite food | 14.9 misbehaves | | | | | |
| I encourage my child to taste each of the foods I serve at mealtimes | 14.10 taste | | | | | |
| I allow my child to wander around during a meal | 14.11 wander | | | | | |
| I encourage my child to try foods that s/he hasn't tasted before | 14.12 try | | | | | |
| I give my child something to eat to make him/herfeel better when s/he has been hurt | 14.13 feel | | | | | |
| I let my child decide when s/he would like to have her meal | 14.14 like | | | | | |
| I give my child something to eat if s/he is feeling bored | 14.15 bored | | | | | |
| I allow my child to decide when s/he has had enough snacks to eat | 14.16 enough | | | | | |
| I decide when it is time for my child to have a snack | 14.17 time | | | | | |
| I use puddings as a bribe to get my child to eat his/her main course | 14.18 course | | | | | |
| I encourage my child to enjoy his/her food | 14.19 enjoy | | | | | |
| I decide the times when my child eats his/her meals | 14.20 meals | | | | | |
| I give my child something to eat if s/he is feeling worried | 14.21 btr_wry | | | | | |
| I reward my child with something to eat when s/he is well behaved | 14.22 reward | | | | | |
| I let my child eat between meals whenever s/he wants | 14.23 between | | | | | |
| I insist my child eats meals at the table | 14.24 insist | | | | | |
| I give my child something to eat to make him/her feel better when s/he is feeling angry | 14.25 angry | | | | | |
| I decide what my child eats between meals | 14.26 eats | | | | | |
| I praise my child if s/he eats a new food | 14.27 new_food | | | | | |

**Food preference questionnaire for adults (20-29 unmarried )**

9.

| Food Item | 1 Dislike a lot | 2. Dislike a little | 3. Neither like or dislike | 4. Like a little | 5. Like a lot |
|---|---|---|---|---|---|
| 9.1 Beef | | | | | |
| 9.2 Beef burgers | | | | | |
| 9.3 Lamb | | | | | |
| 9.4 Chicken | | | | | |
| 9.5 Bacon | | | | | |
| 9.6 Ham | | | | | |
| 9.7 Sausages | | | | | |
| 9.8 White fish (e.g. cod, haddock) | | | | | |
| 9.9 Oily fish (e.g. mackerel, kippers) | | | | | |
| 9.10 Smoked Salmon | | | | | |
| 9.11 Tinned Tuna | | | | | |
| 9.12 Eggs (boiled, scrambled or fried) | | | | | |
| 9.13 Baked beans | | | | | |
| 9.14 Bread or Bread rolls | | | | | |
| 9.15 Bran cereal (e.g. All Bran, Bran Flakes) | | | | | |
| 9.16 Porridge | | | | | |
| 9.17 Plain boiled rice | | | | | |
| 9.18 Sugared cereal (e.g. Frosties, Sugar Puffs) | | | | | |
| 9.19 Hummus | | | | | |
| 9.20 Wheat cereal (e.g. Weetbix, Shredded Wheat) | | | | | |
| 9.21 Potatoes (boiled or mashed) | | | | | |
| 9.22 Chips | | | | | |
| 9.23 Rice or corn cereal (e.g. corn Flakes, Rice Krispies) | | | | | |
| 9.24 Soft cheese (e.g. Camembert, Brie) | | | | | |
| 9.25 Hard cheese (e.g. cheddar) | | | | | |
| 9.26 Cottage Cheese | | | | | |
| 9.27 Plain, Low-fat yoghurt | | | | | |
| 9.28 Oranges | | | | | |
| 9.29 Grapes | | | | | |
| 9.30 Apples | | | | | |
| 9.31 Melon | | | | | |
| 9.32 Peaches | | | | | |
| 9.33 Apricots | | | | | |
| 9.34 Strawberries | | | | | |
| 9.35 Avocadoes | | | | | |
| 9.36 Spinach | | | | | |
| 9.37 Carrots | | | | | |
| 9.38 Green beans | | | | | |
| 9.39 Cucumber | | | | | |
| 9.40 Celery | | | | | |
| 9.41 Mushrooms | | | | | |
| 9.42 Parsnips | | | | | |
| 9.43 Peas | | | | | |
| 9.44 Sweetcorn | | | | | |
| 9.45 Broccoli | | | | | |
| 9.46 Salad leaves (e.g. lettuce) | | | | | |
| 9.47 Red peppers | | | | | |
| 9.48 Raw tomatoes | | | | | |
| 9.49 Beetroot | | | | | |
| 9.50 Brussels sprouts | | | | | |
| 9.51 Butter | | | | | |
| 9.52 Butter-like spreads (e.g. sunflower spread, Flora) | | | | | |
| 9.53 Cream | | | | | |
| 9.54 Mayonnaise | | | | | |
| 9.55 Plain biscuits (e.g. Digestives) | | | | | |
| 9.56 Chocolate biscuits | | | | | |
| 9.57 Cake | | | | | |
| 9.58 Ice cream | | | | | |
| 9.59 Custard | | | | | |
| 9.60 Chocolate | | | | | |
| 9.61 Crisps | | | | | |
| 9.62 Chewy gummy sweets (e.g. Haribo-style sweets, wine gums) | | | | | |

**Food preference questionnaire for adults (19-64) Married**16.

| Food Item | 1 Dislike a | 2. Dislike a | 3.Neither like | 4. Like a little | 5. Like a |
|---|---|---|---|---|---|

| | lot | little | or dislike | | lot |
|---|---|---|---|---|---|
| 16.1 Beef | | | | | |
| 16.2 Beef burgers | | | | | |
| 16.3 Lamb | | | | | |
| 16.4 Chicken | | | | | |
| 16.5 Bacon | | | | | |
| 16.6 Ham | | | | | |
| 16.7 Sausages | | | | | |
| 16.8 White fish (e.g. cod, haddock) | | | | | |
| 16.9 Oily fish (e.g. mackerel, kippers) | | | | | |
| 16.10 Smoked Salmon | | | | | |
| 16.11 Tinned Tuna | | | | | |
| 16.12 Eggs (boiled, scrambled or fried) | | | | | |
| 16.13 Baked beans | | | | | |
| 16.14 Bread or Bread rolls | | | | | |
| 16.15 Bran cereal (e.g. All Bran, Bran Flakes) | | | | | |
| 16.16 Porridge | | | | | |
| 16.17 Plain boiled rice | | | | | |
| 16.18 Sugared cereal (e.g. Frosties, Sugar Puffs) | | | | | |
| 16.19 Hummus | | | | | |
| 16.20 Wheat cereal (e.g. Weetbix, Shredded Wheat) | | | | | |
| 16.21 Potatoes (boiled or mashed) | | | | | |
| 16.22 Chips | | | | | |
| 16.23 Rice or corn cereal (e.g. corn Flakes, Rice Krispies) | | | | | |
| 16.24 Soft cheese (e.g. Camembert, Brie) | | | | | |
| 16.25 Hard cheese (e.g. cheddar) | | | | | |
| 16.26 Cottage Cheese | | | | | |
| 16.27 Plain, Low-fat yoghurt | | | | | |
| 16.28 Oranges | | | | | |
| 16.29 Grapes | | | | | |
| 16.30 Apples | | | | | |
| 16.31 Melon | | | | | |
| 16.32 Peaches | | | | | |
| 16.33 Apricots | | | | | |
| 16.34 Strawberries | | | | | |
| 16.35 Avocadoes | | | | | |
| 16.36 Spinach | | | | | |
| 16.37 Carrots | | | | | |
| 16.38 Green beans | | | | | |
| 16.39 Cucumber | | | | | |
| 16.40 Celery | | | | | |
| 16.41 Mushrooms | | | | | |
| 16.42 Parsnips | | | | | |
| 16.43 Peas | | | | | |
| 16.44 Sweetcorn | | | | | |
| 16.45 Broccoli | | | | | |
| 16.46 Salad leaves (e.g. lettuce) | | | | | |
| 16.47 Red peppers | | | | | |
| 16.48 Raw tomatoes | | | | | |
| 16.49 Beetroot | | | | | |
| 16.50 Brussels sprouts | | | | | |
| 16.51 Butter | | | | | |
| 16.52 Butter-like spreads (e.g. sunflower spread, Flora) | | | | | |
| 16.53 Cream | | | | | |
| 16.54 Mayonnaise | | | | | |
| 16.55 Plain biscuits (e.g. Digestives) | | | | | |
| 16.56 Chocolate biscuits | | | | | |
| 16.57 Cake | | | | | |
| 16.58 Ice cream | | | | | |
| 16.59 Custard | | | | | |
| 16.60 Chocolate | | | | | |
| 16.61 Crisps | | | | | |
| 16.62 Chewy gummy sweets (e.g. Haribo-style sweets, wine gums) | | | | | |